%% file: clac.tex
 \nonstopmode
\newcommand\Short[1]{#1} \newcommand\Long[1]{}
\newcommand{\Longer}[1]{} \newcommand{\Shorter}[1]{#1}

\Longer{
 \documentclass{Paper} 
 \appearedas{International Workshop on Classical Logic and Computation (CL\&C'08), 2008}
}

\Shorter{%%%%%%%%%%%%%%%%%%%%%%%%%%%%%%%%%%%%%
\documentclass{llncs}

\usepackage{mathpazo}\usepackage{times}
 \newskip \point \point1pt\newcommand{\Comment}[1]{}
 
\newif\ifcolour\colourfalse
\newif\ifmycolour\mycolourfalse

\def\arraystretch{1.05}
\arraycolsep 3pt

\spnewtheorem{Theorem}[definition]{Theorem}{\bf}{\it }
 \spnewtheorem{Property}[definition]{Property}{\bf}{\rm }
 \spnewtheorem{Proposition}[definition]{Proposition}{\it}{\rm }
 \spnewtheorem{Example}[definition]{Example}{\it}{\rm}
 \spnewtheorem{Definition}[definition]{Definition}{\bf}{\rm }
 \spnewtheorem{Corollary}[definition]{Corollary}{\it}{\it }
 \spnewtheorem{Lemma}[definition]{Lemma}{\bf}{\it }

 \renewenvironment{theorem}{\begin {Theorem}}{\end {Theorem}}
 \renewenvironment{lemma}{\begin {Lemma}}{\end {Lemma}}
 \renewenvironment{example}{\begin {Example}}{\end {Example}}
 \renewenvironment{definition}{\begin {Definition}}{\end {Definition}}

\newenvironment{lemma*}[1]{\trivlist \itemindent \labelsep \item[{\bf Lemma #1.}]}{}
\newenvironment{theorem*}[1]{\trivlist \itemindent \labelsep \item[{\bf Theorem #1.}]}{}

\def\@listI{\leftmargin\leftmargini \parsep 1.6pt plus 2pt minus 1pt%
\topsep 3pt plus 2pt%
\itemsep 2pt plus 2pt minus 1pt}
\newenvironment{Proof}{\begin{proof}}{\end{proof}}

}% Short %%%%%%%%%%%%%%%%%%%%%%%%%%%%%%%%%%%%%

\input{Xmacros}
\input{PiXmacros}

 \def \picolour{}
 \def \PiPair<#1,#2> {\mbox{\footnotesize $\langle$}{#1}{,}{#2}\mbox{\footnotesize $\rangle$}}%
 \def \PiOverline{\overline}
 \def \forwarder{{\Rightarrow}}
 \def \PiSem[#1]{[\kern-1.75\point[\,#1\,]\kern-1.75\point]}
 \def \SemL#1#2{\PiSem[{#1}]_{#2}}
 \def \PiSemL[#1]#2{\PiSem[{#1}]_{#2}}
 \def \PiC	{$\Red `p$-cal\-cu\-lus}
 
\def \redX {\,{\rightarrow}\kern-2\point_{\cal X}\,}
\renewcommand\ShiftL[2]{#2}

\begin{document}

\bibliographystyle {plain}

\title{From {\X} to $`p$}
\subtitle{Representing the Classical Sequent Calculus in the $`p$-calculus
\Short{\\[3mm]\normalsize Extended Abstract}}

\author{Steffen van Bakel$^1$ \and Luca Cardelli$^2$ \and Maria Grazia Vigliotti$^1$}

\institute{\small $1$: Department of Computing, Imperial College, 180 Queen's Gate, London SW7 2BZ, UK
 \\
$2$: Microsoft Research Cambridge, 7 J J Thomson Avenue, Cambridge, CB3 0FB, UK
 \\
 \email{svb@doc.ic.ac.uk,luca@microsoft.com,mgv98@doc.ic.ac.uk}
}

\date{}

\maketitle

\begin{abstract}
We study the $`p$-calculus, enriched with pairing and non-blocking input, and define a notion of type assignment that uses the type constructor $\arrow$.
We encode the circuits of the calculus $\X$ into this variant of $`p$, and show that all reduction (cut-elimination) and assignable types are preserved.
Since $\X$ enjoys the Curry-Howard isomorphism for Gentzen's calculus {\LK}, this implies that all proofs in {\LK} have a representation in $`p$. 
%We then enrich the logic with the connectors $\And, \Or$ and $\neg$, and show that also these can be represented in $`p$.

\end{abstract}

% \cite{Sangiorgi-Walker'03,Milner'99,Milner'92bis}

%\input{introduction}

\section*{Introduction}

In this paper we present an encoding of proofs of Gentzen's (implicative) {\LK} \cite{GentzenG'35} into the $`p$-calculus \cite{Milner'92} that respects {\Cut}-elimination, and define a new notion of type assignment for $`p$ so that processes will become witnesses for the provable formulae.
The encoding of classical logic into $`p$-calculus is attained by using the intuition of the calculus $\X$, which gives a computational meaning to {\LK} (a first version of this calculus was proposed in \cite{Urban'00,Urban-Bierman'01,Urban'01}; the implicative fragment of $\X$ was studied in \cite{Bakel-Lescanne'08}).

$\X$ enjoys the Curry-Howard property for {\LK}; it achieves the isomorphism by constructing witnesses, called \emph{nets}, for derivable sequents. 
Nets in $\X$ have multiple named inputs and multiple named outputs, that are collectively called \emph{connectors}.
Reduction in $\X$ is expressed via a set of rewrite rules that represent {\Cut}-elimination, eventually leading to renaming of connectors.
%% and gives computational meaning to classical (sequent) proof reduction. 
It is well known that {\Cut}-elimination in {\LK} is not confluent, and, since $\X$ is Curry-Howard for {\LK}, neither is reduction in $\X$. 
These two features --non-confluence and reduction as connection of nets via the exchange of names-- lead us to consider the $`p$-calculus as an alternative computational model for {\Cut}-elimination and proofs in {\LK}. 

The relation between process calculi and classical logic is an interesting and very promising area of research (similar attempts we made in the context of natural deduction \cite{Honda-Yoshida-Berger'04} and linear logic \cite{Bellin-Scott'94}). 
Our aim is to widen further the path to practical application of classical logic in computation by providing an interpretation of classical logic into process algebra, that fully exploits the non-determinism of both {\LK} and $`p$.

The aim of this paper is to link {\LK} and $`p$ via $\X$; the main achievements are:

 \begin{itemize}
 \item 
an encoding of $\X$ into $`p$ is defined, that preserves the operational semantics -- to achieve this result, reduction in $`p$ is generalised;
\item
we define a non-standard notion of type assignment for $`p$ (types do not contain channel information) that encompasses implication;
 \item 
the encoding preserves assignable types, effectively showing that all proofs in {\LK} have a representation in $`p$ -- to represent {\LK}, $`p$ is enriched with pairing \cite{Abadi-Gordon'97}.
\Long{%%%%%%%
 \item 
in addition to \cite{Bakel-Lescanne'08}, we treat the full classical logic, including the connectives $\arrow$, $\neg$, $\And$, and $\Or$, not only for $\X$, but also for $`p$.

\item we represent the connectives $\arrow$, $\neg$, $\And$, and $\Or$ in $`p$-calculus.
}%%% Long

 \end{itemize}

\subsection*{Classical sequents\Short{, {\X}, and $`p$}}
The \emph{sequent calculus} {\LK}, introduced by Gentzen in \cite{GentzenG'35}, is a logical system in which the rules only introduce connectives (but on either side of a sequent), in contrast to \emph{natural deduction} (also introduced in \cite{GentzenG'35}) which uses rules that introduce or eliminate connectives in the logical formulae.
Natural deduction normally derives statements with a single conclusion, whereas {\LK} allows for multiple conclusions, deriving sequents of the form $\Der { A_1,\ldots,A_n }{ B_1,\ldots,B_m }$, where $A_1,\ldots,A_n$ is to be understood as $A_1 {\wedge}\ldots{\wedge} A_n$ and $B_1,\ldots,B_m$ is to be understood as $B_1 {\vee}\ldots{\vee} B_m$. 
The version $G_3$ of Implicative {\LK} has four rules: \emph{axiom}, \emph{left introduction} of the arrow, \emph{right introduction}, and \emph{cut}.
 \[ \begin {array}{rlcrlcrlcrl}
(\Ax): &
 \Inf	{ \Der { `G, A }{ A, `D} }
&\quad &
(\ArrL): &
 \Inf	{ \Der { `G }{ A, `D }
	 \quad
	 \Der { `G, B }{ `D }
	}
	{ \Der { `G , A \Arrow B }{ `D } }
\\[4mm]
(\ArrR): &
 \Inf	{ \Der { `G, A }{ B, `D } }
	{ \Der { `G }{ A \Arrow B, `D } }
&\ &
(\textit{cut}): &
 \Inf	{ \Der { `G }{ A, `D }
	 \quad
	 \Der { `G, A }{ `D }
	}
	{ \Der { `G }{ `D } }
 \end {array} \]

Since {\LK} has only introduction rules, the only way to eliminate a connective is to eliminate the whole formula in which it appears via an application of the $(\Cut)$-rule.
Gentzen defined a procedure that eliminates all applications of the $(\Cut)$-rule from a proof of a sequent, generating a proof in \emph{normal form} of the same sequent, that is, without a cut. 
This procedure is defined via local reductions of the proof-tree, which has --with some discrepancies-- the flavour of term rewriting \cite{Klop'92} or the evaluation of explicit substitutions \cite{DeBruijn'78,Abadi-Cardelli-Curien'91}.
\Long{Indeed, the typing rule of an explicit substitution, say in {\Lx} \cite{Bloo-Rose'95}, is nothing but a variant of the $(\Cut)$-rule, and a lot of work has been done to better understand the connection between explicit substitutions and local cut-reduction procedures.}

%\subsection*{The principle of $\X$}
The calculus $\X$ achieves a Curry-Howard isomorphism\Short{, first discovered for Combinatory Logic~\cite{Curry-Feys'58},} for the proofs in {\LK} by constructing \emph{witnesses} (called \emph{nets}) for derivable sequents, without any notion of application.
In establishing the isomorphism for $\X$, similar to calculi like $\lmu$ \cite{Parigot'92} and $\lmmt$ \cite{Curien-Herbelin'00}, Roman names are attached to formulae in the left context, and Greek names for those on the right, and syntactic structure is associated to the rules.
\Long{%%%%%%%
Names on the left can be seen as inputs to the net, and names to the right as outputs; since multiple formulae can appear on both sides, this implies that a net can not only have more than one input, but also more than one output.
There are two kinds of names (connectors) in $\X$: \emph{sockets} (inputs, with Roman names, that are reminiscent of values) and \emph{plugs} (outputs, with Greek names, that are reminiscent of continuations), that}%% Long
\Short{These} 
correspond to \emph{variables} and \emph{co-variables}, respectively, in \cite{WadlerDual}, or, alternatively, to Parigot's $`l$- and $`m$-variables \cite{Parigot'92} (see also \cite{Curien-Herbelin'00}).

\Long{%%%%%%%%%%
In the construction of the witness, when in applying a rule a premise or conclusion disappears from the sequent, the corresponding name gets bound in the net that is constructed, and when a premise or conclusion gets created, a different free (often new) name is associated to it.
For example, in the creation of the net for right introduction of the arrow
 \[ 
\Inf	{ \derX P : `G,x{:}A |- `a{:}B,`D }
	{ \derX \exp x P `a . `b : `G |- `b{:}A\arr B`D }
%\ExpRule{x}{P}{`a}{`b}{A}{B}{`G}{`D} \\ % &&
 \]
the input $x$ and the output $`a$ are bound, and $`b$ is free.
This case is interesting in that it highlights a special feature of $\X$, not found in other calculi.
In (applicative) calculi related to natural deduction, like the $`l$-calculus, only inputs are named, and the linking to a term that will be inserted is done via $`l$-abstraction and application.
The output (i.e.~result) on the other hand is anonymous; where a term `moves to' carries a name via a variable that acts as a pointer to the positions where the term is to be inserted, but where it comes from is not mentioned, since it is implicit.
Since a net $P$ can have many inputs and outputs, it is unsound to consider $P$ a function; however, fixing \emph{one} input $x$ and \emph{one} output $`a$, we can see $P$ as a function `from $x$ to $`a$'.
We make this limited view of $P$ available via the output $`b$, thereby \emph{exporting} `$P$ is a function from $x$ to $`a$'. 
The types given to the connectors confirm this view.
}%%%% Long
Gentzen's proof reductions by cut-elimination become the fundamental principle of computation in {\X}.
Cuts in proofs are witnessed by $\cut P `a + x Q $ (called the {\Cut} of $P$ and $Q$ via $`a$ and $x$), and the reduction rules specify how to remove them\Long{: a term is in normal form if and only if it has no sub-term of this shape}. 
\Long{%%%%%%
The intuition behind reduction is: the cut $\cut P `a + x Q $ expresses the intention to connect all $`a$s in $P$ and $x$s in $Q$, and reduction will realise this by either connecting all $`a$s to all $x$s (if $x$ 
does not exist in $Q$, $P$ will disappear), or all $x$s to all $`a$s (if $`a$ does not exist in $P$, $Q$ will disappear).
}%%% Long
Since {\Cut}-elimination in {\LK} is not confluent, neither is reduction in $\X$; for example, %as suggested above, 
when $P$ does not contain $`a$ and $Q$ does not contain $x$, reducing $\cut P `a + x Q $ can lead to both $P$ and $Q$\Long{, two different nets}.
\Long{%%%%%%%
 \[ \begin{array}{ccccc}
P &\worra & \cut P `a + x Q & \arrow & Q
 \end{array} \]
}% Long %%%%%%
Reduction in $\X$ boils down to \emph{renaming}: \Long{since the calculus is substitution-free, }during reduction nets are re-organised, creating nets that are similar, but with different connector names inside.
\Long{%%%%%
As an illustration of reduction in {\X}, consider the following reductions (a net $\caps<`o,`o>$ is a witness for the axiom in {\LK}):
 \[ \begin{array}{rcl}
\cut \caps<x,`a> `a + y \caps<y,`b> &\redX& \caps<x,`b>
\\
\cut \caps<x,`a> `g + y \caps<y,`b> &\redX& \caps<x,`a>
\\
\cut \caps<x,`a> `a + z \caps<y,`b> &\redX& \caps<y,`b>
\\
\cut \caps<x,`a> `g + z \caps<y,`b> &\redX& 
 \left \{ \begin{array}{l}
\caps<x,`a>
\\
\caps<y,`b>
 \end{array} \right.
 \end{array} \]
% It is not difficult to see that the reductions above correspond to proof with application of cut in {\LK}.
Notice the change, between the various lines, in connector names involved in the ${\Cut}$, and that the last variant has \emph{two} outcomes, underlining the non-confluence of $\X$.
}%%%% Long

% \subsection*{Capturing $\X$ in $`p$}
$\X$'s notion of multiple inputs and outputs is also found in $`p$, and was the original inspiration for our research. 
The aim of this work is to find a simple and intuitive encoding of {\LK}-proofs in $`p$, and to devise a notion of type assignment for $`p$ so that the types in $\X$ are preserved in $`p$. 
In this precise sense we view processes in $`p$ as giving an alternative computational meaning to proofs in classical logic.
\Long{To achieve this goal, we made full use of the view of $\X$-nets sketched above. }%
Clearly this implies that we had to define a notion of type assignment that uses the type constructor $\arrow$ for $`p$; we managed this without having to linearise the calculus as done in \cite{Honda-Yoshida-Berger'04}, and this is one of the contributions of this paper.

Although the calculi $\X$ and $`p$ are, of course, essentially different, the similarities go beyond the correspondence of inputs and output between nets in $\X$ and processes in $`p$.
Like $\X$, $`p$ is application free, and substitution only takes place on \emph{channel names}, similar to the renaming feature of $\X$, so {\Cut}-elimination is similar to synchronisation.

\Long{As discussed above, when creating a witness for $(\ArrR)$ (the net $\exp x P `a . `b $, called an \emph{export}), the exported interface of $P$ is the functionality of `receiving on $x$, sending on $`a$', which is made available on $`b$.
When encoding this behaviour in $`p$, we are faced with a problem.
It is clearly not sufficient to limit communication to the exchange of single names, since then we would have to separately send $x$ and $`a$, breaking perhaps the exported functionality, and certainly disabling the possibility of assigning arrow types.
We overcome this problem by sending out a pair of names, as in $\Out{a} <{ \PiPair<v,`d> }> $.
Similarly, when interpreting a witness for $(\ArrL)$ (the net $\imp P `a [x] y Q $, called an \emph{import}), the circuit that is to be connected to $x$ is ideally a function whose input will be connected to $`a$, and its output to $y$. 
This means that we need to receive a pair of names over $x$, as in $\In{x}({\PiPair<v,`d>}) . P $.

A cut $\cut P `a + x Q $ in $\X$ expresses two nets that need to be connected via $`a$ and $x$.
If we model $P$ and $Q$ in $`p$, then we obtain one process sending on $`a$, and one receiving on $x$, and we need to link these via $\In {`a} . \Out{x} $.
Since each output on $`a$ in $P$ takes place only once, and $Q$ might want to receive in more than one $x$, we need to replicate the sending; likewise, since each input $x$ in $Q$ takes place only once, and $P$ might have more than one send operation on $`a$, $Q$ needs to be replicated.

Considering the reduction in $\X$ shown above, we are able to show that (where $\bisimilarity$ is a simulation relation, defined in Definition~\ref{bisimilarity}):
 \[ \begin{array}{rcl}
\Scut \caps<x,`a> `a + y \caps<y,`b> &\antibisimilarity & \Scaps<x,`b>
\\
\Scut \caps<x,`a> `g + y \caps<y,`b> &\antibisimilarity & \Scaps<x,`a>
\\
 \end{array}
 \quad\quad
 \begin{array}{rcl}
\Scut \caps<x,`a> `a + z \caps<y,`b> &\antibisimilarity & \Scaps<y,`b>
\\
\Scut \caps<x,`a> `g + z \caps<y,`b> &\antibisimilarity & \Scaps<x,`a>
 \mid \Scaps<y,`b>
 \end{array} \]
The last alternative represents the fact that the two possible reductions in $\X$ are represented by the composition of their translation.  

To the best of our knowledge, this is the first time our result, i.e.~that the $`p$-calculus is an adequate computational model for {\LK}, is presented.
}%%%% Long

\subsection*{Related work}
\Long{%%%% 
The relation between \emph{logic} and \emph{computation} hinges around the Curry-Howard isomorphism (sometimes also attributed to De Bruijn), which expresses the fact that, for certain calculi with a notion of types, one can find a corresponding logic such that it is possible to associate terms with proofs in such a way that types become propositions, and proof contractions become term reductions (or computations).
This phenomenon was first discovered for Combinatory Logic~\cite{Curry-Feys'58}. 
}%%%% Long

In the past, say before Herbelin's PhD \cite{Herbelin'95} and Urban's
PhD \cite{Urban'00}, the study of the relation between computation, programming languages and logic has concentrated mainly on \emph{natural deduction systems} (of course, exceptions exist \cite{LL,Girard'91}).
In fact, these carry the predicate `\emph{natural}' deservedly; in comparison with, for example, \emph{sequent style systems}, natural deduction systems are easy to understand and reason about.
This holds most strongly in the context of \emph{non-classical} logics; for example, the Curry-Howard relation between \emph{Intuitionistic Logic} and the \emph{Lambda Calculus} (with types) is well studied and understood, and has resulted in a vast and well-investigated area of research, resulting in, amongst others, functional programming languages and much further to system~\textsf{F} \cite{Girard'86} and the Calculus of
Constructions~\cite{Coquand-Huet'88}.
Abramsky \cite{Abramsky'93,Abramsky'94} has studied correspondence between multiplicative linear logic and processes, and later moved to the context of game semantics \cite{Abramsky-Jagadeesan'94}.
In fact, all the calculi are \emph{applicative} in that abstraction and application (corresponding to arrow introduction and elimination) are the main constructors in the syntax.
The link between Classical Logic and continuations and control was first established for the $\lambda_C$-Calculus \cite{Griffin'90} (where $C$ stands for Felleisen's $C$ operator).

The introduction-elimination approach is easy to understand and convenient to use, but is also rather restrictive: for example, the handling of negation is not as nicely balanced, as is the treatment of contradiction (normally represented by the type $\Bottom$; for a detailed discussion, see \cite{Summers'08}).
This imbalance can be observed in Parigot's {\lmu}-calculus \cite{Parigot'92}, an approach for representing classical proofs via a natural deduction system in which there is one main conclusion that is being manipulated and possibly several alternative ones.
Adding $\Bottom$ as pseudo-type (only negation, or $A \arrow \Bottom$, is expressed; $\Bottom \arrow A$ is not a type), the {\lmu}-calculus corresponds to \emph{minimal classical logic} \cite{Ariola-Herbelin'03}.

Herbelin has studied the calculus $\lmmt$ as a non-applicative extension of $\lmu$, which gives a fine-grained account of manipulation of sequents \cite{Herbelin'95,Curien-Herbelin'00,Herbelin'05}.
The relation between call-by-name and call-by-value in the fragment of {\LK} with negation and conjunction is studied in the Dual Calculus \cite{WadlerDual}; as in calculi like $\lmu$ and $\lmmt$, that calculus considers a logic with \emph{active} formulae, so these calculi do not achieve a direct Curry-Howard isomorphism with {\LK}.
The relation between $\X$ and $\lmmt$ has been investigated in \cite{vBLL'05,Bakel-Lescanne'08}; there it was shown that it is straightforward to map $\lmmt$-terms into $\X$ whilst preserving reduction, but that it is not possible to do the converse.

The $`p$-calculus is equipped with a rich type theory \cite{Sangiorgi-Walker'03}: from the basic type system for counting the arity of channels to sophisticated linear types in \cite{Honda-Yoshida-Berger'04}, which studies a relation between Call-by-Value $\lmu$ and a linear $`p$-calculus.
Linearisation is used to be able to achieve processes that are functions, by allowing output over one channel name only. 
Moreover, the encoding presented in \cite{Honda-Yoshida-Berger'04} is type dependent, in that, for each term, there are different $`p$-processes assigned, depending on the original type; this makes the encoding quite cumbersome. 
By contrast, our encoding is very simple and intuitive by interpreting the cut operationally as a communication. 
The idea of giving a computational interpretation of the cut as a communication primitive is also used by \cite{Abramsky'94} and \cite{Bellin-Scott'94}. 
In both papers, only a small fragment of Linear Logic was considered, and the encoding between proofs and $`p$-calculus was left rather implicit. 

The type system presented in this paper differs quite drastically from the standard type system presented in \cite{Sangiorgi-Walker'03}: here input and output channels essentially have the type of the data they are sending or receiving, and are separated by the type system by putting all inputs with their types on the left of the sequent, and the outputs on the right. 
In our paper, types give a logical view to the $`p$-calculus rather than an abstract specification on how channels should behave.

\Long{%%%%%%%%%%%%%%
\subsection*{Overview of this paper}
In Section~\ref{sect:Xsyntax} we briefly repeat the definitions of (implicative) $\X$, followed in Section~\ref{types section} by the notion of type assignment which establishes the Curry-Howard isomorphism.
In Section~\ref{pi with pairing} we present the $`p$-calculus with pairing, and in Section~\ref{tas for pi} a notion of type assignment for this calculus.
We then define the encoding from $\X$ to $`p$ in Section~\ref{encoding}, followed in Section~\ref{lambda calculus} by a brief application of our results to the $`l$-calculus
.
Then, in Section~\ref{extension} we look at how to represent the other connectives in $\X$, and study the relation between these representations and reduction.
We conclude in Section~\ref{other into pi} by extending the syntax of names in $`p$ to elegantly represent the other connectives directly in $`p$.
}%%%%%%%% Long

 \def\Peirce{
 \Inf	[\Exp]
	{ \Inf [\Imp]
		{ \Inf [\Exp]
			{ \Inf [\Cap]
				{ \DerX { \caps<y,`d> }{ \stat{y}{A} }{ \stat{`d}{A}, \stat{`h}{B} } }
			}
			{ \DerX { \exp y \caps<y,`d> `h . `a }{ }{ \stat{`a}{A\arr B}, \stat{`d}{A} } }
%		  \quad
		  \Inf	[\Cap]
			{ \DerX { \caps<w,`d> }{ \stat{w}{A} }{ \stat{`d}{A} } }
		}
		{ \DerX { \imp { \exp y \caps<y,`d> `h . `a } `a [z] w \caps<w,`d> }{ \stat{z}{(A \arr B)\arr A} }{ \stat{`d}{A} } }
	}
	{ \DerX {
 \exp z { \imp { \exp y \caps<y,`d> `h . `a } `a [z] w \caps<w,`d> } `d . `g }{ }{ \stat{`g}{((A \arr B)\arr A)\arr A} }
	}
}

\section{The calculus $\X$} \label{sect:Xsyntax}

In this section we will give the definition of the $\X$-calculus which has been proven to be a fine-grained implementation model for various well-known calculi \cite{vBLL'05}, like the $`l$-calculus \cite{Barendregt'84}, $\lmu$ \cite{Parigot'92} and $\lmmt$ \cite{Herbelin'05}.
As discussed in the introduction, the calculus {\X} is inspired by the sequent calculus; the system we will consider in this section has only implication, no structural rules and a changed axiom\Long{; we will consider the other connectives in Section~\ref{other connectives}}.
$\X$ features two separate categories of `connectors',\, \emph{plugs} and \emph {sockets}, that act as input and output channels, and is defined without any notion of substitution or application.

 \begin {definition}[Syntax]
The nets of the $\X$-calculus are defined by the following syntax, where $x,y$ range over the infinite set of {\em sockets}, $`a, `b$ over the infinite set of {\em plugs}.
 \[ \begin{array}{ccccccccc}
P,Q &::=& \caps<x,`a> &\mid& \exp y P `b . `a &\mid& \imp P `b [y] x Q &\mid& \cut P `a + x Q 
\\
&& \emph {capsule} && \emph {export} && \emph {import}  && \emph {cut}
 \end{array} \]
 \end {definition}
The $\hat{`.}$ symbolises that the socket or plug underneath is bound in the net. 
%\Long
{%%%%%
The notion of bound and free connector (free sockets $\FS{P}$, and free plugs $\FP{P}$, respectively, and $\FC{P} =\FS{P}\cup \FP{P}$) is defined as usual, and we will identify nets that only differ in the names of bound connectors, as usual.
We accept Barendregt's convention on names, which states that no name can occur both free \emph{and} bound in a context; $`a$-conversion is supposed to take place silently, whenever necessary.
}%%%%% Long

The calculus, defined by the reduction rules below, explains in detail how cuts are propagated through nets to be eventually evaluated at the level of capsules, where the renaming takes place.
Reduction is defined by specifying both the interaction between well-connected basic syntactic structures, and how to deal with propagating active nodes to points in the net where they can interact. 

It is important to know when a connector is introduced, i.e.\ is connectable, i.e.\ is exposed and unique; this will play an important role in the reduction rules. 
Informally, a net $P$ introduces a socket $x$ if $P$ is constructed from sub-nets which do not contain $x$ as free socket, so $x$ only occurs at the ``top level.''\, 
This means that $P$ is either an import with a middle connector $[x]$ or a capsule with left part $x$. 
Similarly, a net introduces a plug $`a$ if it is an export that ``creates'' $`a$ or a capsule with right part $`a$.

 \begin {definition}%[Introduction]
\begin {description}
 \item [$P$ introduces $x$]
Either $P = \imp Q `b [x] y R $ with $x \notin \FS{Q,R}$, or $P = \caps<x,`a>$.
 \item [$P$ introduces $`a$]
Either $P = \exp x Q `b . `a $ and $`a \notin \FP{Q}$, or $P = \caps<x,`a>$.
 \end {description}
 \end {definition}
 
The principal reduction rules are:

 \begin {definition}[Logical rules]
Let $`a$ and $x$ be introduced in, respectively, the left- and right-hand side of the main cuts below.
 \[ \begin {array}{rrcll}
 (\Cap): &
\cut \caps<y,`a> `a + x \caps<x,`b> &\redX& \caps<y,`b>
\\
 (\Exp): &
\cut { \exp y P `b . `a } `a + x \caps<x,`g> &\redX&
 \exp y P `b . `g \ % & ITRS
% & `a \notin \FP{P}
\\
 (\Imp): &
\cut \caps<y,`a> `a + x { \imp Q `b [x] z R } &\redX&
 \imp Q `b [y] z R \ % & ITRS
% & x \notin \FS{Q,P}
\\
 (\Ins): &
\cut { \exp y P `b . `a } `a + x { \imp Q `g [x] z R }
 &\redX&
\begin{cases}
 \cut Q `g + y { \cut P `b + z R } \\
 \cut { \cut Q `g + y P } `b + z R
\end{cases}
 \end {array} \]
 \end {definition}

The first three logical rules above specify a renaming procedure, whereas the last rule specifies the basic computational step: it links the export of a function, available on the plug $`a$, to an adjacent import via the socket $x$.
The effect of the reduction will be that the exported function is placed in-between the two sub-terms of the import, acting as interface.
Notice that two cuts are created in the result, that can be grouped in two ways; these alternatives do not necessarily share all normal forms (reduction is non-confluent, so normal forms are not unique).

In $\X$ there are in fact two kinds of reduction, the one above, and the one which  defines  how to reduce a cut when one of its sub-nets does not introduce a connector mentioned in the cut. 
This will involve moving the cut inwards, towards a position where the connector \emph{is} introduced.
In case both connectors are not introduced, this search can start in either direction, indicated by the tilting of the dagger.

 \begin {definition}[Active cuts]
The syntax is extended with two {\em flagged} or {\em active} cuts:
 \[ P ::= \ldots \mid \cutL P_1 `a + x P_2 \mid \cutR P_1 `a + x P_2 \]
\Long{Terms constructed without these flagged cuts are called \emph {pure}.}

We define two cut-activation rules.
\[ \begin {array}{llcll}
 (\actL): & \cut P `a + x Q &\redX& \cutL P `a + x Q 
& \textit{if $P$ does not introduce $`a$}
\\
 (\actR): & \cut P `a + x Q &\redX& \cutR P `a + x Q 
&\textit{if $Q$ does not introduce $x$}
 \end {array} \] %$ \\
 \end {definition}

The next rules define how to move an activated dagger inwards.

 \begin {definition}[Propagation rules]
Left propagation:
\[ \kern-1mm \begin {array}{rrcll}
 (\deactL): & \cutL \caps<y,`a> `a + x P &\redX& \cut \caps<y,`a> `a + x P 
\\
 (\Li): & \cutL \caps<y,`b> `a + x P &\redX& \caps<y,`b> & `b \not= `a 
\\
 (\Lii): & \cutL { \exp y Q `b . `a } `a + x P &\redX&
\ShiftL{4}{\cut { \exp y { \cutL Q `a + x P } `b . `g } `g + x P & `g\textit{ fresh} }
\\
 (\Liii): & \cutL { \exp y Q `b . `g } `a + x P &\redX&
	\exp y { \cutL Q `a + x P } `b . `g  & `g \not= `a
\\
 (\Liv): & \cutL { \imp Q `b [z] y R } `a + x P &\redX&
\ShiftL{4}{\imp { \cutL Q `a + x P } `b [z] y { \cutL R `a + x P } }
\\
(\Lv): & \cutL { \cut Q `b + y R } `a + x P &\redX&
\ShiftL{4}{\cut { \cutL Q `a + x P } `b + y { \cutL R `a + x P } }
 \end {array} \]

\noindent Right propagation:
\[ \begin {array}{rlcll}
 (\deactR): & \cutR P `a + x \caps<x,`b> &\redX& \cut  P `a + x \caps<x,`b> 
\\
 (\Ri): & \cutR P `a + x \caps<y,`b> &\redX& \caps<y,`b>  & y \not= x
\\
 (\Rii): & \cutR P `a + x { \exp y Q `b . `g } &\redX& \exp y { \cutR P `a + x Q } `b . `g 
\\
 (\Riii): & \cutR P `a + x { \imp Q `b [x] y R } &\redX&
\\ \multicolumn{4}{r}{
	\cut P `a + z { \imp { \cutR P `a + x Q } `b [z] y { \cutR P `a + x R } },
}
 & z\textit{ fresh}
\\
 (\Riv): & \cutR P `a + x { \imp Q `b [z] y R } &\redX& 
\imp { \cutR P `a + x Q } `b [z] y { \cutR P `a + x R }  & z \not= x 
\\
 (\Rv): & \cutR P `a + x { \cut Q `b + y R } &\redX&
\ShiftL{4}{\cut { \cutR P `a + x Q } `b + y { \cutR P `a + x R } }
 \end {array} \]

We write $\redX$ for the (reflexive, transitive, compatible) reduction relation generated by the logical, propagation and activation rules.
 \end {definition}

The reduction $\redX$ is not confluent; \Long{this comes in fact from the critical pair that activates a cut $\cut P `a + x Q $ in two ways. 
C}%
\Short{c}onfluent sub-\Long{reduction }%
systems are defined in \cite{Bakel-Lescanne'08}. \\

\noindent
\emph{Summarising}, reduction brings all cuts down to logical cuts where both connectors single and introduced, or elimination cuts that are cutting towards a capsule that does not contain the relevant connector.
Cuts towards connectors occurring in capsules lead to renaming ($\cutR P `a + x \caps<x,`b> \redX P[`b/`a] $ and $\cutL \caps<z,`a> `a + x P \redX P[z/x]$), and towards non-occurring connectors leads to elimination ($\cutR P `a + x \caps<z,`b> \redX \caps<z,`b>$ and $\cutL \caps<z,`b> `a + x P \redX$ $\caps<z,`b> $).

\Long{%%%%%%%%%%
In \cite{Bakel-Lescanne'08}, two sub-reduction systems were introduced which explicitly favour one kind of activation whenever the above critical pair occurs:

 \begin {definition}[Call By Name and Call By Value]
We define Call By Name and Call By Value reduction by:
 \begin {itemize} \itemsep0pt
 \item
If a cut can be activated in two ways, the {\CBV} strategy only allows to activate it via $(\actL)$; we write $P \redCBV Q$ in that case.
We can reformulate this as the reduction system obtained by replacing rule $(\actR)$ by:
 \[ \def\arraystretch{1} \begin {array}{llcll}
 (\actR): & \cut P `a + x Q &\redX& \cutR P `a + x Q , &
 \textit{if $P$ introduces $`a$ and }
\Short{ \\&&&&}
 \textit{$Q$ does not introduce $x$}.
\end {array} \]

 \item
The {\CBN} strategy can only activate such a cut via $(\actR)$; like above, we write $P \redCBN Q$.
Likewise, we can reformulate this as the reduction system obtained by replacing rule $(\actL)$ by:
 \[ \def\arraystretch{1} \begin {array}{llcll}
 (\actL): & \cut P `a + x Q &\redX& \cutL P `a + x Q , &
 \textit{if $P$ does not introduce $`a$ }
\Short{ \\&&&&}
\textit{and $Q$ introduces $x$}.
\end {array} \]

 \item
We split the two variants of $(\Ins)$ over the two notions of reduction:
 \[ \begin{array}{rcl}
\cut { \exp y P `b . `a } `a + x { \imp Q `g [x] z R } &\red& \cut Q `g + y { \cut P `b + z R } 
 \end{array} \]
for {\CBV}, and
 \[ \begin{array}{rcl}
\cut { \exp y P `b . `a } `a + x { \imp Q `g [x] z R } &\red& \cut{ \cut Q `g + y P } `b + z R
 \end{array} \]
for \CBN.

 \end {itemize}
This way, we obtain two notions of reduction that are clearly confluent: all rules are left-linear and non-overlapping.

 \end {definition}

In \cite{Bakel-Lescanne'08,vBR'06} some basic properties are shown, which essentially show that the calculus is well-behaved, as well as the relation between $\X$ and a number of other calculi.
These results motivate the formulation of admissible rules:

 \begin {lemma}[Garbage Collection and Renaming \cite{vBR'06}]\label{renaming}
 \[ \begin {array}{rrcl@{\quad}l}
(\gcL): & \cutL P `a + x Q &\redX& P &\textrm{if }`a \not \in \FP{P}
\\ % &\quad&
(\gcR): & \cutR P `a + x Q &\redX& Q &\textrm{if }x \not \in \FS{Q}
\\
(\renL): & \cut P `d + z \caps<z,`a> &\redX& P[`a/`d]
\\ % &&
(\renR): & \cut \caps<z,`a> `a + x + P &\redX& P[z/x]
\end {array} \]
 \end {lemma}
}% Long %%%%%%%%%%%%%%%%%%%%%%%%%%%%

 \section{Typing for $\X$: from {\LK} to \X} \label {types section}

$\X$ offers a natural presentation of the classical propositional calculus with implication, and can be seen as a variant of system \LK.

We first define types and contexts.

 \begin {definition}[Types and Contexts]\label{types}

 \begin {enumerate}

 \item
The set of types is defined by the grammar:
 $ \begin {array}{rcl}
A,B & ::= & \tvar \mid A \arrow B,
 \end {array} $
where $\tvar$ is a basic type of which there are infinitely many.
\Long{The types considered in this paper are normally known as {\em simple} (or {\em Curry}) types.}

 \item
A {\em context of sockets} $`G$ is a\Long{ mapping from sockets to types, denoted as a} 
finite set of {\em statements} $\stat{x}{A}$, such that the {\em subject} of the statements ($x$) are distinct.
We write $`G_1,`G_2$ to mean the union of $`G_1$ and $`G_2$, provided $`G_1$ and $`G_2$ are compatible (if $`G_1$ contains $\stat{x}{A_1}$ and $`G_2$ contains $\stat{x}{A_2}$ then $A_1 = A_2$), and write $`G, \stat{x}{A}$ for $`G, \{\stat{x}{A}\}$.
\Long{So, when writing a context as $`G, \stat{x}{A}$, this implies that $\stat{x}{A} \in `G$, or $`G$ is not defined on $x$. }

\item
Contexts of {\em plugs} $`D$ are defined in a similar way.
 \end {enumerate}

 \end {definition}

The notion of type assignment on $\X$ that we present in this section is the basic implicative system for Classical Logic (Gentzen's system \LK) as described above.
The Curry-Howard property is easily achieved by erasing all term-information.
When building witnesses for proofs, propositions receive names; those that appear in the left part of a sequent receive names like $x, y, z$, etc, and those that appear in the right part of a sequent receive names like $`a, `b, `g$, etc.
When in applying a rule a formula disappears from the sequent, the corresponding connector will get bound in the net that is constructed, and when a formula gets created, a new connector will be associated to it.

 \begin {definition}[Typing for $\X$]

 \begin {enumerate}

 \item
{\em Type judgements} are expressed via a ternary relation $\DerX {P}{`G}{`D}$, where $`G$ is a context of {\em sockets} and $`D$ is a context of {\em plugs}, and $P$ is a net.
We say that $P$ is the {\em witness} of this judgement.

 \item
{\em Type assignment for} $\X$ is defined by the following rules:
 \[ %\def\arraystretch{3}
\kern-7mm \begin {array}{rlcrl}
 \CapRule{y}{`a}{A}{ `G}{ `D} 
&~&
 \ImpRule{P}{`a}{y}{x}{Q}{A}{B}{ `G}{ `D } 
\\[5mm]
 \ExpRule{x}{P}{`a}{`b}{A}{B}{`G}{`D}
&&
 \DagRule{P}{`a}{x}{Q}{A}{ `G}{ `D}
 \end {array}
 \]
\Long{We write $\derX  P : `G |- `D $ if there exists a derivation that has this judgement in the bottom line, and write $\derX \D :: P : `G |- `D $ if we want to name that derivation.}

 \end {enumerate}
 \end {definition}

Notice that $`G$ and $`D$ carry the types of the free connectors in $P$, as unordered sets.
There is no notion of type for $P$ itself, instead the derivable statement shows how $P$ is connectable.

 \begin{example}[A proof of Peirce's Law] \label{peirce example}
The following is a proof for Peirce's Law in {\LK}:
\[ 
{
 \def \Exp{\ArrR}
 \def \Imp{\ArrL}
 \def \Cap{\Ax}
 \def\arr{\Arrow}
 \def \stat#1#2{#2}
 \def \DerX#1#2#3{ \Der {#2}{#3} }
 \begin {array}{ccc}
\Peirce
 \end {array}
}
 \]
Inhabiting this proof in $\X$ gives the derivation:
\[ \Peirce \]
 \end{example}

The following soundness result is proven in  \cite{Bakel-Lescanne'08}:
\Long{The soundness result of simple type assignment with respect to reduction is stated as usual:}

 \begin {theorem}[Witness reduction] 
If $\DerX {P}{`G}{`D}$, and $P \redX Q$, then $\DerX {Q}{`G}{`D}$.
 \end {theorem}

%)]}

%\input{pi}

\section{The asynchronous $`p$-calculus with pairing and nesting} \label{pi with pairing}

The notion of asynchronous $`p$-calculus that we consider in this paper is different from other systems studied in the literature \cite{Honda-Tokoro'91}.
One reason for this change lies directly in the calculus that is going to be interpreted, $\X$: since we are going to model sending and receiving pairs of names as interfaces for functions, we add pairing, inspired by \cite{Abadi-Gordon'97}.
\Long{We take the view that processes communicate by sending data over channels, not just names. }%
The other reason is that we want to achieve a preservation of \emph{full} cut-elimination; to this aim, we need to use \emph{non-blocking} inputs, by adding the reduction rule $(\emph{nesting})$ (see Definition~\ref{pi reduction}).
Without this last addition, we cannot model full cut-elimination; this was, for example, also the case with the interpretations defined by Milner \cite{Milner'92}, Sangiorgi \cite{Sangiorgi-Walker'03}, Honda \emph{et al} \cite{Honda-Yoshida-Berger'04}, and Thielecke \cite{Thielecke'97}, where reduction in the original calculus had to be restricted in order to get a completeness result. 
Notice that this last extension of $`p$ \emph{only} relates to cut-elimination: that all proofs in {\LK} are representable in $`p$ is not affected by this, nor is the preservation of types. 

To ease the definition of the interpretation function of circuits in $\X$ to processes in the \PiC, we deviate slightly from the normal practice, and write either Greek characters $`a, `b, `y, \ldots$ or Roman characters $x, y, z, \ldots$ for channel names; we use $n$ for either a Greek or a Roman name, and `$`o$' for the generic variable\Long{, and $\varv, \varw, \vard, \vare, \ldots$ for $`p$'s `internal' variables when there is need to distinguish them from $\X$'s connector names}.
We also introduce a structure over names, such that not only names but also pairs of names can be sent (but not a pair of pairs). 
In this way a channel may pass along either a name or a pair of names. 
We also introduce the let-construct to deal with inputs of pairs of names that get distributed over the continuation.

 \begin{definition}
Channel names and data are defined by:
 \[ \begin{array}{rcl@{\quad}l}
\achan,\bchan,\chan,\dchan &::=&  x  \mid `a & \textit{names} %\\
 \end{array} \hspace*{3cm} \begin{array}{rcl@{\quad}l}
p & ::= & \achan \mid \PiPair<\achan,\bchan>  & \textit{data}
 \end{array} \]
Notice that pairing is \emph{not} recursive.
Processes are defined by:
 \[ \begin{array}{rrl@{\quad}l}
P,Q & ::= & 0 & \textsl{Nil} 
\\
&\mid& P \Par Q  & \textsl{Composition} 
\\
&\mid& \Rep{P} & \textsl{Replication} 
\\
&\mid& \New{\achan} P & \textsl{Restriction} 
 \end{array} \quad\quad \begin{array}{rrl@{\quad}l}
&\mid& \In \achan(x) . P & \textsl{Input} 
\\
&\mid& \Out a <p> & \textsl{(Asynchronous) Output} 
\\
&\mid& \Let <x,y> = z in P	& \textsl{Let construct}
%
%&\mid& P \Choice Q & \textsl{Choice} 
%\\
 \end{array} \]
 \end{definition}
We abbreviate $\In \achan(x) . \Let <y,z> = x in P $ by $\In {\achan} ( \PiPair<y,z> ) . P $, and $\New m \New n P $ by $\New {m,n} P $.

A (process) context is simply a term with a hole $[\cdot]$.

 \begin{definition}[Congruence]
The structural congruence  is the smallest equivalence relation  closed under  contexts   defined by the following rules:
 \[ \begin{array}{rcll}
P \Par {\bf 0} &\StrCon& P 
\\
P \Par Q &\StrCon	 & Q \Par P
\\
(P \Par Q ) \Par R &\StrCon& P \Par (Q \Par R) 
\\
\New n {\bf 0} &\StrCon& {\bf 0} 
 \end{array} \quad \begin{array}{rcll} 
\New m \New n P &\StrCon& \New n \New m P 
\\
\New n (P \Par Q) &\StrCon& P \Par \New n Q & \textit{if } n \notin \fn(P)
\\
\Rep P &\StrCon& P \Par \Rep P 
\\ 
\Let <x,y> = { \PiPair<\achan,\bchan> } in R &\StrCon& R [\achan/x,\bchan/y]
%\\
%
%P \StrCon Q \StrCon R &\Then& P \StrCon Q
%
 \end{array}\]
 \end{definition}

\Comment{%%%%%%%%%%%%%%%OLD DEFINITION WITH +
 \begin{definition}[Congruence]
The structural congruence  is the smallest equivalence relation  closed under  contexts   defined by the following rules:
 \[ \begin{array}{rcll}
P \Par {\bf 0} &\StrCon& P 
\\
P \Par Q &\StrCon	 & Q \Par P
\\
(P \Par Q ) \Par R &\StrCon& P \Par (Q \Par R) 
\\
\Rep P &\StrCon& P \Par \Rep P 
\\ 
\New n {\bf 0} &\StrCon& {\bf 0} 
 \end{array} \quad \begin{array}{rcll} 
\New m \New n P &\StrCon& \New n \New m P 
\\
\New n (P \Par Q) &\StrCon& P \Par \New n Q & \textit{if } n \notin \fn(P)
\\
P \Choice Q &\StrCon & Q \Choice P 
\\
(P \Choice Q ) \Choice R  &\StrCon & P  \Choice (Q \Choice R) 
\\
\Let <x,y> = \PiPair<\achan,\bchan> in R &\StrCon& R [\achan/x,\bchan/y]
%\\
%P \StrCon Q \StrCon R &\Then& P \StrCon Q
 \end{array}\]
 \end{definition}
}%%%%%%%%%%%%%%%%%

 \begin{definition} \label{pi reduction}
 \begin{enumerate}

\item
The {\em reduction relation} over the processes of the $`p$-calculus is defined by following (elementary) rules: 
 \[ \begin{array}{rrcll}
(\textsl{synchronisation}): & \Short{\tab}
\ShiftL{2}{
%\Out \achan<\bchan> . P   \Par \In \achan(x) . Q 
\Out a <\bchan> \Par \In \achan(x) . Q 
%%%%\Out \achan<\bchan> . P \Choice G  \Par \In \achan(x) . Q \Choice S OLD DEFINITION
}  
%&\redPi& P \Par Q [\bchan/x] 
&\redPi& Q [\bchan/x] 
\\
%(\textsl{ choice}): &
%P \redPi P' &\Then& P \Choice Q \redPi P' 
%\\
%(\textsl{right choice}): &
%Q \redPi Q' &\Then& P \Choice Q \redPi Q' 
%\\
(\textsl{binding}): &
P \redPi P' &\Then& \New n P \redPi \New n P'
\\
%\Short{ \end{array} \] \[ \begin{array}{rrcll}}
 (\textsl{composition}): &
P \redPi P' &\Then& P \Par Q \redPi P' \Par Q
\\
(\textsl{nesting}): &
P \redPi Q &\Then& \In{n}(x) . P \redPi \In{n}(x) . Q  
\\
(\textsl{congruence}): &
\ShiftL{2}{
P \StrCon Q~\And~Q \redPi Q'~\And~Q' \StrCon P'
}
&\Then& P \redPi P'
 \end{array} \]

\item
We write $\tcredPi$ for the  reflexive and transitive closure  of $\redPi$.
 
\item
We write $P \shows n$ if  $P \StrCon  \New {b_1 \ldots b_m} ( \Out n <p> \Par Q ) $  for some $Q$, where $ n\neq b_1 \ldots b_m$.

% or $\alpha = \In \nchan(x)$ 

\item
We write $Q \Shows n$ if there exists $P$ such that $Q \tcredPi P$ and $P \shows n$. 

 \end{enumerate}
 \end{definition} 
Notice that we no longer consider input in $`p$ to be \emph{blocking}; we are aware that this is a considerable breach with normal practice, but this is strongly needed in our completeness result (Theorem \ref{correctness theorem}); without it, we can at most show a partial result.

Moreover, notice that
 \[ \begin{array}{lcl}
\Out a <\mbox{$ \PiPair<\bchan,\chan> $}> \Par \In \achan({ \PiPair<x,y> }) . Q 
\Short{& \tcredPi &}%
\Long{%%%%%%%%%%%%%%%%%%%
& = &
\Out a <{ \PiPair<\bchan,\chan> }> \Par \In \achan(z) . \Let <x,y> = z in Q  \\ 
& \redPi &
\Let <x,y> = \PiPair<\bchan,\chan> in Q  \\
& \StrCon &
}% Long %%%%%%%%%%%%%%%%%
Q [\bchan/x,\chan/y] 
 \end{array} \]

\begin{definition}[\cite{Honda-Yoshida'95}] \label{bisimilarity}
{\em Barbed contextual simulation} is the largest  relation $\bisimilarity $ such that $P \bisimilarity Q$ implies:
\begin{itemize}
\item
for each name $n $, if $P \shows n$ then $Q \Shows n$;
\item
for any context $C$, if $C[P] \redPi P'$, then for some $Q'$, $C[Q] \tcredPi Q'$ 
and $P' \bisimilarity Q'$.
\end{itemize}

%We write $ P \CRP Q$ if there exist $P',Q'$ such that $P\tcredPi P'$, $Q\tcredPi Q'$, and $P'\bisimilarity Q'$.

 \end{definition}

\section{Type assignment} \label{tas for pi}

In this section, we introduce a notion of type assignment for processes in $`p$ that describes the `\emph{input-output interface}' of a process.
This notion is novel in that it assigns to channels the type of the input or output that is sent over the channel; in that it differs from normal notions, that would state:
 \[ 
\Inf{\def \TurnPi{\Turn} \Pider \Out a <b> : `G,b{:}A |- a{:}\textsf{ch}(A),`D }
 \]
In order to be able to encode {\LK}, types in our system will not be decorated with channel information.

As for the notion of type assignment on $\X$ terms, in the typing judgements we always write channels used for input on the left and channels used for output on the right; this implies that, if a channel is both used to send and to receive, it will appear on both sides.

 \begin{definition}[Type assignment]
The types and contexts we consider for the {\PiC} are defined like those of Definition~\ref{types}, generalised to names\Long{, and allowing both Roman and Greek names on both sides}.
Type assignment for {\PiC} is defined by the following sequent system:
 \[ %\def \arraystretch{2.75} 
 \begin{array}{rl}
(0): &
 \Inf	{ \Pider 0 : `G |- `D }
\\[5mm] % \quad
(!): & %~
 \Inf	{ \Pider P : `G |- `D }
	{ \Pider \Rep P : `G |- `D }
\\[5mm]
(`n): &
 \Inf	{ \Pider P : `G,a{:}A |- a{:}A,`D }
	{ \Pider \New a P : `G |- `D }
\\[5mm]
(\mid): &
 \Inf	{ \Pider P : `G |- `D \quad \Pider Q : `G |- `D }
	{ \Pider P \Par Q : `G |- `D }
\\
%(+) : &
% \Inf	{ \Pider P : `G |- `D \quad \Pider Q : `G |- `D }
%	{ \Pider P \Choice Q : `G |- `D }
% \\
\end{array}~%\kern-6mm
\begin{array}{rl}
(\textsl{in}) : &
 \Inf	{ \Pider P : `G, x{:}A |- x{:}A.`D }
	{ \Pider \In{a}(x) . P : `G,a{:}A |- `D }
\\[5mm]
(\textsl{out}) : &
 \Inf	%{ \Pider P : `G,b{:}A |- b{:}A,`D } 
	{ \Pider \Out a <b> %. P 
		: `G,b{:}A |- a{:}A,b{:}A,`D }
\\[5mm]
% \end{array} \]
% \[ \def \arraystretch{2.75} \begin{array}{rl}
(\textsl{pair-out}) : &
 \Inf	%{ \Pider P : `G,b{:}A |- c{:}B,`D }
	{ \Pider \Out a <\mbox{$ \PiPair<b,c> $}> %. P 
		: `G,b{:}A |- a{:}A\arr B,c{:}B,`D }
\\[5mm]
(\Let) : &
 \Inf	{ \Pider P : `G,y{:}B |- x{:}A,`D }
	{ \Pider \Let <x,y> = z in P : `G,z{:}A\arr B |- `D }
 \end{array} \]

\Long{As usual, we write $\Pider P : `G |- `D $ if there exists a derivation using these rules that has this expression in the conclusion, and write $ \Pider \D :: P : `G |- `D $ if we want to name that derivation.}

 \end{definition}

Notice that it is possible to derive $\Pider \Out a <a> : {} |- a{:}A $, although sending a channel name over that channel itself is never produced by our encoding, nor by the reduction of processes created by the encoding.

\Long{The `\emph{input-output interface of a $`p$-process}' property is nicely preserved by all the rules; it also explains how the handling of pairs is restricted by the type system in to the rules $(\Let)$ and $(\textsl{pair-out})$.}

 \begin{example}
We can derive
\[
\Inf	{ \Inf	{ \Pider P : `G,y{:}B |- x{:}A,`D } 
		{ \Pider \Let <x,y> = z in P : `G,z{:}A\arr B |- `D }
	}
	{ \Pider \In a(z) . \Let <x,y> = z in P : `G,a{:}A\arr B |- `D } 
\]
so the following rule is derivable: 
\[ \begin{array}{rl}
(\textsl{pair-in}) : &
 \Inf	{ \Pider P : `G,y{:}B |- x{:}A,`D }
	{ \Pider \In{a}( \PiPair<x,y> ) . P : `G,a{:}A\arr B |- `D }
 \end{array} \]

Notice that the rule $(\textsl{pair-out})$ does not directly correspond to the logical rule $(\ArrR)$, as that $(\textsl{pair-in})$ does not directly correspond to $(\ArrL)$; this is natural, however, seen that the encoding does not map rules to rules, but proofs to type derivations.
This apparent discrepancy is solved by Theorem \ref{encoding preserves types}.

 \end{example}

In fact, this notion of type assignment does not (directly) relate back to {\LK}.
For example, rules $(\mid)$ and $(!)$ do not change the contexts, so do not correspond to any rule in the logic, not even to a $\lmu$-style activation step.
\Long
{%%%%%%%%%%%%%%%%%%%%%%%%%%%%%%%%%
Moreover, rule $(`n)$ just removes a formula; it would perhaps be better to have syntax that expresses sending a private name, as used in \cite{Honda-Yoshida-Berger'04}.
E.g., we could write $\Out a (b) . P $ for the process $\New{b} \Out a (b) . P$; notice the use of `$(~)$' rather than `$\Group<~>$'.
This would justify the derivation rules \[ \begin{array}{rlcrl}
(\textsl{out}') : &
 \Inf	{ \Pider P : `G,b{:}A |- b{:}A,`D } 
	{ \Pider \Out a (b) . P : `G |- a{:}A,`D }
\Short{\\[5mm]}\Long{&\quad&}
(\textsl{pair-out}') : &
 \Inf	{ \Pider P : `G,b{:}A |- c{:}B,`D }
	{ \Pider \Out a (\PiPair<b,c>) . P : `G |- a{:}A\arr B,`D }
 \end{array} \]
which have a clearer logical content.
However, this would not suffice in our case since we cannot encode, for example, $\exp x \caps<x,`a> `a . b $ by $\In{x} . \Out{`a } \Par \Out{`b}( \PiPair<x,`a> ) $, since that would imply that $x$ and $`a$ occur both free and bound.

 \begin {lemma}[Weakening] \label{weakening}
The following rule is admissible: \[ \begin{array}{rl}
(\Weak): & 
 \Inf	[`G' \supseteq `G,`D' \supseteq `D]
	{ \Pider P : `G |- `D }
	{ \Pider P : `G' |- `D' }
 \end{array} \]
 \end {lemma}

This result allows us to be a little less precise when we construct derivations, and allow for rules to join contexts, by using, for example, the rule
 \[ \begin{array}{rl} 
(\mid): &
 \Inf	{ \Pider P : `G_1 |- `D_2 \quad \Pider Q : `G_1 |- `D_2 }
	{ \Pider P \Par Q : `G_1,`G_2 |- `D_1,`D_2 } \\
 \end{array} \]

In what follows, we will use the following abbreviation:
\[
\D_{\In a  . \Out{b}}{:}A :
\Inf	{ \Inf	{ \Inf	%[W]
			%{\Pider 0 : `G |- `D } 
			{\Pider 0 : `G,`o{:}A |- `o{:}A,`D }
		}
		{ \Pider \Out b  : `G,`o{:}A |- b{:}A,`o{:}A,`D }
	}
	{ \Pider \In a  . \Out b  : `G,a{:}A |- b{:}A,`D }
\]

We have a soundness (witness reduction) result, for which we first need to prove a substitution lemma\Long{ and a congruence lemma}.

 \begin{lemma}[Substitution] \label{substitution lemma}
If $\Pider P : `G,x{:}A |- x{:}A,`D $ then also \Short{\\}%
$\Pider P[b/x] : `G,b{:}A |- b{:}A,`D $.
 \end{lemma}

 \begin{Proof}
By induction on the structure of processes; we only show the base cases.

 \begin{description}

\item[$\Out a <x> . P$]
The derivation is shaped like: \[ 
 \Inf	{ \InfBox	{ \Pider P : `G,x{:}A |- x{:}A,`D } }
	{ \Pider \Out a <x> . P : `G,x{:}A |- a{:}A,x{:}A,`D }
 \]
Then, by induction, $\Pider P[b/x] : `G,b{:}A |- b{:}A,`D $, so we can construct \[ 
 \Inf	{ \InfBox	{ \Pider P[b/x] : `G,b{:}A |- b{:}A.`D } }
	{ \Pider \Out a <b> . P[b/x] : `G,b{:}A |- a{:}A,b{:}A,`D }
 \]
Notice that $(\Out a <x> . P)[b/x] = \Out a <b> . P[b/x]$.

\item[$\Out{x}<a> . P$]
Notice that $x$ might be free in $P$. 
The derivation is shaped like: \[ 
 \Inf	{ \InfBox	{ \Pider P : `G,a{:}A,x{:}A |- x{:}A,a{:}A,`D } }
	{ \Pider \Out{x}<a> . P : `G,a{:}A,x{:}A |- x{:}A,a{:}A,`D }
 \]
Then, by induction, $\Pider P[b/x] : `G,a{:}A,b{:}A |- b{:}A,a{:}A,`D $, so we can construct \[ 
 \Inf	{ \InfBox	{ \Pider P[b/x] : `G,a{:}A,b{:}A |- b{:}A,a{:}A,`D } }
	{ \Pider \Out{b}<a> . P[b/x] : `G,b{:}A |- a{:}A,b{:}A,`D }
 \]
Notice that $(\Out{x}<a> . P)[b/x] = \Out{b}<a> . P[b/x]$

\item[$\Out a<{ \PiPair<y,z> }> . P$, with $x = y$ or $x = z$, or $\Out x<{ \PiPair<y,z> }> . P$]
Similar.

 \end{description}

The other cases follow by induction.

% 0 \mid \Rep{P} \mid \New{n} P \mid P \Choice Q \mid \In a(x) . P \mid \In a( \PiPair<x,y> ) . P \mid \Out a<b> . P \mid \Out a<{ \PiPair<b,c> }> . P

 \end{Proof}
}% Long %%%%%%%%%%%%%%%%%%%%%%%%%%%%%%%%%%%%%%%%%

Notice that the cases $ \Pider P : `G |- x{:}A,`D $ and $\Pider P : `G,x{:}A |- `D $ can be generalised by weakening to fit the lemma.

\Long{%%%%%%%%%%%%%%%%%%%%%%%%%%%%%%%%%%%%%%%%%%
 \begin{lemma}[Witness congruence]
If $\Pider P : `G |- `D $ and $P \StrCon Q$, then $\Pider Q : `G |- `D $.
 \end{lemma}

 \begin{Proof}
By easy induction on the congruence relation.
 \end{Proof}
}% Long %%%%%%%%%%%%%%%%%%%%%%%%%%%%%%%%%

We now come to the main soundness result for our notion of type assignment for $`p$.

 \begin{theorem}[Witness reduction] \label{Witness reduction}
If $\Pider P : `G |- `D $ and $P \redPi Q$, then $\Pider Q : `G |- `D $.
 \end{theorem}

\Long{%%%%%%%%%%%%%%%%%%%%%%%%%%%%%%%%%%%%%%%%%%
 \begin{Proof}
By induction on the reduction relation.

 \begin{description}
 
 \item [{$\Out a<b> . P \Par \In a(x) . Q \redPi P \Par Q [b/x]$}]

Then the derivation is shaped like: \[
\Inf	{ \Inf	{ \InfBox	{ \Pider P : `G,b{:}A |- b{:}A,`D } }
		{ \Pider \Out a<b> . P : `G,b{:}A |- a{:}A,b{:}A,`D }
	 \quad
	 \Inf	{ \InfBox	{ \Pider Q : `G,x{:}A |- x{:}A,`D } }
		{ \Pider \In a(x) . Q : `G,a{:}A |- `D }
	}
	{ \Pider \Out a<b> . P \Par \In a(x) . Q : `G,a{:}A,b{:}A |- a{:}A,b{:}A,`D }
 \]

Then, by Lemma~\ref{substitution lemma}, we have $ \Pider Q[b/x] : `G,b{:}A |- b{:}A,`D $ and we can construct: 

\[
\Inf	{ \InfBox	{ \Pider P : `G,b{:}A |- b{:}A,`D } 
	 \quad
	 \InfBox	{ \Pider Q[b/x] : `G,b{:}A |- b{:}A,`D } 
	}
	{ \Pider P \Par Q[b/x] : `G,b{:}A |- b{:}A,`D }
 \]
Now either $a{:}A$ already occurs in $`G$ or $`D$ (notice that $`G,a{:}A$ stands for adding $a{:}A$ to $`G$ when no statement for $a$ occurs in $`G$, and for $`G$ when $a{:}A$ already occurs in $`G$) or it can be added by weakening.

 \item [{$\Out a<{ \PiPair<b,c> }> . P \Par \In a( \PiPair<x,y> ) . Q \redPi P \Par Q [b/x,c/y] $}]
Similar.

% \item [$P \redPi P' \Then P \Choice Q \redPi P' $]
%Case

% \item [$Q \redPi Q' \Then P \Choice Q \redPi Q' $]
%Case

% \item [$P \redPi P' \Then \New n P \redPi \New n P'$]
%Case

% \item [$P \redPi P' \Then P \Par Q \redPi P' \Par Q $]
%Case

% \item [$P \StrCon Q~\And~Q \redPi Q'~\And~Q' \StrCon P' \Then P \redPi P'$] 
%Case

 \end{description}

The other cases follow by induction.

 \end{Proof}

}% Long %%%%%%%%%%%%%%%%%%%%%%%%%%%%%%%%%%%%%%%%%%

%\input{encoding}

% {[(
\section {Interpreting $\X$ into $\pi$} \label{encoding}

In this section, we define an encoding from nets in $\X$ onto processes in $`p$.

The encoding defined below is based on the intuition as formulated in \cite{Bakel-Lescanne'08}: the cut $\cut P `a + x Q $ expresses the intention to connect all $`a$s in $P$ and $x$s in $Q$, and reduction will realise this by either connecting all $`a$s to all $x$s, or all $x$s to all $`a$s. 
Translated into $`p$, this results in seeing $P$ as trying to send at least as many times over $`a$ as $Q$ is willing to receive over $x$, and $Q$ trying to receive at least as many times over $x$ as $P$ is ready to send over $`a$.

As discussed above, when creating a witness for $(\ArrR)$ (the net $\exp x P `a . `b $, called an \emph{export}), the exported interface of $P$ is the functionality of `receiving on $x$, sending on $`a$', which is made available on $`b$.
When encoding this behaviour in $`p$, we are faced with a problem.
It is clearly not sufficient to limit communication to the exchange of single names, since then we would have to separately send $x$ and $`a$, breaking perhaps the exported functionality, and certainly disabling the possibility of assigning arrow types. %  
We overcome this problem by sending out a pair of names, as in $\Out{a} <\mbox{$ \PiPair<v,`d> $}> $.
Similarly, when interpreting a witness for $(\ArrL)$ (the net $\imp P `a [x] y Q $, called an \emph{import}), the circuit that is to be connected to $x$ is ideally a function whose input will be connected to $`a$, and its output to $y$. 
This means that we need to receive a pair of names over $x$, as in $\In{x} ({ \PiPair<v,`d> }) . P $.

A cut $\cut P `a + x Q $ in $\X$ expresses two nets that need to be connected via $`a$ and $x$.
If we model $P$ and $Q$ in $`p$, then we obtain one process sending on $`a$, and one receiving on $x$, and we need to link these via $\In {`a} . \Out{x} $.
Since each output on $`a$ in $P$ takes place only once, and $Q$ might want to receive in more than one $x$, we need to replicate the sending; likewise, since each input $x$ in $Q$ takes place only once, and $P$ might have more than one send operation on $`a$, $Q$ needs to be replicated.

\Long{As mentioned in the introduction, w}\Short{W}%
e added pairing to the $`p$-calculus in order to be able to deal with arrow types.
Notice that using the polyadic $`p$-calculus would not be sufficient: since we would like the interpretation to respect reduction, in particular we need to be able to reduce the interpretation of $\cut {\exp x P `a . `b } `b + z \caps<z,`g> $ \Long{--essentially mapped to $\New `b ({ \New {x,`a} ( \PiSem[P] \Par \Out{`b} <{ \PiPair<x,`a> }> ) \Par \New z ( \In{ `b} . \Out{z} \Par \Picaps<z,`g> ) })$--}% 
to that of $\exp x P `a . `g $ (when $`b$ not free in $P$).
So, choosing to encode the export of $x$ and $`a$ over $`b$ as $\Out{`b} <x,`a> $ would force the interpretation of $\caps<z,`g>$ to receive a pair of names.
But requiring for a capsule to always deal with pairs of names is too restrictive, it is desirable to allow capsules to deal with single names as well.
So, rather than moving towards the polyadic $`p$-calculus, we opt for letting communication send a single item, which is either a name or a pair of names.
This implies that a process sending a pair can also successfully communicate with a process not explicitly demanding to receive a pair.

\begin{definition}[Notation]
In the definition below, we use `$`o$' for the generic variable, to separate plugs and 
sockets (and their interpretation) from the `internal' variables of $`p$.
Also, although the departure point is to view Greek names for outputs and Roman names forinputs, by the very nature of the $`p$-calculus (it is only possible to communicate using the \emph{same} channel for in and output), in the implementation we are forced to use Greek names also for inputs, and Roman names for outputs; in fact, we need to explicitly convert `\emph{an output sent on $`a$ is to be received as input on $x$}' via `$\In{`a} (`o) . \Out{x} <`o>$' (so $`a$ is now also an input, and $x$ also an output channel), which for convenience is abbreviated into $\Eq `a=x $.
\end{definition}

 \begin{definition}
The interpretation of circuits is defined by: 
 \[ \begin{array}{rcl} 
%(\Cap): & 
\Scaps<x,`a> &=& \Picaps<x,`a> 
\\
% (\Exp ): & 
\Sexp y Q `b . `a &=& \Piexp y \PiSem[Q] `b . `a 
 \\
%(\Imp): & 
\Simp P `a [x] y Q &=& 
\Piimp \PiSem[P] `a [x] y \PiSem[Q] 
\\
%(\Cut): & 
% \multicolumn{3}{c}{
\Scut P `a + x Q &=& \ScutL P `a + x Q = \ScutR P `a + x Q =
 \Picut \PiSem[P] `a + x \PiSem[Q] 
% }
 \end{array} \]
\end{definition}

Notice that the interpretation of the inactive cut is the same as that of activated cuts.
This implies that we are, in fact, also interpreting a variant of $\X$ \emph{without} activated cuts, allowing arbitrary movement of cuts over cuts, but with the same set of rewrite rules.
This is very different from Gentzen's original definition -- he in fact does not define a cut-over-cut step, and uses innermost reduction for his \textsl{Hauptsatz} result -- and different from Urban's definition -- allowing only \emph{activated} cuts to propagate is crucial for his Strong Normalisation result.
Also, one could argue that then the reduction rules no longer present a system of \emph{cut-elimination}, since now rule $(\Rv)$ reads: 

 \[ \begin{array}{rcl}
\cut P `a + x { \cut Q `b + y R } &\redX&
\ShiftL{4}{\cut { \cut P `a + x Q } `b + y { \cut P `a + x R } }
 \end {array} \]
in which it is doubtful that a cut has been eliminated; it is also easy to show that this creates loops in the reduction system.
However, this rewriting is still sound with respect to typeability.
Here we can abstract from these aspects, since we only aim to prove a \emph{simulation} result, for which the encoding above will be shown adequate.

 \begin{example}
The encoding of the witness of Peirce's law becomes: 
\[ \begin{array}{l}
\Sexp z { \imp{ \exp y \caps<y,`d> `h . `a } `a [z] w \caps<w,`d> } `d . `g = \\
%\Short{%%%%%%%%%%%%%%%%%%%%%%%
\Piexp z { \Piimpdl { \Piexp y \Picaps<y,`d> `h . `a } `a [z] w \Picaps<w,`d> } `d . `g 
% }
%}% Short %%%%%%%%%%%%%%%%%%%%%
%\Long{
% \Piexp z { \Piimp { \Piexp y \Picaps<y,`d> `h . `a } `a [z] w \Picaps<w,`d> } `d . `g 
%}
 \end{array} \]
That this process is a witness of $((A\arr B)\arr A)\arr A $ is a straightforward application of Theorem~\ref{encoding preserves types} below.
 
\Comment{%%%%%%%%%%%%%%%%%%%%%%%%%%%%%%%%%%%%%%%%%
\[
\Inf	{\vbox to 5cm{~}}
	{ \Pider \Piexp z { \Piimp { \Piexp y \Picaps<y,`d> `h . `a } `a [z] w \Picaps<w,`d> } `d . `g : {} |- `g{:}((A\arr B)\arr A)\arr A }
\]
}% End \Comment %%%%%%%%%%%%%%%%%%%%%%%%%%%%%%%%%%%%%
 \end{example}

\Long{%%%%%%%%%%%%%%%%%
One of the main goals we aimed for with our interpretation was: if $`a$ does not occur free in $P$, and $x$ does not occur free in $Q$, then both $\PiSem[ \cut P `a + x Q ] \redPi \PiSem[P]$ and $\PiSem[ \cut P `a + x Q ] \redPi \PiSem[Q]$.
However, we have not achieved this; we can at most show that $\PiSem[ \cut P `a + x Q ]$ reduces to a process that contains $\PiSem[P] \Par \PiSem[Q]$.
}% Long %%%%%%%%%%%%%%
 
The correctness result for the encoding essentially states that the image of the encoding in $`p$ contains some extra behaviour that can be disregarded. 
\Long{%%%%%%%%%%%%%%%%%%
As the examples below show, this is mainly due to the presence of replicated processes in the translation of the cut. 
The precise formulation of the correctness result is: 
In any case, we can show:% (where $=$ stands for contextual equivalence):
}% Long %%%%%%%%%%%%%%%%%%%

 \begin{theorem} \label{correctness theorem}
If $P \redX P' $, then for some $Q$, \, $\PiSem[P] \tcredPi Q$ and $\PiSem[P'] \bisimilarity Q$.
 \end{theorem}
 
\Long{%%%%%%
 \begin{proof}
By induction on the length of the reduction sequence, focussing on the first step, spread out over the following lemmae.
\QED
 \end{proof}
}%%%%% Long

This result might appear weak at first glance, but it would be a mistake to dismiss the encoding on such an observation. 

Our result states that the encoding of $\X$ into $`p$ contains more behaviour than the original term. 
In part, the extra behaviour is due to replicated processes, which can be easily discharged; but, more importantly, $`p$ has no notion of \emph{erasure} of processes: the cut $\cut P `a + x Q $, with $`a$ not in $P$ and $x$ not in $Q$, in $\X$ erases either $P$ or $Q$, but $\PiSem[ \cut P `a + x Q ]$ then runs to $\PiSem[P] \Par \PiSem[Q]$.
The result presented in \cite{Honda-Yoshida-Berger'04} is stronger, but only achieved for Call-by-Value $`l`m$, and at the price of a very intricate translation that depends on types. 
Also $\PiSem[P]$ essentially contains all normal forms of $P$ in parallel; since $\lmu$ is confluent, there is only one normal form, so the problem disappears.
Moreover, restricting to either (confluent) call-by-name or call-by-value restrictions, also then the problem disappears.

\Long{%%%%%%%%%%%%%%%%%%%%%%%%%%%%%%%%%
 \begin{example}
We check \Short{two }\Long{three of the four }% 
examples from the introduction.
%
%\cut \caps<x,`a> `a + y \caps<y,`b> &\redX& \caps<x,`b>
%
%\cut \caps<x,`a> `g + y \caps<y,`b> &\redX& \caps<x,`a>
%
%\cut \caps<x,`a> `a + z \caps<y,`b> &\redX& \caps<y,`b>
%
% \New `a ({ \Scaps<x,`a> \Par \Rep { \New y ( \Eq `a=y \Par \Scaps<y,`b> ) } }) 
%\tab \Choice \quad \\ \hfill 
% \New y ({ \Rep { \New `a ( \Scaps<x,`a> \Par \Eq `a=y ) } \Par \Scaps<y,`b> }) 
%

 \begin{itemize}
\item $\cut \caps<x,`a> `a + y \caps<y,`b> \redX \caps<x,`b>$:
 \[ \kern-4mm \begin{array}{lcl}
 \Scut \caps<x,`a> `a + y \caps<y,`b> 
&\ByDef &
\\
 \Picutdl \Picaps<x,`a> `a + y \Picaps<y,`b> 
&\StrCon &
\\
 \New `a ({ \Picaps<x,`a> \Par \New y ( \Eq `a=y \Par \Picaps<y,`b> ) \Par \\
\tab \Rep { \New y ( \Eq `a=y \Par \Picaps<y,`b> ) } }) \quad \Choice \\ 
\tab \tab \New y ({ \Rep { \New `a ( \Picaps<x,`a> \Par \Eq `a=y ) } \Par \\
\tab \tab \tab \New `a ( \Picaps<x,`a> \Par \Eq `a=y ) \Par \Picaps<y,`b> }) 
& \antibi &
\\
\Picaps<x,`b> \Choice \Picaps<x,`b> 
 & \antibi &
 \Scaps<x,`b> 
 \end{array} \]

\item
$\cut \caps<x,`a> `g + y \caps<y,`b> \redX \caps<x,`a>$:
 \[ \begin{array}{lcl}
\Scut \caps<x,`a> `g + y \caps<y,`b> 
&\ByDef &
\\
 \Picutdl \Scaps<x,`a> `g + y \Scaps<y,`b> 
&\StrCon &
\\
 \Scaps<x,`a> \Par \New `g ({ \Rep { \New y ( \Eq `g=y \Par \Scaps<y,`b> ) } }) \quad \Choice \\ 
\tab \New y ({ \Rep { \New `g ( \Scaps<x,`a> \Par \Eq `g=y ) } \Par \\
\tab \tab \Scaps<x,`a> \Par \New `g (\Eq `g=y ) \Par \Scaps<y,`b> }) 
& \StrCon &
\\
\Scaps<x,`a> \Par \Rep { \New y ( \Eq `g=y \Par \Scaps<y,`b> ) } \quad \Choice \\ 
\tab \New y ({ \Rep { \New `g ( \Scaps<x,`a> \Par \Eq `g=y ) } \Par \\
\tab \tab \Scaps<x,`a> \Par \New `g ( \Eq `g=y ) \Par \Scaps<y,`b> }) 
% & \Bydef &
% \\
 & \antibi &
\\
\Picaps<x,`a> \Choice \Picaps<x,`a> 
 & \antibi &
 \Scaps<x,`a> 
 \end{array} \]

\Long{\item
$\cut \caps<x,`a> `g + z \caps<y,`b> \redX \caps<x,`a>$ and $ \redX \caps<y,`b>$:
 \[ \begin{array}{lcl}
 \Scut \caps<x,`a> `g + z \caps<y,`b> 
&\ByDef &
\\
 \Picutdl \Scaps<x,`a> `g + z \Scaps<y,`b> 
&\ByDef &
\\
 \New `g ({ \Scaps<x,`a> \Par \Rep { \New z ( \Eq `a=x ) \Par \Scaps<y,`b> } }) 
\hfill \Choice \quad \\ \hfill 
 \New x ({ \Rep { \Scaps<x,`a> \Par \New `a ( \Eq `a=x ) } \Par \Scaps<y,`b> }) 
&\StrCon &
\\
 \Scaps<x,`a> \Par \New `g ({ \Rep { \New z ( \Eq `a=x ) \Par \Scaps<y,`b> } }) 
\tab \Choice \quad \\ \hfill 
 \New z ({ \Rep { \Scaps<x,`a> \Par \New `a ( \Eq `a=x ) } })\Par \Scaps<y,`b> 
&\StrCon &
\\
 \Scaps<x,`a> \Par \Rep { \Scaps<y,`b> } 
\Choice 
 \Rep { \Scaps<x,`a> } \Par \Scaps<y,`b> 
 \end{array} \]
}
 \end{itemize}
 \end{example}
}% Long %%%%%%%%%%%%%%%%%%%%%%%%%%%%%%%%%

The following theorem states one of the main results of this paper: it shows that the encoding preserves types.
 
 \begin{Theorem}\label{encoding preserves types}
If $\derX P : `G |- `D $, then $\Pider \PiSem[P] : `G |- `D $.
 \end{Theorem}

\Short{Notice that this theorem links proofs in {\LK} to type derivations in $\TurnPi$\Comment{; for details of the proof, see the appendix.}}

\Long{%%%%%%%%%%%%%%%%%%%%%%%%%%%%%%%%%

 \begin{Proof}
By induction on the structure of circuits in $\X$.

 \begin{description}

 \item[$\caps<x,`a>$]
Then $\PiSem[\caps<x,`a>] = \Picaps<x,`a>$, and the $\X$-derivation: % is shaped like:
\[
\Inf { \Pider \caps<x,`a> : `G,x{:}A |- `a{:}A,`D }
\]
Notice that %$\D_{\In{x} . \Out{`a} }:A :: { \Pider \Picaps<x,`a> : `G,x{:}A |- `a{:}A,`D }$.
 \[ 
\Inf	{ \Inf	{ \Inf	{\Pider 0 : `G,`o{:}A |- `o{:}A,`D }
		}
		{ \Pider \Out{`a} : `G,`o{:}A |- `a{:}A,`o{:}A,`D }
	}
	{ \Pider \In{x}.\Out{`a} : `G,x{:}A |- `a{:}A,`D }
 \]

 \item [$\exp x P `a . `b $]
Then the $\X$-derivation is shaped like:
\[
\Inf	{ \InfBox{ \Pider P : `G,x{:}A |- `a{:}B,`D } }
	{\Pider \exp x P `a . `b : `G |- `b{:}A\arr B,`D }
\]
Then, by induction, $ \Pider \PiSem[P] : `G,x{:}A |- `a{:}B,`D $, and we can construct:
\[ \kern-1cm
\Inf	{ \Inf	{ \Inf	{ {\InfBox{ \Pider \PiSem[P] : `G,x{:}A |- `a{:}B,`D } }
			 \quad
\Short{\raise 44\point \hbox to 25mm\bgroup\kern-2cm}
			 \Inf	{ \Inf	{}
					{ \Pider 0 : `G,x{:}A |- `a{:}B,`D }
				}
				{ \Pider \Out{`b} <{ \PiPair<x,`a> }> : `G,x{:}A |- `a{:}B,`b{:}A\arr B,`D }
\Short{\egroup\multiput(-30,0)(0,8){6}{.}}
			}
			{ \Pider \PiSem[P]\Par \Out{`b} <{ \PiPair<x,`a> }> : `G,x{:}A |- `a{:}B,`b{:}A\arr B,`D }
		}
		{ \Pider \New{`a} (\PiSem[P]\Par \Out{`b} <{ \PiPair<x,`a> }>) : `G,x{:}A |- `b{:}A\arr B,`D }
	}
	{ \Pider \Piexp x \PiSem[P] `a . `b : `G |- `b{:}A\arr B,`D }
\]

 \item[${\imp P `a [y] x Q }$]
Then the $\X$-derivation is shaped like:
\[
\Inf	{ \InfBox{ \Pider P : `G |- `a{:}A,`D } 
	 \qquad
	 \InfBox{ \Pider Q : `G,x{:}B |- `D } 
	}
	{\Pider \imp P `a [y] x Q : `G,y{:}A\arr B |- `D }
\]
Then, by induction, we have derivations for $ \Pider \PiSem[P] : `G |- `a{:}A,`D $ and $ \Pider \PiSem[Q] : `G,x{:}B |- `D $, and we can construct:
\[ \kern-5mm
\Inf	{ \Inf	{ \Inf	{ \Inf	{\Inf	{ \InfBox { \Pider \PiSem[P] : `G |- `a{:}A,`D }
					\quad
					 \InfBox	{\D_{\Eq `a={\varv} }:A}
						{ \Pider \Eq `a={\varv} : `G,`a{:}A |- \varv{:}A,`D }
					}
					{ \Pider \PiSem[P] \Par \Eq `a={\varv} : `G,`a{:}A |- `a{:}A,\varv{:}A,`D }
				}
				{ \Pider \Rep{ \PiSem[P] \Par \Eq `a={\varv} } : `G,`a{:}A |- `a{:}A,\varv{:}A,`D }
			}
			{ \Pider \New{`a} \Rep{ \PiSem[P] \Par \Eq `a={\varv} } : `G |- \varv{:}A,`D }
		~% \quad
\Short{\raise 110\point \hbox to 15mm\bgroup\kern-4.5cm}
		 \Inf	{ \Inf	{ \Inf	{ \InfBox	{\D_{\Eq {\vard}=x }:B}
						{ \Pider \Eq {\vard}=x : `G,\vard{:}B |- x{:}B,`D }
					 \quad
					 \InfBox { \Pider \PiSem[Q] : `G,x{:}B |- `D }
					}
					{ \Pider \Eq {\vard}=x \Par \PiSem[Q] : `G,\vard{:}B,x{:}B |- x{:}B,`D }
				}
				{ \Pider \Rep{ \Eq {\vard}=x \Par \PiSem[Q] } : `G,\vard{:}B, |- x{:}B,`D }
			}
			{ \Pider \New{x} \Rep{ \Eq {\vard}=x \Par \PiSem[Q] } : `G,\vard{:}B |- `D }
\Short{\egroup\multiput(-30,0)(0,8){14}{.}}
		}
		{ \Pider \New{`a} \Rep{ \PiSem[P] \Par \Eq `a={\varv} } \Par \New{x} \Rep{ \Eq {\vard}=x \Par \PiSem[Q] } : `G,\vard{:}B |- \varv{:}A,`D }
	}
	{ \Pider \begin{array}{c} \Piimpdl \PiSem[P] `a [y] x \PiSem[Q] \end{array} : `G,y{:}A\arr B |- `D }
\]
% Perhaps the last two step should be made into one.

 \item[$\cutL P `a + x Q $]
Then the $\X$-derivation is shaped like:
\[ 
\Inf	{ \InfBox{ \Pider P : `G |- `a{:}A,`D } 
	 \qquad
	 \InfBox{ \Pider Q : `G,x{:}A |- `D } 
	}
	{\Pider \cutL P `a + x Q : `G |- `D }
\]
By induction, we have derivations for both $ \Pider \PiSem[P] : `G |- `a{:}A,`D $ and $ \Pider \PiSem[Q] : `G,x{:}A |- `D $, and we can construct:
\[ \kern-2cm
\Inf	{ \Inf	{ {\InfBox { \Pider \PiSem[P] : `G |- `a{:}A,`D }}
		 \quad
\Short{\raise 50\point \hbox to 25mm\bgroup\kern-2cm}
		 \Inf	{ \Inf	{ \Inf	{ \InfBox	{\D_{\Eq `a=x }:A}
						{ \Pider \Eq `a=x : `G,`a{:}A |- x{:}A,`D }
					 \quad
					 \InfBox { \Pider \PiSem[Q] : `G,x{:}A |- `D }
					}
					{ \Pider \Eq `a=x \Par \PiSem[Q] : `G,`a{:}A,x{:}A |- x{:}A,`D }
				}
				{ \Pider \New{x} ( \Eq `a=x \Par \PiSem[Q] ) : `G,`a{:}A |- `D }
			}
			{ \Pider \Rep { \New{x} ( \Eq `a=x \Par \PiSem[Q] ) } : `G,`a{:}A |- `D }
\Short{\egroup\multiput(-30,0)(0,8){6}{.}}
		}
		{ \Pider \PiSem[P] \Par \Rep{ \New{x} ( \Eq `a=x \Par \PiSem[Q] ) } : `G,`a{:}A |- `a{:}A,`D }
	}
	{ \Pider \PicutL \PiSem[P] `a + x \PiSem[Q] : `G |- `D }
\]

 \item[$\cutR P `a + x Q $]
Then the $\X$-derivation is shaped like:
\[
\Inf	{ \InfBox{ \Pider P : `G |- `a{:}A,`D } 
	 \qquad
	 \InfBox{ \Pider Q : `G,x{:}A |- `D } 
	}
	{\Pider \cutR P `a + x Q : `G |- `D }
\]
Then both $ \Pider \PiSem[P] : `G |- `a{:}A,`D $ and $ \Pider \PiSem[Q] : `G,x{:}A |- `D $ follow by induction, and we can construct:
\[ \kern1cm
\Inf	{ \Inf	{
\Short{\multiput(30,0)(0,8){6}{.}
\raise 50\point \hbox to 25mm\bgroup\kern-2cm}
		 \Inf	{ \Inf	{\Inf	{ \InfBox { \Pider \PiSem[P] : `G |- `a{:}A,`D }
					 \quad
					 \InfBox	{\D_{\Eq `a=x }:A}
						{ \Pider \Eq `a=x : `G,`a{:}A |- x{:}A,`D }
					}
					{ \Pider \PiSem[P] \Par \Eq `a=x : `G,`a{:}A |- `a{:}A,x{:}A,`D }
				}
				{ \Pider \Rep{ \PiSem[P] \Par \Eq `a=x } : `G,`a{:}A |- `a{:}A,x{:}A,`D }
			}
			{ \Pider \New{`a} \Rep{ \PiSem[P] \Par \Eq `a=x } : `G |- x{:}A,`D }
\Short{\egroup}
		 \quad
		 \InfBox { \Pider \PiSem[Q] : `G,x{:}A |- `D }
		}
		{ \Pider \New{`a} ( \Rep{ \PiSem[P] \Par \Eq `a=x } ) \Par \PiSem[Q] : `G,x{:}A |- x{:}A,`D }
	}
	{ \Pider \PicutR \PiSem[P] `a + x \PiSem[Q] : `G |- `D }
\]

 \item[$\cut P `a + x Q $]
The $\X$-derivation is shaped like:
\[
\Inf	{ \InfBox{ \Pider P : `G |- `a{:}A,`D } 
	 \qquad
	 \InfBox{ \Pider Q : `G,x{:}A |- `D } 
	}
	{\Pider \cut P `a + x Q : `G |- `D }
\]
and both $ \Pider \PiSem[P] : `G |- `a{:}A,`D $ and $ \Pider \PiSem[Q] : `G,x{:}A |- `D $ follow by induction.
Since we have both \[ \Pider \PicutL \PiSem[P] `a + x \PiSem[Q] : `G |- `D \] and \[ \Pider \PicutR \PiSem[P] `a + x \PiSem[Q] : `G |- `D \] 
by the previous two parts, we can construct:
\[ \kern-1cm
\Inf	{{ \InfBox { \Pider \PicutL \PiSem[P] `a + x \PiSem[Q] : `G |- `D }}
	 \qquad
\Short{\raise 50\point \hbox to 10mm\bgroup\kern-4cm}
	 \InfBox { \Pider \PicutR \PiSem[P] `a + x \PiSem[Q] : `G |- `D }
\Short{\egroup
\multiput(-30,0)(0,8){6}{.}}
	}
	{ \Pider \Short{\begin{array}{c}\Picutdl }{\Long \Picut }\PiSem[P] `a + x \PiSem[Q] \end{array} 
		\Long{\Picutdl \PiSem[P] `a + x \PiSem[Q] } : `G |- `D }
\]

 \end{description}

 \end{Proof}

Essentially in this section we have shown how we can achieve a compositional encoding of $\X$ into $`p$ that preserves the types. 
}% Long %%%%%%%%%%%%%%%%%%%%%%%%%%%%%%%%%

\section{The Lambda Calculus} \label{lambda calculus}

We assume the reader to be familiar with the $`l$-calculus; we just repeat the definition of (simple) type assignment.

 \begin {definition}[Type assignment for the \LC]
 \[ \begin {array}{rlcrlcrl}
 (\Ax):&
	\Inf{\vspace*{\proofrulebaseline}}{ \derL `G ,x{:}A |- x : A }
&\quad&
 (\arrI):&
	\Inf
	{ \derL `G,x{:}A |- M : B }
	{ \derL `G |- `lx . M : A\arr B }
\Short{ \end {array} \]
 \[ \begin {array}{rl}} 
\Long{&\quad&}
 (\arrE):&
	\Inf{ \derL `G |- M : A\arr B \qquad \derL `G |- N : A }
	{ \derL `G |- MN : B }		
 \end {array} \]

 \end {definition}

The following was already defined in \cite{Bakel-Lescanne'08}:

 \begin {definition}[Interpretation of the {\LC} in \X]\label{def:lc to x}
 \[ \begin {array}{rcll}
\SemL{x}{`a} & \ByDef & \caps<x,`a>
\\
\SemL{`lx.M}{`a} & \ByDef & \Exp{x}{ \SemL{M}{`b} }{`b}{`a} & `b\textit{ fresh}
\\
\SemL{MN}{`a} & \ByDef & \Cut{ \SemL{M}{`g} }{`g}{x}{ \Imp{ \SemL{N}{`b}
}{`b}{x}{y}{ \caps<y,`a> } } & `g,`b,x,y\textit{ fresh}
 \end {array} \]
 \end {definition}
Observe that every sub-net of $\SemL{M}{`a}$ has exactly one free
plug, and that this is precisely $`a$.
Moreover, notice that, in the $`l$-calculus, the output (i.e.~result) is anonymous; where an operand `moves' to carries a name via a variable, but where it comes from is not mentioned, since it is implicit.
Since in $\X$, a net is allowed to return a result in more than one way, in order to be able to connect outputs to inputs we have to name the outputs; this forces a name on the output of an interpreted $`l$-term $M$ as well, carried in the sub-script of $\SemL{M}{`a}$; this name $`a$ is also the name of the current continuation, i.e.~the name of the hole in the context in which $M$ occurs.

Combining the interpretation of $`l$ into $\X$ and $\X$ into $`p$, we get yet another encoding of the $`l$-calculus into $`p$ \cite{Milner'99,Milner'92}, one that preserves assignable simple types; as usual, the interpretation is parametric over a name.

 \begin {definition}[Interpretation of the {\LC} in $`p$ via $\X$]\label{def:lc to pi}~
The mapping $\PiSemL[`. ]{`.} : `L \arrow `p$ is defined by: 
$\PiSemL[M]{`a} = \PiSem [ \SemL{M}{`a} ] $
\Long{%%%%%%%%%%%%%%%%%%%%%%%%%%%%%
i.e.
\[ \begin {array}{rcl}
\PiSemL[x]{`a}  
 & \ByDef & 
\PiSem[ \SemL{x}{`a} ] 
\\ & \ByDef & 
\PiSem[ \caps<x,`a> ] 
\\ & \ByDef & 
\Picaps<x,`a> 
\\%[2mm]
\PiSemL{`lx.M}{`a} 
 & \ByDef & 
\PiSem{ \SemL{`lx.M}{`a} } 
\\ & \ByDef &
\PiSem {\exp x \SemL{M}{`b} `b . `a }
\\ & \ByDef &
\Piexp x \PiSemL{M}{`b} `b . `a ,~ `b\textit{ fresh}
\\%[2mm]
\PiSemL{MN}{`a} 
& \ByDef & 
\PiSem{ \SemL{MN}{`a} } 
\\ & \ByDef & 
\PiSem{ \cut \SemL{M}{`g} `g + x { \imp \SemL{N}{`b} `b [x] y \caps<y,`a> } } 
\\ & \ByDef &
\\
	\multicolumn{3}{l}{ 
\begin{array}[b]{l}
\Picutdl \PiSemL{M}{`g } `g + x { \Piimp \PiSemL{N}{`b} `b [x] y \Picaps<y,`a> } 
\end{array}
	} ~ \StrCon 
\\
	\multicolumn{3}{l}{ 
\begin{array}[t]{l}
\Picut \PiSemL{M}{`g } `g + x { \In{x} ( \PiPair<\varv,\vard> ) . \New{`b} \Rep { \PiSemL{N}{`b} \Par \Eq `b={\varv} \Par \Eq {\vard}={`a} } } 
\end{array}, ~ `g,`b,x,y\textit{ fresh}
	} ~ \StrCon 
 \end {array} \]
}
 \end {definition}

Since in \cite{Bakel-Lescanne'08} it is shown that the interpretation $\SemL{`.}{`.} $ preserves both %{\CBN}- and {\CBV}-
reduction and types, the following result is immediate:

 \begin{Corollary}[Simulation of the Lambda Calculus]~
\begin{enumerate}
\item If $M\bred N $ then $\PiSemL[M]{`g} \antibi \PiSemL[N]{`g} $.
%\item If $M\redCBV N $ then $\PiSemL{M}{`g} \CRP \PiSemL{N}{`g} $.
\item If $\derL `G |- M : A $, then $\Pider \PiSemL[M]{`a} : `G |- `a{:}A $.
\end{enumerate}
 \end{Corollary}

{}% End Comment %%%%%%%%%%%%%%%%%%%%%%%%%%%%%%%%%%%%%%%%%%

% )]}

\Comment{ \input{extension} }

\section*{Conclusion}
We studied how to give the computational meaning to classical proofs via the $`p$-calculus. Our results have been achieved in two steps: (1) we have encoded $\X$ into $`p$ enriched with pairing and non-blocking input, and showed that the encoding preserves interesting semantic properties; (2) we have defined a novel and `unusual' type system for $`p$ and proved that types are preserved by the encoding. 

The caveat of the paper was to find the right intuition to reflect the computational meaning of {\Cut}-elimination in $`p$. 
Essentially we have interpreted the input in $`p$ as `witness' for the formulae on the left-hand side of the turnstyle in {\LK}, and outputs as `witnesses' for the right-hand side. 
Arrow-right in {\LK} corresponds to an output channel that sends a pair of names, while arrow-left corresponds to a channel that inputs a pair of names (via the let constructor). 
The {\Cut}-elimination procedure is then interpreted as a forwarder that connects an input and an output via private channels that have the same type. 
%\Comment
{%%%%
Essentially, if we take the view that input are witnesses for fomulae on the left-hand side of the turnstyle in {\LK} and output are witnesses for fomulae on the right-hand side of the turnstye in {\LK} then the cut eliminates the same formulue on the right and on the left of the turnstyle. 
Thus the representation of a cut in $`p$ has to guarantee that the input's and the output's witness of formulae on the right and left-hand side of the turnstyle can communicate. 
This is achieved by using the concept of forwarder, that connects two processes with different inputs and outputs. 
}%%% Long

The work that naturally compares with ours is \cite{Honda-Yoshida-Berger'04}, where the encoding of $\CBV$-$\lmu$ is presented. 
In that paper, full abstraction is proved, but for natural deduction rather than for the sequent calculus as treated in this paper. In order to achieve the full abstraction result, the authors have to introduce a notion of typed equivalence of Call-by-Value $`l`m$.
% Moreover, only the implicative fragment has been encoded. 
By contrast, we have tried to give a simple, intuitive compositional encoding of {\LK} in $`p$ and we leave for future work to consider a restriction of $`p$ in order to make our result stronger.
$\X$ is a calculus without application and substitution that is much easier to interpret in $`p$; notice that we needed no continuation-style encoding to achieve our results. 

In \cite{Bellin-Scott'94} an intuive relation between fragments of linear logic and $`p$-calculus was studied; the results there do not compare with ours. 
The notion of correctness presented in that paper is not between the logical rules and $`p$, but between $`p$ and the `\emph{cut algebra}' which is essentially a dialect of $`p$. Note also that they encode the linear logic as opposed to the implicative fragment of Classical Logic.
In other work \cite{Abramsky'93}, the relationship with linear logic and game semantics is studied.
Both linear logic and game semantics are outside the scope of this paper, yet we leave for future work the study of the relation of linear $\X$ (with explicit weakening and contraction) \cite{Zunic'07}, and relate that with both game semantics and $`p$ without replication. 
\Comment{%%%%%
We have not considered the $\CBV$ or $\CBN$ strategy in $\X$ in this paper. 
Both the strategy induce two confluent subcalculi, on which standard equivalence relations - such as applicative bisimulation -can be defined. It will be of great interest to compare such equivalce relations on $\X$ with the different variant of bi-smilarity defined in $`p$-calculus \cite{Sangiorgi-Walker'01}.
}%%% Long

One of the main goals we aimed for with our interpretation was: if $`a$ does not occur free in $P$, and $x$ does not occur free in $Q$, then both $\PiSem[ \cut P `a + x Q ] \redPi \PiSem[P]$ and $\PiSem[ \cut P `a + x Q ] \redPi \PiSem[Q]$.
However, we have not achieved this; we can at most show that $\PiSem[\cut P `a + x Q ]$ reduces to a process that contains $\PiSem[P] \Par \PiSem[Q]$.
It is as yet not clear what this say about either $\X$, or \LK, or $`p$, or simply about the encoding.
The problem is linked to the fact that $`p$ does not have an automatic \emph{cancellation}: since communication is based on the exchange of channel names, processes that do not communicate with each other just `sit next to each other'.
In $\X$, a process that wants to be `heard', but is not `listened' to, disappears; this corresponds to a proof contracting to a proof, not to two non-connected proofs for the same sequent.
But, when moving to \emph{linear} {\X}, or $*\x$, studied in \cite{Zunic'07}, this all changes.
Since there reduction can generate non-connected nets, it seems promising to explore an encoding of $*\x$ in $`p$.

\Comment{\small \bibliography {../bieb/references} }

\Short
{%\small

}
\onecolumn

\Comment{\newpage \input{appendix} }

 \end{document}

%% file: Xmacros.tex
\def \scsc#1{\mbox{\scriptsize \sc #1}}

\def \xsind{\mbox{\scriptsize {\bf x}{\sc s}}}
\def \xhind{\mbox{\scriptsize {\bf x}{\sc h}}}
\def \xind{\mbox{\scriptsize {\bf x}}}
\def \sind{\mbox{\scriptsize {\sc s}}}
\def \hind{\mbox{\scriptsize {\sc h}}}

\def \slangle {\raise.5\point\hbox{\small $\langle$}}
\def \srangle {\raise.5\point\hbox{\small $\rangle$}}

\def \stackrel#1#2{\setbox155=\hbox{$#1$}\setbox156=\hbox{$#2$}%
\mathrel{\copy156\kern-.5\wd156\kern-.5\wd155\raise4.5\point\hbox{\copy155}\kern-.5\wd155\kern.5\wd156}}

\newfont{\itbf}{ptmbo at 10.25pt}
\newfont{\itsf}{phvro at 10.25pt}
\newfont{\sfbf}{phvb at 10.25pt}
\newfont{\oldmath}{cmsy10 at 11pt}
\newfont{\largeoldmath}{cmsy10 at 15pt}

\def \mysl#1{\mbox{\rm\textsl{#1}}\kern0\point}
\def \mytiny{\fontsize{6.5}{9}\selectfont}
\def \tinyfont{\fontsize{6pt}{6.6pt}\selectfont}

\makeatletter 

\newif\if@om
\def \omfalse{\let\if@om\iffalse}
\def \omtrue{\let\if@om\iftrue}

\newif\if@lom
\def \lomfalse{\let\if@lom\iffalse}
\def \lomtrue{\let\if@lom\iftrue}

\newif\if@Xom
\def \Xomfalse{\let\if@Xom\iffalse}
\def \Xomtrue{\let\if@Xom\iftrue}
\Xomtrue

\newskip \adjustablepoint \adjustablepoint .9\point

\usepackage{graphicx} 
\usepackage{qsymbols}
\usepackage{inference}
\usepackage{relsize}

\Comment{
}% Comment

\def \Cont[#1]{\mbox{\small %\large 
\sf C}[\ssk{#1}\ssk]}
\def \ContH[#1]{\mbox{\small %\large 
\sf C}_{\scsc{h}}[\ssk{#1}\ssk]}

\def \typeout#1{\@latex@warning{#1}}

\def \ifnextchar{\@ifnextchar}

\def \qskp {\hspace*{.25\point}}
\def \hskp {\hspace*{.5\point}}
\def \skp {\hspace*{1\point}}
\def \psk {\hspace*{1\point}}
\def \skipper {\hspace*{1.5\point}}
\def \Bogskipleft	{~\kern-1\point~}
\def \Bogskipright	{\ \kern-1\point\ }

\def \dquad {\quad \quad}
\def \tquad {\quad \quad \quad}
\def \qquad {\quad \quad \quad \quad}

 \def \psset#1{}

\def \BoUndnccurve[#1]#2#3{\psset{linecolor=red}%
\nccurve[#1]{#2}{#3}\psset{linecolor=ruleyellow}}

\def \It#1{\mbox{\rmfamily \it \Itcol #1\hspace*{1\point}}}%{\mbox{\it\Itcol #1}\hspace*{1\point}}

\def \Sc#1{\textsc{\Purple #1}}

\def \mynewrgbcolor#1#2{%
\ifmycolour\newrgbcolor{#1}{#2}\else \newrgbcolor{#1}{0 0 0}\fi}

\def \mynewrgbcolor#1#2{}
\def \newrgbcolor#1#2{}

\mynewrgbcolor{ltermcol}{.8 0 .8}
\mynewrgbcolor{termcol}{1 0 0}
\mynewrgbcolor{typecol}{0 .50 0}
\mynewrgbcolor{rulegreencol}{.1 1 .1}
\mynewrgbcolor{DarkGreencol}{0.1 .5 0.1}
\mynewrgbcolor{Greencol}{0.1 .5 0.1}
\mynewrgbcolor{Yellowcol}{.9 .9 0}
\mynewrgbcolor{Itcol}{.9 .5 0}
\mynewrgbcolor{Blackcol}{0 0 0}
\mynewrgbcolor{Bluecol}{0.35 0.35 1}
\mynewrgbcolor{Orangecol}{1 0.35 0.25}
\mynewrgbcolor{Purplecol}{.75 0 .75}
\mynewrgbcolor{Whitecol}{1 1 1}
\mynewrgbcolor{Greencol}{0 .25 0}
\mynewrgbcolor{yellowcol}{1 1 0}
\mynewrgbcolor{orangecol}{1 0.75 0.5}
\mynewrgbcolor{bluecol}{0 0 1}
\mynewrgbcolor{Redcol}{1 0.05 0.05}
\mynewrgbcolor{lightgreencol}{0.75 1 0.75}
\mynewrgbcolor{lightyellowcol}{1 1 .75}
\mynewrgbcolor{lightbluecol}{.75 .75 1}
\mynewrgbcolor{lightredcol}{1 .5 .5}
\mynewrgbcolor{purplecol}{.9 0.1 .9}
\mynewrgbcolor{whitecol}{1 1 1}
\mynewrgbcolor{blackcol}{0 0 0}
\mynewrgbcolor{neutralcol}{0 0 0}

\newrgbcolor{Redcol}{1 0.05 0.05}
\newrgbcolor{Bluecol}{0.05 0.05 1}

\mynewrgbcolor{ruleyellow}{1 1 .5}
\mynewrgbcolor{rulegreen}{.1 1 .1}
\mynewrgbcolor{foreground}{0 0 0}
\mynewrgbcolor{background}{1 1 1}

\def \ltermcolour{\ifmycolour\ltermcol\fi}
\def \termcolour{\ifmycolour\termcol\fi}
\def \Yellow{\ifmycolour\Yellowcol\fi}
\def \typecolour{\ifmycolour\typecol\fi}

\def \Green{\ifmycolour\Greencol\fi}
\def \Yellow{\ifmycolour\Yellowcol\fi}
\def \Black{\ifmycolour\Blackcol\fi}
\def \Orange{\ifmycolour\Orangecol\fi}
\def \Purple{\ifmycolour\Purplecol\fi}

\def \green{\ifmycolour\greencol\fi}

\def \orange{\ifmycolour\orangecol\fi}
\def \blue{\ifmycolour\bluecol\fi}
\def \Red{\ifmycolour\Redcol\fi}

\def \black{\ifmycolour\blackcol\fi}

\def \specialcol{\termcolour}

 \def \thiscol{}
\def\psline{}

\def \mysemcolour {\mynewrgbcolor{thiscol}{1 0 1}\thiscol}
\ifmycolour \mynewrgbcolor{semcolour}{.9 0 0}\else \mynewrgbcolor{semcolour}{0 0 0}\fi

\psset{linecolor=ruleyellow,linewidth=0.3\point,arrows=->,nodesep=0.05}
\def \PSFrame(#1,#2){\mbox{\psframe[unit=\unitlength,linewidth=.5\point,framearc=.5%,linecolor=yellow
](#1,#2)}}
\def \PSBox(#1,#2){\mbox{\psframe[unit=\unitlength,linewidth=.5\point,framearc=0%,linecolor=blue
](#1,#2)}}

\def \Pr[#1]{~[\,#1\,]}

\def \subssymb {(\tvar \mapsto C)}
\def \subsone#1{\subssymb\,(#1)}
\def \subs{\@ifnextchar\bgroup{\subsone}{\subssymb}}

\def \unifytwo#1#2{\setbox33=\hbox{$#1$}\setbox34=\hbox{$#2$}
\ifdim\wd33<12\point\ifdim\wd34<12\point {\It {unify}}\ #1\ #2
\else {\It {unify}}\ #1\ (#2) \fi
\else \ifdim\wd34<12\point {\It {unify}\ (#1)\ #2}
\else {\It {unify}}\ (#1)\ (#2) \fi \fi }

\def \unify{\@ifnextchar{\bgroup}{\unifytwo}{\It {unify}}}

\def \eqA	{\mathrel{{\sim}\kern-2\point_{\cal A}}}
\def \eqAmu	{\mathrel{{\sim}\kern-2\point_{{\cal A}_{`m}}}}
\def \eqAx	{\mathrel{{\sim}\kern-2\point_{{\cal A}_{\xhind}}}}
\def \eqAxs	{\mathrel{{\sim}\kern-2\point_{{\cal A}_{\xsind}}}}

\def \And	{\mathrel{\&}}
\def \Or	{\mathrel{\vee}}

\def \Conj	{\mathord{\wedge}}

\def \longarrow#1{\psset{unit=1\point,linewidth=0.35\point}%
\psline{cc-cc}(0,0)(#1,0)%
\hspace*{#1}%
\pscurve{cc-cc}(0,0)(-1.1,.4)(-2.5,2.2)%
\pscurve{cc-cc}(0,0)(-1.1,-.4)(-2.5,-2.2)}

\def \Rel#1#2{\setbox155=\hbox{\scriptsize $#1$}\setbox156=\hbox{\scriptsize $#2$}%
\,\raise5.5pt\copy155\kern-1\wd155%
\raise3pt\hbox{\longarrow{1\wd155}}%\kern-.1\wd155%
\kern-.5\wd155\kern-.5\wd156\raise-2pt\copy156\kern-.5\wd156\kern.5\wd155\,}

\def \Skip	{\hspace*{2\point}}

\def \pCsymb{\It{pC}}
\def \pCarg#1{\pCsymb\psk{\termcolour #1}}
\def \pC{\@ifnextchar\bgroup{\pCarg}{\pCsymb}}

\def \pCgsymb{\It{pC}\kern-1pt_{\mbox{\mysemcolour \scriptsize \sc g}}}
\def \pCgarg#1{\pCgsymb\,({\termcolour #1})}
\def \pCg{\@ifnextchar\bgroup{\pCgarg}{\pCgsymb}}

\def \SetAppBr(#1){{\cal A}(#1)}
\def \SetApp{\@ifnextchar({\SetAppBr}{{\cal A}}}
\def \AppMu	{{\termcolour {\cal A}_{`m}}}
\def \SetAppMuBr(#1){\AppMu(#1)}
\def \SetAppMu{\@ifnextchar({\SetAppMuBr}{\AppMu}}
\def \AppS	{{\termcolour \cal A}_{\sind}}
\def \SetAppSBr(#1){\AppS(#1)}
\def \SetAppS{\@ifnextchar({\SetAppSBr}{\AppS}}
\def \AppXS	{{\termcolour \cal A}_{\xsind}}
\def \SetAppXSBr(#1){\AppXS(#1)}
\def \SetAppXS{\@ifnextchar({\SetAppSBr}{\AppXS}}
\def \AppX	{{\termcolour {\cal A}_{\xhind}}}
\def \SetAppXBr(#1){\AppX(#1)}
\def \SetAppX{\@ifnextchar({\SetAppXBr}{\AppX}}

\def \AppS	{{\termcolour \cal A}_{\sind}}

\def \bred	{\mathrel{\Red \bredr}}

\def \bredr {\mbox	{\Red $\rightarrow_{`b}$}}

\def \dbredr	{\arrow \kern-7\point\bredr}

\def \eqbmu	{=_{`b\kern-.5\point`m}}
\def \eqbmx	{=_{`b\kern-.5\point`m\xsind}}
\def \LC	{\Sc{lc}}

\def \diverges {\,{\raise1\point\hbox{\Red \small $\uparrow$}}}
\def \converges {\,{\raise1\point\hbox{\Red \small $\downarrow$}}}
\def \divergesPi {\,{\raise1\point\hbox{\Red \small $\uparrow_{`p}$}}}
\def \convergesbeta {\,{\raise1\point\hbox{\Red \small $\downarrow_{`b}$}}}
\def \convergesbmu {\,{\raise1\point\hbox{\Red \small $\downarrow_{`b`m}$}}}
\def \divergesbeta {\,{\raise1\point\hbox{\Red \small $\uparrow_{`b}$}}}
\def \divergesxs {\,{\raise1\point\hbox{\Red \small $\uparrow$}_{\kern-.5\point\xsind}}}
\def \divergess {\,{\raise1\point\hbox{\Red \small $\uparrow$}_{\kern-.5\point\sind}}}
\def \divergesh {\,{\raise1\point\hbox{\Red \small $\uparrow$}_{\kern-.5\point\hind}}}
\def \divergesxh {\,{\raise1\point\hbox{\Red \small $\uparrow$}_{\kern-.5\point\xhind}}}
\def \divergesS {\,{\raise1\point\hbox{\Red \small $\uparrow_{\sind}$}}}
\def \convergesS {\,{\raise1\point\hbox{\Red \small $\downarrow_{\sind}$}}}
\def \convergesxs {\,{\raise1\point\hbox{\Red \small $\downarrow_{\xsind}$}}}

\def \Bottom {\mbox{\typecolour $\perp$}}
\def \Types	{{\Red \cal T}}

\def \Tproper	{\Types\kern-3\point_{\textit{\scriptsize \Yellow p}}}

\def \dom(#1){\textit{dom}\,(#1)}

\def \arrow	{\mathord{\rightarrow}}
\def \prod	{\mathord{\times}}

\def \arr {\arrow}
\def \Arrow	{{\Rightarrow}}
\def \tvar	{\varphi}

\def \intsymb	{\mbox {\small $\cap$}} 
\def \inter {\mathord{\hskp{\intsymb}\hskp}}

\def \intTwo#1(#2){\intsymb_{\ul{#1}}\qskp(#2)}
\def \intOne#1{\@ifnextchar({\intTwo{#1}}{\intsymb_{\ul{#1}}\qskp}}
\def \int{\@ifnextchar\bgroup{\intOne}{\inter}}

\def \unsymb	{{\cup}} 
\def \union {\mathord{\hskp\type {\unsymb} \hskp}}
\def \compunion	{\setbox172=\hbox{$\cup$}%
\copy172\kern-.75\wd172^c\kern.25\wd172}

\def \unTwo#1(#2){\unsymb\kern-.5\point_{\ul{#1}}\qskp({#2})}
\def \unOne#1{\@ifnextchar({\unTwo{#1}}{\unsymb\kern-.5\point_{\ul{#1}}\qskp}}
\def \un{\@ifnextchar\bgroup{\unOne}{\union}}

\def \BuildInt({\@ifnextchar{n}{\BuildIntn(}{\@ifnextchar{m}{\BuildIntm(}{\BuildIntAux(}}}
\def \BuildIntAux(#1)#2{\int{#1}{#2}_l}
\def \BuildIntn(#1)#2{\int{n}{#2}_i}
\def \BuildIntm(#1)#2{\int{m}{#2}_j}
\def \BuildUn({\@ifnextchar{n}{\BuildUnn(}{\@ifnextchar{m}{\BuildUnm(}{\BuildUnA(}}}
\def \BuildUnA(#1)#2{\un{#1}{#2}_l}
\def \BuildUnn(#1)#2{\un{n}{#2}_i}
\def \BuildUnm(#1)#2{\un{m}{#2}_j}

\def \GUnd_#1{`G\kern-1.5pt_{#1}}
\def \GCom,{`G\kern-1.5pt,}
\def \G{\@ifnextchar_{\GUnd}{\@ifnextchar,}{\GCom}{`G}}
\def \Ga_#1{`G'\kern-1.5pt_{#1}}

\def \Ax {{\Red \mbox{\sl Ax}}}
\def \arrI	{{\Red \arrow \mbox{\sl I}}}
\def \arrE	{{\Red \arrow \mbox{\sl E}}}

\def \D {{\orange \mathcal D}}%{\textrm{D}}
\def \dcol {\,{::}\,}

\def \ArrL	{\Arr\textsl{L}}
\def \ArrR	{\Arr\textsl{R}}

\def \Weak	{\textsl{W}}

\def \Iff	{\mathbin{\,{\Leftarrow}\kern-6\point{\Rightarrow}\,}}
\def \Thenone#1{\mathrel{{\Rightarrow}\,(#1)}}
\def \Then{\@ifnextchar{\bgroup}{\Thenone}{\mathrel{\Rightarrow}}}

\def \Implyone#1{\mathrel{\Rightarrow\,(#1)}}
\def \Imply{\@ifnextchar{\bgroup}{\Implyone}{\mathrel{\Rightarrow}}}

\def \X {{\Red \mbox{$\mathcal{X}$}}}%{\Red \mathcal X} %{\textbf{X}}
\def \Xi {{\Red{\mathcal X}^{\textit{\footnotesize i}}}} %{\textbf{X}}
\def \AstX {{\Red^*\mathcal X}} %{\textbf{X}}
\def \astX {\AstX}
\def \diagX {{\Red ^d \kern-.5\point \mathcal X}} %{\textbf{X}}
\def \copyX {{\Red ^{\scriptsize \copyright} \kern-.5\point \mathcal X}} %{\textbf{X}}
\setbox134=\hbox{\leavevmode\raise0.5\point\hbox{${\mysemcolour	\lceil}\kern-3\point{\mysemcolour \lceil}\kern-.5pt$}}
\setbox135=\hbox{\raise0.5\point\hbox{$\kern-.5pt{\mysemcolour	\rfloor}\kern-3\point{\mysemcolour	\rfloor}$}}

\def \LeftTop#1{{\hspace{1.5\point}%\leavevmode\kern2\point
\setbox66=\hbox{$#1$}%
\ifdim\ht66>5\point \setlength{\unitlength}{.1\point} 
	\psset{unit=.125\ht66,linewidth=0.4\point,linecolor=black}%
\else \setlength{\unitlength}{.2\point} 
	\psset{unit=.8\point,linewidth=0.4\point,linecolor=black}%
\fi
\ifmycolour\psset{linecolor=semcolour}\fi
\setbox67=\hbox{\put(-3,2)%
{\psline{c-c}(0,0)(3,0)%
 \psline{cc-cc}(0,0)(0,-8)%
 \psline{cc-cc}(1.5,0)(1.5,-6)%
}}%
\raise7.25\point\box67%\fi 
\kern2.4\point}}

\def \RightBot{{\kern2.75\point
\psset{unit=1\point,linewidth=0.4\point,linecolor=black}%
\ifmycolour\psset{linecolor=semcolour}\fi
\hbox{\put(0,-2)%
{\psline{c-c}(0,0)(-3,0)%
 \psline{cc-cc}(-1.35,0)(-1.35,6)%
 \psline{cc-cc}(0,0)(0,8)%
}}\kern2\point}}

\def \Rightbot{{\kern3.25\point
\psset{unit=1\point,linewidth=0.4\point,linecolor=black}%
\ifmycolour\psset{linecolor=semcolour}\fi
\hbox{\put(0,-2)%
{\psline{c-c}(0,0)(-3,0)%
\psline{cc-cc}(-1.35,0)(-1.35,5)%
\psline{cc-cc}(0,0)(0,5)%
}}\kern1\point}}

\def \OldSem[#1]{{\semcolour [\kern-1.75\point[}\kern.6\point \lterm{#1} \kern.6\point{\semcolour ]\kern-1.75\point]} }

\def \x {\X}
\def \semEl#1{\LeftTop{#1}{#1}\RightBot}
\def \semThree#1#2#3{\setbox136=\hbox{{\scriptsize \Yellow $#2$}}\setbox137=\hbox{#3}%
	(#1)_{\copy136}\kern-\wd136%
	\raise5\point\box137}
\def \SemThree#1#2#3{%
	\setbox136=\hbox{{\scriptsize \termcolour ${#2}$}}%
	\setbox137=\hbox{#3}%
	\semEl{#1}_{\kern-1\point\termcolour \copy136}
	\ifdim\wd136>1pt\kern-\wd136\raise5\point\copy137
	\ifdim\wd136>\wd137 \kern-\wd137\kern\wd136\fi
	\fi}
\def \SemTwo#1#2{\@ifnextchar{\bgroup}{\SemThree{#1}{#2}}{\semEl{#1}\kern-1\point_{#2}}}
\def \Sem#1{\@ifnextchar{\bgroup}{\SemTwo{#1}}{\semEl{#1}}}
\def \SemL#1#2{\SemThree{{\ltermcolour #1}}{#2}{\mbox{\mysemcolour	\scriptsize $\lambda$}}}
\def \semL [#1] #2 {\SemL{#1}{#2}}

\def \SemGr#1{\Sem{#1}\kern-1\point_{\mbox{\scriptsize \sc g}}}

\def \ftn{\mbox{\Red \textsc{\scriptsize n}}}
\def \ftv{\mbox{\Red \textsc{\scriptsize v}}}

\def \SemN#1{\semEl{#1}\kern-1\point_{\ftn}}
\def \SemV#1{\semEl{#1}\kern-1\point_{\ftv}}

\def \SemN#1{\semEl{#1}\kern-1\point_{\ftn}}
\def \SemV#1{\semEl{#1}\kern-1\point_{\ftv}}

 \def\NF#1{\mbox{\itbf #1}}

 \def \LmuNFa'{\NF{N}\skp'}
 \def \LmuNF{\@ifnextchar{'}{\LmuNFa}{\NF{N}}}

 \def \LmuHNFa'{\NF{H}\skp'}
 \def \LmuHNF {\@ifnextchar{'}{\LmuHNFa}{\NF{H}}}

 \def \Id<#1>{#1}
 \def \VarBrTerm<(#1)>{#1}
 \def \VarTermAux<#1>{\@ifnextchar({\VarBrTerm<#1>}{\Id<#1>}}%)
 \def \VarTerm<#1>{\if@om {\global \omfalse} \VarTermAux<#1> {\global \omtrue} \else \Id<#1>\fi}

 \def \LTerm<#1>{\LTermAux<#1>}

 \def\RemBr<(#1)>{\LTermAux<#1>}

 \def \NoBrLTermAux<{\@ifnextchar({\RemBr<}{\LTermAux<}} %
 \def \LTermAux<%
{\ltermcolour 
	 \@ifnextchar l{\LTermAbs<}%
	{\@ifnextchar a{\LTermapp<}%
	{\@ifnextchar X{\LTermXpp<}%
	{\@ifnextchar d{\LTermDpp<}%
	{\@ifnextchar e{\LTermEpp<}%
	{\@ifnextchar t{\LTermTpp<}%
	{\@ifnextchar v{\LTermVar<}%
	{\@ifnextchar S{\LTermSub<}%
	{\@ifnextchar s{\LTermsub<}%
	{\@ifnextchar ({\LTermBr<}%
	{\@ifnextchar X{\LTermMultiSub<}%
	{\@ifnextchar C{\LTermContSub<}%
	{\@ifnextchar c{\LTermContsub<}%
	{\@ifnextchar n{\LTermName<}%
	{\@ifnextchar m{\LTermMu<}%
	{\@ifnextchar B{\LTermBot<}%
		{\VarTerm<}%
}}}}}}}}}}}}}}}}%}

 \def \LTermapp<a #1 #2>{\LTermAux<#1> \LTermAux<#2>}
 \def \LTermXpp<X #1 #2>{\LTermAux<#1> \LTermAux<#2>}
 \def \LTermDpp<d #1 #2>{\LTermAux<#1> \LTermAux<#2>}
 \def \LTermEpp<e #1 #2>{\LTermAux<#1> \LTermAux<#2>}
 \def \LTermTpp<t #1 #2>{\LTermAux<#1> \LTermAux<#2>}
 \def \LTermAbs<l #1 . #2>{`l#1 . \LTermAux<#2> }
 \def \LTermBr<(#1)>{(\LTermAux<#1>)}
 \def \LTermVar<v #1>{#1}
 \def \LTermSub<S #1 #2 := #3>{\LTermAux<#1> \hskp \slangle{#2}{\,:=\,}\LTermAux<#3>\srangle } %\def \LTermsub<s #1 #2 := #3>{\LTermAux<#1> \hskp \slangle{#2}{\,:=\,}\LTermAux<#3>\srangle }
\def \LTermsub<s #1 #2 := #3>{\LTermAux<#1> \exsub #2 := {\LTermAux<#3>} }
 \def \LTermContSub<C #1 #2 := #3 . #4>{\LTermAux<#1> \hskp \slangle{#2}{\,:=\,}\LTermAux<#3>{`.}#4\srangle }
 \def \LTermContsub<c #1 #2 := #3 . #4>{\LTermAux<#1> \hskp \slangle{#2}{\,:=\,}\LTermAux<#3>{`.}#4\srangle }
 \def \LTermName<n #1 #2>{ [#1] \LTermAux<#2> }
 \def \LTermMu<m #1 . #2>{ `m #1 . \LTermAux<#2> }
 \def \LTermBot<B>{\bot}

\newbox\@indexbox

\newcounter{myindbeta} 
\def \newmybeta{\ifnum\themyindbeta=-1$b'$\else \ifnum\themyindbeta=0{$`b$}\else${`b}_{\themyindbeta}$\fi\fi\stepcounter{myindbeta}}
\newcounter{myindgamma} 
\def \newmygamma{\ifnum\themyindgamma=-1$`g'$\else \ifnum\themyindgamma=0{$`g$}\else${`g}_{\themyindgamma}$\fi\fi\stepcounter{myindgamma}}
\newcounter{myinddelta} 
\def \newmydelta{\ifnum\themyinddelta=-1$`d'$\else \ifnum\themyinddelta=0{$`d$}\else${`d}_{\themyinddelta}$\fi\fi\stepcounter{myinddelta}}
\newcounter{myindrho} 
\def \newmyrho{\ifnum\themyindrho=-1$`r'$\else \ifnum\themyindrho=0{$`r$}\else${`r}_{\themyindrho}$\fi\fi\stepcounter{myindrho}}

 \def \resetLvariables{%
\setcounter{myindbeta}{0}%
\setcounter{myindgamma}{0}%
\setcounter{myinddelta}{0}%
\setcounter{myindrho}{0}%
}

 \def \XHLSem[#1] #2 {\setbox199=\hbox{$#2$}%
{\Sem{\LTerm<#1>}}\kern-1\point\raise5\point\hbox{\tiny $\cal X$}\kern-6\point_{#2} %{\box199}
\kern2\point}

 \def \SHTerm [#1] #2 {\XHLSem{\LTerm<#1>} {#2} }

 \def \XHLTerm [#1] #2 {\if@Xom \Xomfalse \XHLTermAux[#1] #2 \Xomtrue \resetLvariables 
 \else \XHLTermAux[#1] #2 \fi}

 \def \XHLTermAux[%]
	{\@ifnextchar L{\XHLTermAbs[}%]
	{\@ifnextchar A{\XHLTermApp[}%]
	{\@ifnextchar D{\XHLTermDpp[}%]
	{\@ifnextchar V{\XHLTermVar[}%]
	{\@ifnextchar S{\XHLTermSub[}%]
	{\@ifnextchar T{\XHLTermTub[}%]
	{\@ifnextchar({\XHLTermBr[}%])
	{\@ifnextchar<{\XHLTermLTerm[}%])
		{\XHLSem[}%]
}}}}}}}}

 \def \XHLTermApp[A #1 #2] #3 {%A %
\@ifnextchar<{\XHLTermAppPar [A {#1} {#2}] {#3} }{\XHLTermAppNoPar [A {#1} {#2}] {#3} }}

 \def \XHLTermAppNoPar[A #1 #2] #3 {%A %
\cut { \XHLTermAux [#1] `g } `g + x {\imp { \XHLTermAux [#2] `b } `b [x] z \caps<z,#3> } }

 \def \XHLTermAppPar[A #1 #2] #3 <#4>{P%A %
\cut { \XHLTermAux [#1] `g } `g + x {\imp { \XHLTermAux [#2] {#4} } {#4} [x] z \caps<z,#3> } }

 \def \XHLTermDpp[D #1 #2] #3 {%D %
\cutdl { \XHLTermAux [#1] `g } `g + x { \imp { \XHLTermAux [#2] `b } `b [x] z \caps<z,#3> } }

 \def \XHLTermAbs[L #1 . #2] #3 {%E %
\@ifnextchar<{\XHLTermAbsPar [L {#1} . {#2}] {#3} }{\XHLTermAbsNoPar [L {#1} . {#2}] {#3} }}

 \def \XHLTermAbsNoPar[L #1 . #2] #3 {%E %
\exp #1 {\XHLTermAux [#2] `d } `d . #3 }

 \def \XHLTermAbsPar[L #1 . #2] #3 <#4>{%E %
\exp #1 {\XHLTermAux [#2] #4 } #4 . #3 }

 \def \XHLTermVar[V #1] #2 {%V 
 \caps<#1,#2> }

 \def \XHLTermSub[S #1 #2 := #3] #4 {% S M x := N 
 \cut {\XHLTermAux [#3] `b } `b + {#2} {\XHLTermAux [#1] #4 } }

 \def \XHLTermBr[(#1)] #2 {%B 
\XHLTermAux [#1] #2 }

 \def \XHLTermLTerm[<#1>] #2 {%L
\XHLSem [#1] #2 }

\def \Neg#1{\overline{#1}} 
\def \NegBr(#1){({#1)^o}}
\def \NegNoBr#1{{{#1}^o}}
\def \Neg{\@ifnextchar({\NegBr}{\NegNoBr}}

\def \SemGN#1{\slangle{#1}\srangle\kern-1\point_{\ftn}^{l}}
\def \SemDN#1{\slangle{#1}\srangle\kern-1\point_{\ftn}^{r}}
\def \SemGV#1{\slangle{#1}\srangle\kern-1\point_{\ftv}^{l}}
\def \SemDV#1{\slangle{#1}\srangle\kern-1\point_{\ftv}^{r}}

\def \SemLx[#1] #2 {\SemThree{#1}{#2}{\mbox{\mysemcolour	\tiny $`l$\sc x}}}

\def \LmmtSemN#1{\semEl{#1}\kern-1\point_{\ftn}}
\def \LmmtSemV#1{\semEl{#1}\kern-1\point_{\ftv}}

\newcounter{sind}0
\newcounter{tind}0
\def \newsigma{\if@om\omfalse`s\setcounter{sind}0\setcounter{tind}0\else\stepcounter{sind}`s_{\thesind}\fi}
\def \newtau{\ifnum\thetind=0\stepcounter{tind}`t\else`t_{\thetind}\fi}

\def \lefttop {{\leavevmode \kern2\adjustablepoint
\psset{unit=.9\adjustablepoint,linewidth=0.4\adjustablepoint}%{,linecolor=Redcol}%
\ifmycolour\psset{linecolor=semcolour}\fi
\hbox{\psline{c-c}(0,9)(3,9)\psline{cc-cc}(1.5,9)(1.5,-2)\psline{cc-cc}(0,9)(0,-2)\psline{c-c}(0,-2)(3,-2)}\kern3.5\adjustablepoint}}

\def \righttop{{\kern2\adjustablepoint
\psset{unit=.9\adjustablepoint,linewidth=0.4\adjustablepoint}%{,linecolor=Redcol}%
\ifmycolour\psset{linecolor=semcolour}\fi
\hbox{\psline{c-c}(1.5,9)(-1.5,9)\psline{cc-cc}(0,9)(0,2)\psline{cc-cc}(1.5,9)(1.5,2)}\kern1\adjustablepoint}}

\def \leftsem {\leavevmode \kern1\point
\psset{unit=.9\point,linewidth=0.4\point,linecolor=black}%
\ifmycolour\psset{linecolor=semcolour}\fi
\hbox{\psline{c-c}(0,9)(3,9)\psline{cc-cc}(1.5,9)(1.5,-2)\psline{cc-cc}(0,9)(0,-2)\psline{c-c}(0,-2)(3,-2)}\kern3.5\point}

\def \rightsem{\kern2\point
\psset{unit=.9\point,linewidth=0.4\point,linecolor=black}%
\ifmycolour\psset{linecolor=semcolour}\fi
\hbox{\psline{c-c}(1.5,9)(-1.5,9)\psline{cc-cc}(0,9)(0,-2)\psline{cc-cc}(1.5,9)(1.5,-2)\psline{c-c}(1.5,-2)(-1.5,-2)}\kern1\point}

\def \ContSub #1 #2 := #3 . #4 { {#1} \hskp \slangle{#2}{\,:=\,}{#3}{`.}{#4}\srangle }
\def \Sub #1 #2 := #3 { {#1} \hskp \slangle{#2}{\,:=\,}{#3}\srangle }
\def \VSub #1 #2 := #3 {{#1} `<\Vec{{#2}{\,:=\,}{#3}}`> }
\def \exsub #1 := #2 {\lterm{ `<{#1}{\,:=\,}{#2}`> }}
\def \imsub #1/#2 {[#1/#2]}

\def \Lx {\mbox{$\Red `l${\bf\Red x}}}

\def \LK{\textsc{lk}}

\setbox139=\hbox{{\Yellow \raise0\point\hbox{\rotatebox{20}{$\downarrow$}}\kern-2.5pt\raise1.72\point\hbox{\rotatebox{-20}{$\downarrow$}}}}
 \def \eqX	{\stackrel{\mbox{\scriptsize $\cal X$}}{=}}
\def \eqXCBN	{\mathbin{\eqX\kern-2pt_{\ftn}}}
\def \eqXCBV	{\mathbin{\eqX\kern-2pt_{\ftv}}}

\def \crXCBN	{\mathbin{\downarrow}\kern-.5\point_{\ftn}}
\def \crXCBV	{\mathbin{\downarrow}\kern-.5\point_{\ftv}}

\def \openL{`L\kern-1\point^{\mbox{\scriptsize \sc o}}}
\def \closedL{`L\kern-1\point^{\mbox{\scriptsize \sc c}}}
\def \BothTerm{\mathrel{\stackrel{\raise 1.5\point \hbox{\tiny{$\bullet$}}}{\approx}}}

\def \mutilde	{\tilde{`m}}
\def \mt	{\mutilde}

\def \lmmt	{{\Red \overline{`l}`m\mutilde}}

\def \lmu	{{\Red `l`m}}

\def \redLK {\mathrel{{\rightarrow}\kern-1\point_{\cal LK}}}
\def \rtcredLK {\mathrel{{\rightarrow}\kern-1pt^*\kern-3\point_{\scriptLK}}}
\def \eqX	{\stackrel{{\cal X}}{ = }}
\def \redX {\mathrel{{\rightarrow}\kern-2\point_{\cal X}}}

\def \rtcredX {\mathrel{{\rightarrow}\kern-1pt^*\kern-3\point_{\cal X}}}
\def\scriptLK{\mbox{\tiny $\cal LK$}}
\def \red {\mathrel{\Red \arrow}}

\def \rtcred {\mathrel{\Red \arrow\kern-1\point^*}}
\def \dred {\Red \rightarrow\kern-.75em\rightarrow}

\def \redCBV {\mathrel{\Red \red_{\ftv}}}
\def \redCBN {\mathrel{\Red \red_{\ftn}}}

\def \redlmu {\mathrel{\arrow\kern-1\point_{`l`m}}}

 \def \redbmu {\mathrel{\Red \arrow_{`b\kern-.5\point`m}}}
 \def \rtcredbmu {\mathrel{\Red \arrow^{\ast}_{`b\kern-.5\point`m}}}
 \def \redlazy {\mathrel{\Red \arrow_{\kern-.5pt\mbox{\scriptsize \sc l}}}}
 \def \redom {\mathrel{\Red \arrow_{\kern-.5pt\mbox{\scriptsize \sc om}}}}
 \def \rtcredlazy {\mathrel{\Red \arrow^{\ast}_{\kern-.5pt\mbox{\scriptsize \sc l}}}}
 \def \reds {\mathrel{\Red \arrow_{\kern-.5\point\sind}}}
 \def \redh {\mathrel{\Red \arrow_{\kern-.5\point\hind}}}
 \def \rtcredh {\mathrel{\Red \arrow^{*}_{\kern-.5\point\mbox{\hind}}}}
 \def \rtcredsub {\mathrel{\Red \arrow^{*}_{\kern-.5\point\mbox{\scriptsize {\rm :=}}}}}
 \def \redxh {\mathrel{\Red \arrow_{\kern-.5\point\mbox{\xhind}}}}
 \def \tcredxh {\mathrel{\Red \arrow_{\kern-.5\point\xhind}}}
 \def \rtcredxh {\mathrel{\Red \arrow^{\ast}_{\kern-.5\point\xhind}}}
 \def \eqxh {\mathrel{\Red =_{\kern-.5\point\xhind}}}
 \def \redx {\mathrel{\Red \arrow_{\kern-.5\point\mbox{\scriptsize {\bf x}}}}}
 
 \def \redxsub {\mathrel{\Red \arrow_{\kern-.5\point\mbox{\scriptsize {\rm :=}}}}}
 \def \tcredx {\mathrel{\Red \arrow^{+}_{\kern-.5\point\mbox{\scriptsize {\bf x}}}}}
 \def \rtcredx {\mathrel{\Red \arrow^{*}_{\kern-.5\point\mbox{\scriptsize {\bf x}}}}}
 \def \rtcredxsub {\mathrel{\Red \arrow^{*}_{\kern-.5\point\mbox{\scriptsize {\rm :=}}}}}
 \def \redxV {\mathrel{\Red \arrow_{\kern-.5\point\mbox{\scriptsize {\bf x}{\sc v}}}}}
 \def \redxl {\mathrel{\Red \arrow_{\kern-.5\point\mbox{\scriptsize {\bf x}{\sc l}}}}}
 \def \redxs {\mathrel{\Red \arrow_{\kern-.5\point\xsind}}}
 \def \eqxs {\mathrel{\Red =_{\kern-.5\point\xsind}}}
 \def \crxs {\mathbin{\downarrow_{\kern-.5\point\xsind}}}

 \def \rtcredxs {\mathrel{\Red \arrow^{*}_{\kern-.5\point\xsind}}}
 \def \redxsa {\mathrel{\Red \arrow'_{\kern-.5\point\xsind}}}
 \def \tcreds {\mathrel{\Red \arrow^{+}_{\kern-.5\point\sind}}}
 \def \tcredl {\mathrel{\Red \arrow^{+}_{\kern-.5\point\mbox{\scriptsize {\sc l}}}}}
 \def \tcredxl {\mathrel{\Red \arrow^{+}_{\kern-.5\point\mbox{\scriptsize {\bf x}{\sc l}}}}}
 \def \tcredxs {\mathrel{\Red \arrow^{+}_{\kern-.5\point\xsind}}}
 \def \rtcreds {\mathrel{\Red \arrow^{\ast}_{\kern-.5\point\sind}}}
 \def \rtcredl {\mathrel{\Red \arrow^{\ast}_{\kern-.5\point\mbox{\scriptsize {\sc l}}}}}
 \def \rtcredx {\mathrel{\Red \arrow^{\ast}_{\kern-.5\point\mbox{\scriptsize {\bf x}}}}}
 \def \rtcredxl {\mathrel{\Red \arrow^{\ast}_{\kern-.5\point\mbox{\scriptsize {\bf x}{\sc l}}}}}
 \def \rtcredxs {\mathrel{\Red \arrow^{\ast}_{\kern-.5\point\xsind}}}
\def \redes {\mathrel{\Red \arrow_{\kern-.5\point\mbox{\scriptsize \sc es}}}}
 \def \dreds {\mathrel{\Red \arrow\kern-8\point\arrow_{\kern-.5\point\sind}}}

\def \Pair<#1,#2>{`<{#1},{#2}`>}

\def \Group<#1>{\slangle{#1}\srangle}

\def \bp(#1){{\typecolour \It{bp}({\termcolour #1})}}
\def \bs(#1){{\typecolour \It{bs}({\termcolour #1})}}
\def \BV#1{\textit{bv}(#1)}
\def \bv(#1){\BV{#1}}

\def \FC#1{\fc(#1)}
\def \FP#1{\fp({#1})}
\def \FS#1{{\fs({#1})}}
\def \fp(#1){\textrm{\it fp}({\termcolour #1})}
\def \fs(#1){\textrm{\it fs}\skp({\termcolour #1})}
\def \fc(#1){\textrm{\it fc}\skp({\termcolour #1})}
\def \FV#1{\textrm{\it fv}\skp({\termcolour #1})}
\def \fv(#1){\FV{#1}}
\def \HN#1{\textrm{\it hn}\skp({\termcolour #1})}
\def \hn(#1){\HN{#1}}
\def \HV#1{\textrm{\it hv}\skp({\termcolour #1})}
\def \hv(#1){\HV{#1}}

\def \Mid%{\mid}%
{\hspace*{2.5pt}{\black \mid}\hspace*{2.5pt}}
\def \Langle{\kern1\adjustablepoint
\psset{unit=1\adjustablepoint,linewidth=0.4\adjustablepoint,linecolor=ltermcol}%
\put(0,2.5){\psline{c-c}(0,0)(1.7,4.9)%
\pscurve{c-c}(1.7,4.9)(.7,0)(1.7,-4.9)
\psline{c-c}(0,0)(1.7,-4.9)
}\kern3\adjustablepoint}
\def \Rangle{\psset{unit=1\adjustablepoint,linewidth=0.4\adjustablepoint,linecolor=ltermcol}%
\kern3\adjustablepoint
\put(0,2.5){\psline{c-c}(0,0)(-1.7,4.9)%
\pscurve{c-c}(-1.7,4.9)(-.8,0)(-1.7,-4.9)+
\psline{c-c}(0,0)(-1.7,-4.9)
}\kern\adjustablepoint}
\def \PiCell#1#2{\setbox171=\hbox{$#1$}\setbox172=\hbox{$#2$}
{\ltermcolour
\Langle 
	\ifdim\wd171>24\point\skp{\box171}\skp\else{\box171}\fi
\kern1pt{\mid}\kern1pt
	\ifdim\wd172>24\point\skp{\box172}\skp\else{\box172}\fi
\Rangle 
}}
\def \Cell#1#2{\slangle#1\skp{\mid}\skp#2\srangle}
\def \cell<#1|#2>{\Cell{#1}{#2}}

\def \repl #1[#2/#3]{\lterm{#1}{\skp\Black [}\lterm{#2}{\Black /}\lterm{#3}{\Black ]} }
\def \term#1 {{\termcolour #1}}
\def \ass#1:#2 {{\termcolour #1}{\black :}{\typecol #2} }
\def \lass#1:#2 {\lterm {#1}{\black :}{\typecol #2} }

\def \type #1 {{\typecolour #1}}
\def \lterm#1{{\ltermcolour #1}}
\def \lambdaterm #1 {{\ltermcolour #1}}
\def \Rule(#1){{\blue (#1) }}

\def \Stat#1:#2{{\ltermcolour #1}{:}{\typecolour #2}}
\def \stat#1#2{{\ltermcolour #1}{:}{\typecolour #2}}
\def \lstat#1#2{{\ltermcolour #1}\,{:\,}{\typecolour #2}}

\def \turn	{{\black {\vdash}}}
\def \Turn {\mathrel{\turn}}
\def \TurnSimple {\mathrel{\turn}}
\def \TurnD {\mathrel{\turn\kern-3\point_{\mbox{\tinyfont $\textit{\black D}$}}}}
\def \TurnI {\mathrel{\turn\kern-3\point_{\mbox{\tinyfont $\black \cap$}}}}
\def \TurnL {\mathrel{\turn\kern-.5\point_{`l}}}
\def \TurnLK {\mathrel{\Yellow \turn\kern-2\point_{\ftsc{lk}}}}
\def \TurnLKarr {\mathrel{\Yellow \turn\kern-2\point_{\ftsc{lk}({\rightarrow})}}}
\def \TurnR {\mathrel{\Yellow \turn\kern-2\point_{\ftsc{r}}}}
\def \TurnS {\mathrel{\turn\kern-2\point_{\ftsc{s}}}}
\def \TurnLN {\mathrel{\Yellow \turn\kern-2\point_{\ftsc{ln}}}}
\def \TurnLV {\mathrel{\Yellow \turn\kern-2\point_{\ftsc{lv}}}}

\def \TurnLmmt {\mathrel{\black \turn_{\sclmmt}}}
\def \TurnLmmtSN {\mathrel{\turn_{\sclmmt}^{\ftsc{sn}}}}
\def \TurnLmmtN {\mathrel{\turn\kern-2\point_{\black \ftsc{n}}}}
\def \TurnLmmtV {\mathrel{\turn\kern-2\point_{\black \ftsc{v}}}}
\def \TurnLmmtmin {\mathrel{\turn\kern-2\point_{\black \ftsc{m}}}}
\def \TurnLmmtmin {\mathrel{\turn}}
\def \TurnLmux {\mathrel{\turn\kern-\point_{\xind}}}
\def \TurnLmu {\mathrel{\turn\kern-\point_{`l`m}}}
\def \TurnLmuIU {\mathrel{\,\turn\kern-2\point_{`l`m}^{\,\raise 1\point\hbox{\tinyfont $\cap\cup$}}\,}}
\def \TurnLx {\mathrel{\turn_{`l\mbox{\scriptsize \sf x}}}}
\def \TurnMCL {\mathrel{\Yellow \turn\kern-2\point_{\ftsc{mcl}}}}
\def \TurnN {\mathrel{\Yellow \turn\kern-2\point_{\ftsc{n}}}}
\def \TurnNf {\mathrel{\Yellow \turn\kern-2\point_{\ftrm{n}}}}
\def \TurnV {\mathrel{\Yellow \turn\kern-2\point_{\ftsc{v}}}}
\def \TurnVf {\mathrel{\Yellow \turn\kern-2\point_{\ftrm{v}}}}
\def \TurnX {\mathrel{\turn\kern-2\point_{\cal X}}}
\def \TurnP {\mathrel{\turn\kern-3\point_p}}
\def \TurnM {\mathrel{\turn\kern-4\point_{\mbox{\tinyfont $\cal M$}}}}
\def \TurnMc {\mathrel{\turn\kern-4\point_{\mbox{\tinyfont $\cal M^{\tinysc{c}}$}}}}
\def \TurnMe {\mathrel{\turn\kern-4\point_{\mbox{\tinyfont $\cal M^{\tinysc{e}}$}}}}
\def \TurnMIU {\mathrel{\turn\kern-4\point_{\mbox{\tinyfont $\cal M^{\cap\cup}$}}}}
\def \TurnMI {\mathrel{\turn\kern-4\point_{\mbox{\tinyfont $\cal M^{\cap}$}}}}

\def \sclmmt {\mbox{%\scriptsize
\tiny $\overline{`l}`m\mt$}}

\def \Der#1#2{\@ifnextchar{\bgroup}{\DerThree{#1}{#2}}{\DerTwo{#1}{#2}}}
\def \DerTwo#1#2{{\typecolour #1 \Turn #2}}
\def \DerThree#1#2#3{{\typecolour #2 \Turn #3}}
\def \Stoup#1#2{\lstat{#1}{#2}}
\def \StoupR#1#2{\lstat{#1}{#2} \Mid}
\def \StoupR#1#2{\framebox{$#1\,{:}\,#2$}~}
\def \StoupR#1#2{\StoupFrame{\lstat{#1}{#2}}}

\def \Stoup#1#2{\lstat{#1}{#2}}

\def \StoupR#1#2{\lstat{#1}{#2}\Skip\Mid}
\def \DerLmuThree#1#2#3{#1 \TurnLmu \Stoup{#2}{#3} }
\def \DerLmuFour#1#2#3#4{#1 \TurnLmu \StoupR{#2}{#3} #4}
\def \DerLmu#1#2#3{\@ifnextchar{\bgroup}{\DerLmuFour{#1}{#2}{#3}}{\DerLmuThree{#1}{#2}{#3}}}
\def \derLmuFour #1 |- #2 : #3 | #4 {\type{#1} \TurnLmu \Stoup{#2}{#3} \Mid \type{#4} }
\def \derLmuNormal #1 |- #2 : #3 {\@ifnextchar|{\derLmuFour #1 |- #2 : #3 }{#1 \TurnLmu \Stoup{#2}{#3} }}
\def \derLmuCmd #1 |- #2 | #3 {\type{#1} \TurnLmu \lterm{#2} \Mid \type{#3} }
\def \derLmuBasis #1 |- #2 {\@ifnextchar:{\derLmuNormal #1 |- #2 }{\derLmuCmd #1 |- #2 }}
\def \derLmu {\@ifnextchar|{\derLmuBasis {\emptyset} }{\derLmuBasis }}

\def \derLmuIU #1 {\def \TurnLmu {\TurnLmuIU}\derLmu {#1} }
\def \derLmux #1 {\def \TurnLmu {\TurnLmux}\derLmu {#1} }
\def \derLmuP  #1 {\def \TurnLmu {\skipper\TurnP\skipper}\derLmu {#1} }
\def \derLmuS  #1 {\def \TurnLmu {\TurnS}\derLmu {#1} }

\def \derMCLFour #1 |- #2 | #3 {#1 \TurnMCL {#2} \mid {#3} }
\def \derMCLNormal #1 |- #2 {\@ifnextchar|{\derMCLFour #1 |- #2 }{#1 \TurnMCL {#2} }}
\def \derdeMCL #1 |- | #2 {#1 \TurnMCL \mid {#2} }
\def \derMCL #1 |- {\@ifnextchar|{\derdeMCL #1 |- }{\derMCLNormal {#1} |- }}
\def \DerLmmt#1#2{\@ifnextchar{\bgroup}{\DerLmmtThree{#1}{#2}}{\DerLmmtTwo{#1}{#2}}}
\def \DerLmmtTwo#1#2{{\typecolour #1 \TurnLmmt #2}}
\def \DerLmmtThree#1#2#3{#1~{:}~{\typecolour #2 \TurnLmmt #3}}

\def \derLmmtTwo #1 |- #2 {{\typecolour {#1} \TurnLmmt {#2}}}
\def \derLmmtCommand #1 : #2 |- #3 {{\ltermcolour {#1}}~{\black :}~{\typecolour {#2} \TurnLmmt {#3}}}
\def \derLmmtTerm #1 |- #2 : #3 | #4 {{\typecolour {#1} \TurnLmmt {\ltermcolour #2}\skipper{\black :}\skipper{#3} ~{\black \mid}~ {#4} }}
\def \derLmmtContext #1 | #2 : #3 |- #4 {{\typecolour {#1}~{\black \mid}~{\ltermcolour #2}\skipper{\black :}\skipper{#3} \TurnLmmt {#4} }}
\def \derLmmtActivated #1 |{\@ifnextchar-{\derLmmtTerm {#1} |}{\derLmmtContext {#1} | }}
\def \derLmmtNamedDer #1 :: {{#1} \dcol \derLmmt }
\def \derLmmtColon #1 :{\@ifnextchar:{\derLmmtNamedDer #1 :}{\derLmmtCommand {#1} : }}
\def \derLmmt #1 {\@ifnextchar:{\derLmmtColon {#1} }{\derLmmtActivated {#1} }}
\def \derLmmtV #1 {\def \TurnLmmt{\TurnLmmtV}\derLmmt {#1} }
\def \derLmmtN #1 {\def \TurnLmmt{\TurnLmmtN}\derLmmt {#1} }
\def \derLmmtSN #1 {\def \TurnLmmt{\TurnLmmtSN}\derLmmt {#1} }
\def \derLmmtmin #1 {\def \TurnLmmt{\TurnLmmtmin} \derLmmt {#1} }
\def \derLmmtM #1 {\def \TurnLmmt{\TurnM} \derLmmt {#1} }
\def \derLmmtMc #1 {\def \TurnLmmt{\TurnMc} \derLmmt {#1} }

\def \derL #1 |- #2 : #3 {{\typecolour #1 \TurnL {\ltermcolour #2}\,{\black :}\,#3}}
\def \derLx #1 |- #2 : #3 {{\typecolour #1 \TurnLx {\ltermcolour #2}\,{\black :}\,#3}}
\def \derI #1 |- #2 : #3 {{\typecolour #1 \TurnI {\ltermcolour #2}\,{\black :}\,#3}}
\def \der #1 |- #2 : #3 {{\typecolour #1 \TurnSimple {\ltermcolour #2}\,{\black :}\,#3}}

\def \DerX#1#2#3{\PDer{#1}{#2}{#3}}

\def \derLK #1 |- #2 {{\typecolour #1}	\TurnLK	{\typecolour #2}}
\def \derLKarr #1 |- #2 {{\typecolour #1}	\TurnLKarr	{\typecolour #2}}
\def \deriv #1 |- #2 {{\typecolour #1}	\Turn	{\typecolour #2}}

\setbox161=\hbox{$\cdot$}
\setbox162=\hbox{{\neutral ~\raise-2.5\point\copy161\kern-\wd161\raise2.5\point\copy161\kern.5\point \raise0\point\copy161~}}
\def \DP {\copy162}
\def \DPa {{\neutral \raise-.7\point\hbox{~\large :~}}}

\def \PDer#1#2#3{{\termcolour #1} \DP {\typecolour #2} \Turn {\typecolour #3}}
\def \pDer#1:#2 |- #3 {{\termcolour #1} \DP {\typecolour {#2} \Turn {#3}}}
\def \InfPDerFour#1#2#3#4{ \InfBox{#1}{ \PDer{#2}{#3}{#4} } }
\def \InfPDer#1#2#3{\@ifnextchar{\bgroup}{\InfPDerFour{#1}{#2}{#3}}{\InfPDerFour{}{#1}{#2}{#3}}}
\def \PDerN#1#2#3{\typecolour {\termcolour #1 }\DP {\typecolour #2} \TurnN {\typecolour #3}}

\def \derXaux #1 : #2 |- #3 {{\termcolour #1 } \DP {\typecolour #2}	\TurnX	{\typecolour #3}}
\def \derX #1 : {\@ifnextchar|{\derXaux {#1} : {} }{\derXaux {#1} : }}

\def \derXR #1 : #2 |- #3 {{\termcolour #1 } \DP {\typecolour #2}	\TurnR	{\typecolour #3}}
\def \derXRN #1 : #2 |- #3 {{\termcolour #1 } \DP {\typecolour #2}	\TurnLN	{\typecolour #3}}
\def \derXRV #1 : #2 |- #3 {{\termcolour #1 } \DP {\typecolour #2}	\TurnLV	{\typecolour #3}}
\def \derXN #1 : #2 |- #3 {{\termcolour #1 } \DP {\typecolour #2}	\TurnN	{\typecolour #3}}
\def \derXNf #1 : #2 |- #3 {{\termcolour #1 } \DP {\typecolour #2}	\TurnNf	{\typecolour #3}}
\def \derXV #1 : #2 |- #3 {{\termcolour #1 } \DP {\typecolour #2}	\TurnV	{\typecolour #3}}
\def \derXVf #1 : #2 |- #3 {{\termcolour #1 } \DP {\typecolour #2}	\TurnVf	{\typecolour #3}}
\def \PDerV#1#2#3{\typecolour {\termcolour #1 }\DP {\typecolour #2} \TurnV {\typecolour #3}}
\def \InfPDerFourN#1#2#3#4{ \InfBox{#1}{ \PDerN{#2}{#3}{#4} } }
\def \InfPDerN#1#2#3{\@ifnextchar{\bgroup}{\InfPDerFourN{#1}{#2}{#3}}{\InfPDerFour{}{#1}{#2}{#3}}}
\def \InfPDerFourV#1#2#3#4{ \InfBox{#1}{ \PDerV{#2}{#3}{#4} } }
\def \InfPDerV#1#2#3{\@ifnextchar{\bgroup}{\InfPDerFourV{#1}{#2}{#3}}{\InfPDerFour{}{#1}{#2}{#3}}}

\def \derPure #1 : #2 |- #3 {{\termcolour #1 } \DP {\typecolour #2}	\Turn_{\kern-2pt\ftsc{p}}	{\typecolour #3}}
\def \derPureN #1 : #2 |- #3 {{\termcolour #1 } \DP {\typecolour #2}	\Turn_{\kern-2pt\ftsc{pn}}	{\typecolour #3}}
\def \derPureV #1 : #2 |- #3 {{\termcolour #1 } \DP {\typecolour #2}	\Turn_{\kern-2pt\ftsc{pv}}	{\typecolour #3}}

\def \derDP #1 : #2 |- #3 {{\termcolour #1 } \DP {\typecolour #2}	\Turn	{\typecolour #3}}
\def \LogicDer #1 |- #2 { {\typecolour #1}	\Turn	{\typecolour #2}}

\def \InfDerX #1 :: #2 : #3 |- #4 {\InfPDer{#1}{#2}{#3}{#4}}
\def \InfDer #1 :: #2 |- #3 : #4 {\InfLDer{#1}{#2}{#3}{#4}}

\def \XIn#1#2{{\widehat{#1}}\hspace*{.75pt}#2}
\def \XOut#1#2{#1\hspace*{1pt}{\widehat{#2}}}

\def \Dagger	{\mathrel{\mbox{\oldmath\symbol{121}}}}
\def \RedDagger	{\mathrel{\raise-1\point\hbox{%\lightredcol
 \largeoldmath \symbol{121}}}}
\def \DaggerA{{\Yellow \Dagger\kern-3\point_{\mbox{\scriptsize \sc a}}}}
\def \DaggerLog{\kern-1\point{\Yellow \Dagger\kern-1.5\point_{\mbox{\scriptsize \sc l}\kern-1.5\point}}}

\def \DaggerLeft{\mathop{\kern-1\point{\specialcol \raise2\point\hbox{\rotatebox{-30}{$\Dagger$}}}\kern-1\point}}
\def \DaggerRight{\mathop{\kern-1\point{\specialcol \raise0\point\hbox{\rotatebox{30}{$\Dagger$}}}\kern-1\point}}

\def \DaggerV{\mathop{\,\Yellow \Dagger_{\mbox{\scriptsize \sc v}}}}
\def \DaggerN{\mathop{\,\Yellow \Dagger_{\mbox{\scriptsize \sc n}}}}
\def \DaggerL{\DaggerLeft}
\def \DaggerR{\DaggerRight}

\def \expT#1#2#3#4{\XOut{\XIn{#1}{#2}}{#3}\kern.2mm{`.}\kern.2mm#4}
\def \medT#1#2#3#4#5{\setbox111=\hbox{$#1$}\setbox112=\hbox{$#5$}%
\ifdim\wd111<10pt%
	\ifdim\wd112<10pt%
		\XOut{\box111}{#2}\,[#3]\,\XIn{#4}{\box112}%
	\else
		\XOut{\box111}{#2}~[#3]~\XIn{#4}{\box112}
	\fi
\else
	\XOut{\box111}{#2}~[#3]~\XIn{#4}{\box112}
\fi}
\def \medTdl#1#2#3#4#5{{\termcolour %
	\XOut{(#1)}{#2}~[#3] }\quad \\ \hfill {\termcolour \XIn{#4}{(#5)} }
}

\def \cutT#1#2#3#4{\XOut{#1}{#2} \Dagger \XIn{#3}{#4}}
\def \ColcutT#1#2#3#4{\XOut{#1}{#2} \RedDagger \XIn{#3}{#4}}
\def \cutlog#1#2#3#4{\XOut{#1}{\orange #2}{\orange \Dagger \XIn{#3}{#4}}}
\def \cutA#1#2#3#4{\XOut{#1}{#2}\DaggerA \XIn{#3}{#4}}
\def \invLT#1#2#3{{#1}\kern.2mm{`.}\kern.2mm\XOut{#2}{#3}}
\def \invRT#1#2#3{\XIn{#1}{#2}\kern.2mm{`.}\kern.2mm{#3}}
\def \negL #1 . #2 #3 {\NegL{#1}{#2}{#3}}
\def \negR #1 #2 . #3 {\NegR{#1}{#2}{#3}}
\def \negLT#1#2#3{{#1}\kern.2mm{`.}\kern.2mm\XOut{#2}{#3}}
\def \negRT#1#2#3{\XIn{#1}{#2}\kern.2mm{`.}\kern.2mm{#3}}
\def \pairT#1#2#3#4#5{\Group<{#1}\widehat{#2},{#3}\widehat{#4}>\kern.2mm{`.}\kern.2mm{#5}}
\def \orLT#1#2#3#4#5{{#1}\kern.2mm{`.}\kern.2mm (\widehat{#2}{#3}\kern.2mm{+}\kern.2mm\widehat{#4}{#5}) }
\def \inLT#1#2#3#4{{#1}(\widehat{#2}\kern.2mm{+}\kern.2mm{#3})\kern.2mm{`.}\kern.2mm{#4}}
\def \inRT#1#2#3#4{{#1}\,({#2}\kern.2mm{+}\kern.2mm\widehat{#3})\kern.2mm{`.}\kern.2mm{#4}}
\def \orRPT#1#2#3#4{{#1}\,(\widehat{#2}\kern.3mm{\mid}\kern.3mm\widehat{#3})\kern.2mm{`.}\kern.2mm{#4}}

\def \newconRT#1#2#3#4{\Group<{#1},{#2}>{#3}\kern.2mm{`.}\kern.2mm{#4}}
\def \newconLT#1#2#3#4#5{#1\kern.2mm{`.}\kern.2mm\Group<{#2}\widehat{#3},{#4}\widehat{#5}>}

\def \CBV{\Sc{cbv}}
\def \CBN{\Sc{cbn}}

\def \Weak{\mysl{W}}

\omtrue
\def \InvLterm#1#2#3{{\termcolour \if@om \omfalse\invLT{#1}{#2}{#3}\omtrue
	\else	\skp(\invLT{#1}{#2}{#3})\fi}}
\def \InvRterm#1#2#3{{\termcolour \if@om \omfalse\invRT{#1}{#2}{#3}\omtrue
	\else	\skp(\invRT{#1}{#2}{#3})\fi}}
\def \NegL#1#2#3{{\termcolour \if@om \omfalse\negLT{#1}{#2}{#3}\omtrue
	\else	\skp(\negLT{#1}{#2}{#3})\fi}}
\def \NegR#1#2#3{{\termcolour \if@om \omfalse\negRT{#1}{#2}{#3}\omtrue
	\else	\skp(\negRT{#1}{#2}{#3})\fi}}
\def \NewConR#1#2#3#4{{\termcolour \if@om \omfalse\newconRT{#1}{#2}{#3}{#4}\omtrue	\else	(\newconRT{#1}{#2}{#3}{#4})\fi}}
\def \NewConL#1#2#3#4#5{{\termcolour \if@om \omfalse\newconLT{#1}{#2}{#3}{#4}{#5}\omtrue	\else	(\newconLT{#1}{#2}{#3}{#4}{#5})\fi}}

\def \nonconnectable{\mbox{\tiny $\Box$}}
\def \nonconnectable{\diamond}
\def \newconR < {\@ifnextchar{o}{\newconRR < }{\newconRL < }}
\def \newconRL < #1 , #2 > #3 . #4 {\NewConR{\widehat{#1}}{\nonconnectable}{#3}{#4}}
\def \newconRR < #1 , #2 > #3 . #4 {\NewConR{\nonconnectable}{\widehat{#2}}{#3}{#4}}

\def \newconL #1 . < #2 #3 , #4 #5 > {\NewConL{#1}{#2}{#3}{#4}{#5}}

\newlength\harpoonlength

\def \longharpoon#1{\psset{unit=.7\point,linewidth=0.4\point,linecolor=ltermcol}%
\harpoonlength#1\addtolength{\harpoonlength}{-1\point}%
\psline{cc-cc}(-1,1.25)(-\harpoonlength,1.25)%
\pscurve{cc-cc}(-0.7,1.25)(-1.5,1.44)(-2,1.8)(-2.5,2.3)(-3,3.1)
\vbox to 2\point{}%
}

\def \Vec#1{\setbox155=\hbox{$#1$}%
\leavevmode\copy155\raise\ht155\hbox{\longharpoon{\wd155}}}

\def \longarrow#1{\psset{unit=1\point,linewidth=0.35\point}%
\psline{cc-cc}(0,0)(#1,0)%
\hspace*{#1}%
\pscurve{cc-cc}(0,0)(-1.1,.4)(-2.5,2.2)%
\pscurve{cc-cc}(0,0)(-1.1,-.4)(-2.5,-2.2)}

\def \Rel#1#2{\setbox155=\hbox{\scriptsize $#1$}\setbox156=\hbox{$#2$}%
\mathrel{\raise5.5pt\copy155\kern-1\wd155%
\kern-.5\wd155\kern-.5\wd156\raise-2pt\copy156\kern-.5\wd156\kern.5\wd155}}

\def \PiOverline#1{%
 \ifmycolour\psset{unit=1\point,linewidth=0.4\point,linecolor=picolour}%
 \else\psset{unit=1\point,linewidth=0.35\point,linecolor=black}
 \fi
 \setbox155=\hbox{$#1$}\leavevmode
 \raise\ht155\hbox{\psline{c-c}(.05\wd155,1.25)(.95\wd155,1.25)}\box155% 
}
\def \Underline#1{%
 \ifmycolour\psset{unit=1\point,linewidth=0.4\point,linecolor=yellow}%
 \else\psset{unit=1\point,linewidth=0.35\point,linecolor=black}
 \fi
 \setbox156=\hbox{${#1}$}\leavevmode
 \raise-\dp156\hbox{\psline{c-c}(.05\wd156,-0.75)(.95\wd156,-0.75)}\box156 }

\def \ul#1{\Underline{\mbox{\scriptsize $#1$}}}

\def \CH#1{\kern1\point\lefttop{\mbox{\Purple $#1$}}\righttop\kern1\point}
\def \LtoLMMT#1{\kern1\point\lefttop{#1}\righttop\kern1\point}
\def \CH#1{\raise1.5\point\hbox{\footnotesize ${\mid}\kern-1.5\point[$}%
#1
\raise1.5\point\hbox{\footnotesize $]\kern-1.5\point{\mid}_{`l}$}}

\def \CutGraph{\@ifnextchar[{\CutGraphN}{\CutGraphN[\GCut]}}%]

\def \CutGraphN[#1]#2#3#4#5{%
\begin{psmatrix}[rowsep=1ex,colsep=.05ex]
&&&[name=cutroot]&\\[3ex]
&&&[name=cut]{#1}&\\
&[name=P]#2&&&&[name=Q]#5& \\
&&[name=a]#3&&[name=x]#4\\&
\end{psmatrix}
 \ncline{cutroot}{cut}
\ncline{cut}{P}
\ncline{cut}{a}
\ncline{cut}{x}
\ncline{cut}{Q}
}

\def \V#1{\setbox71=\hbox{#1}\setbox72=\hbox{\v{}}%
\copy71\kern-.5\wd71\kern-.3\wd72\raise\ht71\hbox{\raise.2em\hbox{\lower\ht72\hbox{\copy72}}}\kern-.7\wd72\kern.5\wd71}

\setbox138=\hbox{$=$}
\setbox139=\hbox{{\scriptsize $\Delta$}}%
\setbox149=\hbox{\raise-1.5\point\copy138\kern-.5\wd138\kern-.5\wd139\raise4\point\hbox{\copy139}\kern-.5\wd139\kern.5\wd138}
\def \ByDef {\mathrel{\copy149}}

\def \Arr{{\Rightarrow}}

\def \Cut{\@ifnextchar{\bgroup}{\CutTwo}{{\mysl{\Red cut}}}}

\def \Exp	{\@ifnextchar{\bgroup}{\ExpT}{{\mysl{exp}}}}
\def \Imp	{\@ifnextchar{\bgroup}{\MedT}{{\mysl{imp}}}}
\def \Med	{\@ifnextchar{\bgroup}{\MedT}{{\mysl{med}}}}
\def \Cap	{\@ifnextchar{\bgroup}{\CapT}{{\mysl{cap}}}}

\def \Ins {\mysl{exp-imp}}

\def \actR	{{\Yellow \DaggerR}\mysl{a}}

\def \deactR	{{\Yellow \DaggerR}\mysl{d}}

\def \renR	{{\termcolour {\DaggerR}\mysl{ren}}}

\def \gcR {{\DaggerR}\mysl{gc}}

\def \Ri {{\DaggerR}\mysl{cap}}
\def \Rii {{\DaggerR}\mysl{exp}}
\def \Riii {{\DaggerR}\mysl{imp-outs}}
\def \Riv {{\DaggerR}\mysl{imp-ins}}
\def \Rv {{\DaggerR}\mysl{cut}}

\def \actL	{\mysl{a}{\Yellow \DaggerL}}

\def \deactL	{\mysl{d}{\Yellow \DaggerL}}

\def \renL	{{\termcolour \mysl{ren}{\DaggerL}}}

\def \gcL {\mysl{gc}{\DaggerL}}

\def \Li {\mysl{cap}{\DaggerL}}
\def \Lii {\mysl{exp-outs}{\DaggerL}}
\def \Liii {\mysl{exp-ins}{\DaggerL}}
\def \Liv {\mysl{imp}{\DaggerL}}
\def \Lv {\mysl{cut}{\DaggerL}}

\def \CutL#1#2{\@ifnextchar{\bgroup}{\CutLeft{#1}{#2}}{\widehat{#1} \DaggerL \widehat{#2}}}
\def \CutR#1#2{\@ifnextchar{\bgroup}{\CutRight{#1}{#2}}{\widehat{#1} \DaggerR \widehat{#2}}}
\def \CutTwo#1#2{\@ifnextchar{\bgroup}{\CutT{#1}{#2}}{\widehat{#1} \Dagger \widehat{#2}}}
\def \ColCut#1#2{\@ifnextchar{\bgroup}{\ColCutT{#1}{#2}}{\widehat{#1} \Dagger \widehat{#2}}}
\def \Cutdl#1#2#3#4{\XOut{#1}{#2} \Dagger\\ \tab \XIn{#3}{#4}}
\def \CutDl#1#2#3#4{\XOut{#1}{#2} \Dagger\\ \tab\tab \XIn{#3}{#4}}

\def \notin {\,{\not\in}\,}

\def \TrTwo#1#2{\Sem{\ltermcolour #1}{\termcolour #2}^{\sclmmt}}%{(#1)_{#2}}
\def \Tran#1{\@ifnextchar{\bgroup}{\TrTwo{#1}}{\TrTwo{#1}{}}}

\def \CapRule#1#2#3#4#5{ % termvar, contextvar, Type, Gamma, Delta,
 (\Cap): &
 \Inf	{ \PDer{ \Cap{#1}{#2} }{#4,\stat{#1}{#3} }{\stat{#2}{#3}, {#5} } }
}

\def \DagRule#1#2#3#4#5#6#7{ % term, context, variable, term, type, Gamma, Delta
 (\Cut):&
 \Inf	{ \PDer{#1}{#6}{ \stat{#2}{#5}, {#7} } \quad \PDer{#4}{{#6}, \stat{#3}{#5} }{#7} }
	{ \PDer{ \Cut{#1}{#2}{#3}{#4} }{#6}{#7} }
}

\def \ImpRule#1#2#3#4#5#6#7#8#9{% 
 (\Imp):&
 \Inf	{ \PDer{#1}{#8}{\stat{#2}{#6}, #9} \quad \PDer{#5}{ #8,\stat{#4}{#7} }{#9} }
	{ \PDer{ \Med{#1}{#2}{#3}{#4}{#5} }{#8, \stat{#3}{{#6}\arr{#7}} }{#9}}
}

\def \ExpRule#1#2#3#4#5#6#7#8{ % termvar, term, contextvar, context, typeA, typeB, Gamma, Delta
 (\Exp):&
 \Inf	{ \PDer{#2}{ {#7},\stat{#1}{#5} }{\stat{#3}{#6}, #8} }
	{ \PDer{ \Exp{#1}{#2}{#3}{#4} }{#7}{ \stat{#4}{#5\arr{#6}}, #8} }
}

\newdimen \GeneWidth \newdimen \GeneW \newdimen \HGeneW
\newdimen \GeneHeight \newdimen \GeneH \newdimen \HGeneH
\newcount \PictW \newcount \HPictW
\newcount \PictH \newcount \HPictH
\newcount \BoxH \newcount \HBoxH
\newcount \BoxW \newcount \HBoxW
\newcount \HOffset \newcount \VOffset

\def \StoupFrameR#1{\setbox159=\hbox{$#1$}%
\PictW\wd159\divide\PictW\unitlength\advance\PictW3%
\PictH\ht159`mltiply\PictH18\divide\PictH10\divide\PictH\unitlength
\,\raise-4\unitlength\hbox{\PSFrame(\PictW,\PictH)}%
\kern1\point\box159\,,}

\def \StoupFrameL#1{\setbox159=\hbox{$#1$}%
\PictW\wd159\divide\PictW\unitlength\advance\PictW3%
\PictH\ht159`mltiply\PictH18\divide\PictH10\divide\PictH\unitlength
,\,\raise-4\unitlength\hbox{\PSFrame(\PictW,\PictH)}%
\kern1\point\box159\,}

\def \Gene#1{\unitlength.09\point\setbox140=\hbox{$#1$}%
\GeneHeight\ht140\advance\GeneHeight55\unitlength%
\GeneWidth\wd140\advance\GeneWidth50\unitlength%
\PictW\GeneWidth\divide\PictW\unitlength%
\PictH\GeneHeight\divide\PictH\unitlength%
\ifdim\wd140>10\point
\begin{picture}(\PictW,\PictH)%
 \put(0,0)	{\makebox(\PictW,\PictH){\box140}}%
 \put(0,0)	{\PSFrame(\PictW,\PictH)}%
\end{picture}
\else \box140
\fi}

\def \InvGene#1{\unitlength.09\point%
\setbox140=\hbox{$#1$}%
\GeneHeight\ht140\advance\GeneHeight60\unitlength
\GeneWidth\wd140\advance\GeneWidth50\unitlength
\PictW\GeneWidth\divide\PictW\unitlength
\PictH\GeneHeight\divide\PictH\unitlength
\begin{picture}(\PictW,\PictH)
 \put(0,0)	{\makebox(\PictW,\PictH){\box140}}
 \put(0,0)	{\PSFrame(0,0)}
\end{picture}}

\newcommand{\DiagCap}[2]{\unitlength.09\point%
\begin{picture}(360,120)(-180,-55)
	\put(-180,0)	{\Yellow \vector(1,0){140}}
	\put(-110,50)	{\makebox(0,0){\small $\termcolour #1$}}
	\put(-40,-30)	{\PSBox(80,80)}
	\put(110,50)	{\makebox(0,0){\small $\termcolour #2$}}
	\put(40,0)	{\Yellow	\vector(1,0){140}}
 \end{picture}}
\def \Diagcaps<#1,#2>{\DiagCap{#1}{#2}}

\newcommand{\DiagExp}[4]{\unitlength.09\point%
\setbox143=\hbox{{\termcolour \small $#2$}}%
\GeneWidth\wd143
	\advance\GeneWidth320\unitlength
	\BoxW\GeneWidth \divide\BoxW\unitlength \HBoxW\BoxW \divide\HBoxW2%
	\advance\GeneWidth200\unitlength \PictW\GeneWidth \divide\PictW\unitlength \HPictW\PictW \divide\HPictW2%
\GeneHeight\ht143 \advance\GeneHeight 80\unitlength
	\BoxH\GeneHeight \divide\BoxH\unitlength \HBoxH\BoxH \divide\HBoxH2%
	\advance\GeneHeight70\unitlength
\PictH\GeneHeight \divide\PictH\unitlength \HPictH\PictH \divide\HPictH2%
\HOffset\wd143 \divide\HOffset\unitlength \divide\HOffset2%
\VOffset\HPictW \divide\VOffset\unitlength \divide\VOffset2 \advance\VOffset20%
 \def \ContentExp {\begin{picture}(\PictW,\PictH)(-\HPictW,-\HPictH)
 \put(-50,0){%
 \VOffset\ht143\divide\VOffset\unitlength\divide\VOffset2\advance\VOffset50%
 \makebox(0,0){\Gene{\begin{picture}(\BoxW,\BoxH)(-\HBoxW,20)
		\put(-\HOffset,\VOffset)	{
		\put(-160,-25) {\Yellow \vector(1,0){150}}
		\put(-85,22)	{\makebox(0,0){\termcolour \small $\widehat{#1}$}}
	}%
	\put(0,\VOffset)	{\makebox(0,0){\box143}}
		\put(\HOffset,\VOffset) {
		\put(10,-25)	{\Yellow \vector(1,0){150}}
		\put(75,20)	{\makebox(0,0){\termcolour \small $\widehat{#3}$}}
	}
 \end{picture}}}}
 \HOffset\HBoxW\advance\HOffset-40
 \put(\HOffset,-15)	{\Yellow \vector(1,0){130}}
 \advance\HOffset65
 \put(\HOffset,35) {\makebox(0,0){\termcolour \small $#4$}}
 \end{picture}}
\if@om \omfalse 
\ContentExp 
\omtrue \else \Gene{\ContentExp} \fi
}

\newdimen \MedW \newdimen \MedH
\newcommand{\DiagMed}[5]{\unitlength.09\point%
\setbox144=\hbox{{\termcolour \small $#1$}}%
\setbox145=\hbox{{\termcolour \small $#5$}}%
\setbox146=\hbox{\termcolour \small $#3$}%
\MedW \wd144 \advance \MedW \wd145 %\advance \MedW \wd146 
\advance \MedW 480\unitlength
\ifdim \ht144 > \ht145 \MedH \ht144 \else \MedH \ht145 \fi \advance \MedH 70\unitlength
\BoxW\MedW\divide\BoxW\unitlength\HBoxW\BoxW\divide\HBoxW2%
\BoxH\MedH\divide\BoxH\unitlength\HBoxH\BoxH\divide\HBoxH2%
\GeneWidth \MedW \advance \GeneWidth 240\unitlength
\GeneHeight \MedH \advance \GeneHeight 70\unitlength
\PictW \GeneWidth \divide \PictW \unitlength \HPictW \PictW \divide \HPictW 2%
\PictH \GeneHeight \divide \PictH \unitlength \HPictH \PictH \divide \HPictH 2%
\def \ContentMed{\begin{picture}(\PictW,\PictH)(-\HPictW,-\HPictH)
 \put(75,0){%
	\makebox(0,0){%
	\VOffset \HPictH \advance \VOffset -3%
	\HOffset \MedW 
		\advance \HOffset \wd144
		\advance \HOffset -\wd145 
	\divide \HOffset \unitlength \divide \HOffset 2
 \VOffset \HBoxH \advance \VOffset 0
 \Gene{\begin{picture}(\BoxW,\BoxH)(-\HOffset,-\VOffset)%(-\HOffset,-\HBoxH)%(-\HBoxW,-\HBoxH)%
	\put(-220,0) {\makebox(0,0)[r]{\box144}}
	\put(-180,-20) {\Yellow \vector(1,0){120}}
	\put(-130,30)	{\makebox(0,0){\termcolour \small $\widehat{#2}$}}
	\put(0,0)	{\makebox(0,0){${\Yellow [~]}$}}
	\put(120,30)	{\makebox(0,0){\termcolour \small $\widehat{#4}$}}
	\put(60,-20)	{\Yellow \vector(1,0){120}}
	\put(220,0)	{\makebox(0,0)[l]{\box145}}
	\end{picture}}}}
 \HOffset\HBoxW\advance\HOffset120
 \put(-\HOffset,-15) {\Yellow \vector(1,0){140}}
 \advance\HOffset-70
 \put(-\HOffset,35) {\makebox(0,0){\termcolour $#3$}}
 \end{picture}}
\ContentMed 
}
\def \Diagmed #1 #2 [#3] #4 #5 {\DiagMed{#1}{#2}{#3}{#4}{#5}}
 
\newcommand{\DiagDMed}[5]{\unitlength.09\point%
\setbox144=\hbox{{\termcolour $#1$}}\setbox145=\hbox{{\termcolour $#5$}}\setbox146=\hbox{\termcolour $#3$}%
\ifdim \wd144 > \wd145 \MedW \wd144 \else \MedW \wd145 \fi %\advance \MedW \wd146 
\advance \MedW 300\unitlength
\MedH \ht144 \advance \MedH \ht145 \advance \MedH 100\unitlength
\BoxW\MedW\divide\BoxW\unitlength\HBoxW\BoxW\divide\HBoxW2%
\BoxH\MedH\divide\BoxH\unitlength\HBoxH\BoxH\divide\HBoxH2%
\GeneWidth \MedW \advance \GeneWidth 240\unitlength
\GeneHeight \MedH \advance \GeneHeight 70\unitlength
\PictW \GeneWidth \divide \PictW \unitlength \HPictW \PictW \divide \HPictW 2%
\PictH \GeneHeight \divide \PictH \unitlength \HPictH \PictH \divide \HPictH 2%
\def \ContentDMed{\begin{picture}(\PictW,\PictH)(-\HPictW,-\HPictH)%
 \put(80,0){%
	\makebox(0,0){%
	\VOffset \HPictH \divide \VOffset 2%
	\ifdim \wd144 > \wd145 \HOffset \wd144
	\else \HOffset \wd145 \fi
	\divide \HOffset \unitlength \advance \HOffset 50%
 \Gene{\begin{picture}(\BoxW,\BoxH)(-\HOffset,-\HBoxH)
	\put(0,\VOffset){%
	\put(0,-20) {\makebox(0,0)[r]{\box144}}
	\put(40,-30) {\Yellow \vector(1,0){120}}
	\put(95,18)	{\makebox(0,0){\termcolour $\widehat{#2}$}}
	}
	\put(-\HOffset,-\VOffset){%
	\put(80,0)	{\makebox(0,0){${\Yellow [~]}$}}
	\put(175,18)	{\makebox(0,0){\termcolour $\widehat{#4}$}}
	\put(125,-30)	{\Yellow \vector(1,0){120}}
	\put(250,0)	{\makebox(0,0)[l]{\box145}}
	}
	\end{picture}}}}%
 \HOffset\HBoxW\advance\HOffset120%
 \put(-\HOffset,-20) {\Yellow \vector(1,0){140}}%
 \advance\HOffset-70%
 \put(-\HOffset,30) {\makebox(0,0){\termcolour $#3$}}%
 \end{picture}}%
 \ContentDMed 
}

\newcommand{\DiagCutA}[5]{\unitlength.09\point%
\setbox141=\hbox{\small $#1$}%
\setbox142=\hbox{\small $#4$}%
\GeneWidth \wd141 \advance \GeneWidth \wd142 \advance \GeneWidth 400\unitlength
\ifdim \ht141 > \ht142 \GeneHeight \ht141 \else \GeneHeight \ht142 \fi \advance \GeneHeight 30\unitlength
\PictW \GeneWidth \divide \PictW \unitlength \HPictW \PictW \divide \HPictW 2%
\PictH \GeneHeight \divide \PictH \unitlength \HPictH \PictH \divide \HPictH 2 \advance \PictH 20%
\HOffset \GeneWidth
\ifdim \wd142 > \wd141 \advance \HOffset -\wd142 \advance \HOffset \wd141
\else \advance \HOffset \wd141 \advance \HOffset -\wd142 \fi
\divide \HOffset \unitlength \divide \HOffset 2
\VOffset \HPictH \advance \VOffset 0
\def \ContentCut{ \begin{picture}(\PictW,\PictH)(-\HOffset,-\VOffset)
	\put(-178,0)	{\makebox(0,0)[r]{\termcolour \box141}}
	\put(-140,-20)	{\Yellow \vector(1,0){160}}
	\put(-70,32)	{\makebox(0,0){\termcolour $\widehat{#2}$}}
	\put(0,-100)	{\makebox(0,0){\mbox{\Yellow \footnotesize \sc #5}}}
	\put(70,32)	{\makebox(0,0){\termcolour $\widehat{#3}$}}
	\put(0,-20)	{\Yellow \line(1,0){140}}
	\put(180,0)	{\makebox(0,0)[l]{\termcolour \box142}}
 \end{picture}}
\Gene{ \ContentCut }
}

\newcount\tempcount
\newcommand{\XNat}[1]{\tempcount#1\raise-4\point\hbox{\DiagCap{x}{`a}}\loop{\ifnum\tempcount>0%
\,[f]\,\raise-4\point\hbox{\DiagCap{x}{`a}}%
\advance\tempcount by-1 }\repeat}

\def \typeofargs#1#2{{\It {typeof}}\ #1\ #2}
\def \typeof{\@ifnextchar\bgroup{\typeofargs}{\It {typeof}}}

\def \worra {{\leftarrow}}

\def \ftsc#1{\textsc{\scriptsize #1}}
\def \tinysc#1{\textsc{\tiny #1}}
\def \ftrm#1{\textrm{\scriptsize #1}}

\def \unifyContextsargs#1#2{\setbox31=\hbox{$#1$}\setbox32=\hbox{$#2$}
\ifdim\wd31<12\point\ifdim\wd32<12\point {\It {unifyCont}}\ #1\ #2
\else {\It {unifyCont}}\ #1\ (#2) \fi
\else \ifdim\wd32<12\point {\It {unifyCont}\ (#1)\ #2}
\else {\It {unifyCont}}\ (#1)\ (#2) \fi \fi }

\def \unifyContexts{\@ifnextchar\bgroup{\unifyContextsargs}{\It {unifyCont}}}

\def \GCut {\textsf{ Cut}}

\def\Unr#1{\textsl{Unrv}~#1}
\def \Unravel{\@ifnextchar\bgroup{\Unr}{\textsl{Unrv}}}

\def \LC	{\mbox{$\Red `l$-calculus}}
\def \PiC	{\mbox{$\Red `p$-calculus}}
\def \redPi	{\mbox{\black\,$\rightarrow_{`p}$\,}}

\def \DBs {\textsf{\small DB}}
\def \DBf#1#2#3{\DBs(#1,#2,#3)}
\def \DB {\@ifnextchar{\bgroup}{\DBf}{\DBs}}

\def \Pred[#1]{~[\skp#1\skp]}

\omtrue
\def \expT#1#2#3#4{{\typecolour \XOut{\XIn{#1}{#2}}{#3}\cdot #4}}
\def \medT#1#2#3#4#5{{\typecolour \XOut{#1}{#2}\,[#3]\,\XIn{#4}{#5}}}

\def \expT#1#2#3#4{\XOut{\XIn{#1}{#2}}{#3}\kern.2mm{`.}\kern.2mm#4}

\def \medT#1#2#3#4#5{\setbox111=\hbox{$#1$}\setbox112=\hbox{$#5$}%
\ifdim\wd111<10pt%
	\ifdim\wd112<10pt%
	\XOut{\box111}{#2}\,[#3]\,\XIn{#4}{\box112}%
	\else
	\XOut{\box111}{#2}~[#3]~\XIn{#4}{\box112}
	\fi
\else
	\XOut{\box111}{#2}~[#3]~\XIn{#4}{\box112}
\fi}

\def \cutV#1#2#3#4{\XOut{#1}{#2} \DaggerV \XIn{#3}{#4}}
\def \cutN#1#2#3#4{\XOut{#1}{#2} \DaggerN \XIn{#3}{#4}}
\def \cutleft#1#2#3#4{\XOut{#1}{#2} \DaggerL \XIn{#3}{#4}}
\def \cutright#1#2#3#4{\XOut{#1}{#2} \DaggerR \XIn{#3}{#4}}
\def \cutA#1#2#3#4{\XOut{#1}{#2}\DaggerA \XIn{#3}{#4}}
\def \negLT#1#2#3{{#1}\kern.2mm{`.}\kern.2mm\XOut{#2}{#3}}
\def \negRT#1#2#3{\XIn{#1}{#2}\kern.2mm{`.}\kern.2mm{#3}}

\def \orLT#1#2#3#4#5{{#1}\kern.2mm{`.}\kern.2mm\Group<\widehat{#2}{#3}\kern.3mm{\mid}\kern.3mm\widehat{#4}{#5}>}
\def \orRLT#1#2#3#4{{#1}\Group<\widehat{#2}\kern.2mm{\mid}\kern.2mm{#3}>\kern.2mm{`.}\kern.2mm{#4}}
\def \orRRT#1#2#3#4{{#1}\Group<{#2}\kern.2mm{\mid}\kern.2mm\widehat{#3}>\kern.2mm{`.}\kern.2mm{#4}}
\def \orRPT#1#2#3#4{{#1}\Group<\widehat{#2}\kern.2mm{\mid}\kern.2mm\widehat{#3}>\kern.2mm{`.}\kern.2mm{#4}}

\def \Caps(#1,#2){{\termcolour \slangle#1{`.}#2\srangle}}
\def \CapT#1#2{\Caps(#1,#2)}
\def \ExpT#1#2#3#4{{\termcolour \if@om \omfalse\expT{#1}{#2}{#3}{#4}\omtrue
	\else (\expT{#1}{#2}{#3}{#4}) \fi}}
\def \Exps#1#2#3.#4{{\termcolour \if@om \omfalse\expT{#1}{#2}{#3}{#4}\omtrue
	\else (\expT{#1}{#2}{#3}{#4})\fi}}
\def \MedT#1#2#3#4#5{{\termcolour \if@om \omfalse\medT{#1}{#2}{#3}{#4}{#5}\omtrue
	\else (\medT{#1}{#2}{#3}{#4}{#5})\fi}}
\def \Meddl#1#2#3#4#5{\if@om \omfalse\medTdl{#1}{#2}{#3}{#4}{#5}\omtrue
	\else (\medTdl{#1}{#2}{#3}{#4}{#5})\fi}
\def \Meds#1#2[#3]#4#5{{\termcolour \if@om \omfalse\medT{#1}{#2}{#3}{#4}{#5}\omtrue \else (\medT{#1}{#2}{#3}{#4}{#5})\fi}}
\def \CutT#1#2#3#4{{\termcolour \if@om \omfalse\cutT{#1}{#2}{#3}{#4}\omtrue
	\else	(\cutT{#1}{#2}{#3}{#4})\fi}}
\def \ColCutT#1#2#3#4{{\termcolour \if@om \omfalse\ColcutT{#1}{#2}{#3}{#4}\omtrue
	\else	(\ColcutT{#1}{#2}{#3}{#4})\fi}}
\def \CutLog#1#2#3#4{{\termcolour \if@om \omfalse\cutlog{#1}{#2}{#3}{#4}\omtrue
	\else	(\cutlog{#1}{#2}{#3}{#4})\fi}}
\def \CutLeft#1#2#3#4{{\termcolour \if@om \omfalse\cutleft{#1}{#2}{#3}{#4}\omtrue \else	(\cutleft{#1}{#2}{#3}{#4})\fi}}
\def \CutRight#1#2#3#4{{\termcolour \if@om \omfalse\cutright{#1}{#2}{#3}{#4}\omtrue \else	(\cutright{#1}{#2}{#3}{#4})\fi}}
\def \CutV#1#2#3#4{{\termcolour \if@om \omfalse\cutV{#1}{#2}{#3}{#4}\omtrue
	\else	(\cutV{#1}{#2}{#3}{#4} )\fi}}
\def \CutN#1#2#3#4{{\termcolour \if@om \omfalse\cutN{#1}{#2}{#3}{#4}\omtrue
	\else	(\cutN{#1}{#2}{#3}{#4} )\fi}}
\def \CutA#1#2#3#4{{\termcolour \if@om \omfalse\cutA{#1}{#2}{#3}{#4}\omtrue
	\else	(\cutA{#1}{#2}{#3}{#4})\fi}}

\def \NegL#1#2#3{{\termcolour \if@om \omfalse\negLT{#1}{#2}{#3}\omtrue
	\else	(\negLT{#1}{#2}{#3})\fi}}
\def \NegR#1#2#3{{\termcolour \if@om \omfalse\negRT{#1}{#2}{#3}\omtrue
	\else	(\negRT{#1}{#2}{#3})\fi}}
\def \AndL#1#2#3#4{{\termcolour \if@om \omfalse \andLPT{#1}{#2}{#3}{#4} \omtrue \else \skp (\andLPT{#1}{#2}{#3}{#4}) \skp \fi}}
\def \AndLL#1#2#3#4{{\termcolour \if@om \omfalse \andLLT{#1}{#2}{#3}{#4} \omtrue \else \skp (\andLLT{#1}{#2}{#3}{#4}) \skp\fi}}
\def \AndLR#1#2#3#4{{\termcolour \if@om \omfalse\andLRT{#1}{#2}{#3}{#4}\omtrue	\else	(\andLRT{#1}{#2}{#3}{#4})\fi}}
\def \AndLP#1#2#3#4{{\termcolour \if@om \omfalse\andLPT{#1}{#2}{#3}{#4}\omtrue	\else	(\andLPT{#1}{#2}{#3}{#4})\fi}}
\def \AndR#1#2#3#4#5{{\termcolour \if@om \omfalse\andRT{#1}{#2}{#3}{#4}{#5}\omtrue	\else	(\andRT{#1}{#2}{#3}{#4}{#5})\fi}}
\def \OrL#1#2#3#4#5{{\termcolour \if@om \omfalse\orLT{#1}{#2}{#3}{#4}{#5}\omtrue	\else	(\orLT{#1}{#2}{#3}{#4}{#5})\fi}}
\def \OrRL#1#2#3#4{{\termcolour \if@om \omfalse\orRLT{#1}{#2}{#3}{#4}\omtrue	\else	(\orRLT{#1}{#2}{#3}{#4})\fi}}
\def \OrRR#1#2#3#4{{\termcolour \if@om \omfalse\orRRT{#1}{#2}{#3}{#4}\omtrue	\else	(\orRRT{#1}{#2}{#3}{#4})\fi}}
\def \OrRP#1#2#3#4{{\termcolour \if@om \omfalse\orRPT{#1}{#2}{#3}{#4}\omtrue	\else	(\orRPT{#1}{#2}{#3}{#4})\fi}}

\def \NewConR#1#2#3#4{{\termcolour \if@om \omfalse\newconRT{#1}{#2}{#3}{#4}\omtrue	\else	(\newconRT{#1}{#2}{#3}{#4})\fi}}
\def \NewConL#1#2#3#4#5{{\termcolour \if@om \omfalse\newconLT{#1}{#2}{#3}{#4}{#5}\omtrue	\else	(\newconLT{#1}{#2}{#3}{#4}{#5})\fi}}

\def \caps<#1,#2>{\Cap{#1}{#2}}
\def \med #1 #2 [#3] #4 #5 {\Med{#1}{#2}{#3}{#4}{#5}}
\def \imp #1 #2 [#3] #4 #5 {\Med{#1}{#2}{#3}{#4}{#5}}
\def \impdl #1 #2 [#3] #4 #5 {\Meddl{#1}{#2}{#3}{#4}{#5}}
\def \exp #1 #2 #3 . #4 {\Exp{#1}{#2}{#3}{#4}}
\def \expl #1 #2 #3 #4 . #5 {{#1}:\Exp{#2}{#3}{#4}{#5}}
\def \cut #1 #2 + #3 #4 {\Cut{#1}{#2}{#3}{#4}}
\def \cutdl #1 #2 + #3 #4 {\Cutdl{#1}{#2}{#3}{#4}}
\def \cutDl #1 #2 + #3 #4 {\CutDl{#1}{#2}{#3}{#4}}
\def \cutLog #1 #2 + #3 #4 {\CutLog{#1}{#2}{#3}{#4}}
\def \cutL #1 #2 + #3 #4 {\CutL{#1}{#2}{#3}{#4}}
\def \cutR #1 #2 + #3 #4 {\CutR{#1}{#2}{#3}{#4}}
\def \colcut #1 #2 + #3 #4 {\ColCut{#1}{#2}{#3}{#4}}
\def \colcutR #1 #2 + #3 #4 {\ColCut{#1}{#2}{#3}{#4}}
\def \colcutL #1 #2 + #3 #4 {\ColCut{#1}{#2}{#3}{#4}}
\def \negL #1 . #2 #3 {\NegL{#1}{#2}{#3}}
\def \negR #1 #2 . #3 {\NegR{#1}{#2}{#3}}

\def \andLL #1 . < #2 , #3 > #4 {\AndLL{#1}{#2}{\nonconnectable}{#4}}
\def \andLP #1 . < #2 , #3 > #4 {\AndLP{#1}{#2}{#3}{#4}}
\def \andLR #1 . < #2 , #3 > #4 {\AndLR{#1}{\nonconnectable}{#3}{#4}}
\def \andLLaux #1 . < #2 , {\@ifnextchar{o}{\andLL #1 . < #2 , }{\andLP #1 . < #2 , }}
\def \andL #1 . < {\@ifnextchar{o}{\andLR {#1} . < }{\andLLaux {#1} . < }}
\def \andLL#1#2#3#4{{#1}\kern.2mm{`.}\kern.2mm\Group<\widehat{#2},{#3}>{#4}}
\def \andLR#1#2#3#4{{#1}\kern.2mm{`.}\kern.2mm\Group<{#2},\widehat{#3}>{#4}}
\def \andLLT#1#2#3#4{{#1}\kern.2mm{`.}\kern.2mm\Group<\widehat{#2},{#3}>{#4}}
\def \andLRT#1#2#3#4{{#1}\kern.2mm{`.}\kern.2mm\Group<{#2},\widehat{#3}>{#4}}
\def \andLPT#1#2#3#4{{#1}\kern.2mm{`.}\kern.2mm\Group<\widehat{#2},\widehat{#3}>{#4}}
\def \andRT#1#2#3#4#5{\Group<{#1}\widehat{#2},{#3}\widehat{#4}>\kern.2mm{`.}\kern.2mm{#5}}
\def \andRrule{\Conj\textit{R}}
\def \andRlog < #1 #2 , #3 #4 > . #5 {\AndR{#1}{#2}{#3}{#4}{#5}}
\def \andR{\@ifnextchar<{\andRlog}{\andRrule}}
\def \invL #1 . #2 #3 {\InvLterm{#1}{#2}{#3}}
\def \invR #1 #2 . #3 {\InvRterm{#1}{#2}{#3}}

\def \orR #1 < {\@ifnextchar{o}{\orRR {#1} < }{\orRLaux {#1} < }}
\def \orRLaux #1 < #2 | {\@ifnextchar{o}{\orRL #1 < #2 | }{\orRP #1 < #2 | }}
\def \orRL #1 < #2 | #3 > . #4 {\OrRL{#1}{#2}{\nonconnectable}{#4}}
\def \orRR #1 < #2 | #3 > . #4 {\OrRR{#1}{\nonconnectable}{#3}{#4}}
\def \orRP #1 < #2 | #3 > . #4 {\OrRP{#1}{#2}{#3}{#4}}
\def \projLT#1#2#3#4{{#1}\kern.2mm{`.}\kern.2mm\Group<\widehat{#2},{#3}>{#4}}
\def \projLR #1 . < #2 , #3 > #4 {\AndL{#1}{#2}{#3}{#4} }
\def \projL #1 . < #2 , #3 > #4 {\AndLL{#1}{#2}{\nonconnectable}{#4}}
\def \projLaux #1 . < #2 , {\@ifnextchar{o}{\projL {#1} . < {#2} , }{\projLR {#1} . < {#2} , }}
\def \projR #1 . < #2 , #3 > #4 {\AndLR{#1}{\nonconnectable}{#3}{#4}}
\def \proj #1 . < {\@ifnextchar{o}{\projR {#1} . < }{\projLaux {#1} . < }}
\def \pair < #1 #2 , #3 #4 > . #5 {\AndR{#1}{#2}{#3}{#4}{#5}}
\def \inP #1 < {\@ifnextchar{o}{\inR {#1} < }{\inLaux {#1} < }}
\def \inLaux #1 < #2 | {\@ifnextchar{o}{\inL #1 < #2 | }{\inRP #1 < #2 | }}
\def \inL #1 < #2 | #3 > . #4 {\OrRL{#1}{#2}{\nonconnectable}{#4}}
\def \inR #1 < #2 | #3 > . #4 {\OrRR{#1}{\nonconnectable}{#3}{#4}}
\def \inRP #1 < #2 | #3 > . #4 {\OrRP{#1}{#2}{#3}{#4}}
\def \choice #1 . < #2 #3 | #4 #5 > {\OrL{#1}{#2}{#3}{#4}{#5}}

\def \QED{%
 \unskip
 \hfill 
 \begingroup
 \unitlength\point
 \linethickness{.5\point}%
 \framebox(7,7){}%
 \endgroup
 \kern 10\point
}

 \newenvironment{AuxCases}[1]{\left \{ \def\arraystretch{1.1} \begin{array}{#1} }%}
 { \end{array} \right . }
 \renewenvironment{cases}{\@ifnextchar\bgroup{\begin{AuxCases}}{\begin{AuxCases}{@{}lclll}}}{\end{AuxCases}}

\def \Natural{%\hbox{\psset{unit=1\point,linewidth=.5\point}%
{\sf I\kern-1\point N}}

\def \mybullet{\vspace*{2mm} ~ \kern-5mm $\bullet$}

\def \import{\textsl{import}}

\makeatother

\def \ellips(#1,#2){\put (0,0)	{\psellipse[unit=\unitlength, linewidth=1.25pt](#1,#2)}}

\newskip \myitemindent \myitemindent0mm

\def \myitem[#1]{\setbox55=\hbox{#1} \item[{#1}] \leavevmode $
\kern-\wd55 \kern5.25mm \kern-\myitemindent 
 \begin{array}[t]{@{}lclclclclcl} \kern-3mm
\kern\wd55 \kern\myitemindent}

\def\renewfont#1#2{\font#1=#2\relax}

%% file: clac.bbl
\begin{thebibliography}{10}

\bibitem{Abadi-Cardelli-Curien'91}
M.~Abadi, L.~Cardelli, P.-L. Curien, and J.-J. L\'evy.
\newblock Explicit substitutions.
\newblock {\em JFP}, 1(4), 1991.

\bibitem{Abadi-Gordon'97}
M.~Abadi and A.~Gordon.
\newblock A {C}alculus for {C}ryptographic {P}rotocols: {T}he {S}pi {C}alculus.
\newblock In {\em 4th CCCS}, ACM Press, 1997.

\bibitem{Abramsky'93}
S.~Abramsky.
\newblock Computational interpretations of linear logic.
\newblock {\em TCS}, 111(1{\&}2), 1993.

\bibitem{Abramsky-Jagadeesan'94}
S.~Abramsky and R.~Jagadeesan.
\newblock Games and full completeness for multiplicative linear logic.
\newblock {\em JSL}, 59(2), 1994.

\bibitem{Abramsky'94}
S.~ Abramsky.
\newblock Proofs as processes.
\newblock {\em TCS}, 135(1), 1994.

\bibitem{Ariola-Herbelin'03}
Z.~M. Ariola and H.~Herbelin.
\newblock Minimal classical logic and control operators.
\newblock In {\em ICALP'03}, {\em LNCS} 2719, 2003.

\bibitem{vBLL'05}
S.~van Bakel, S.~Lengrand, and P.~Lescanne.
\newblock The language \x: circuits, computations and classical logic.
\newblock In {\em ICTCS'05}, {\em LNCS} 3701, 2005.

\bibitem{Bakel-Lescanne'08}
S.~van Bakel and P.~Lescanne.
\newblock Computation with classical sequents.
\newblock {\em MSCS}, 2008.

\bibitem{Barendregt'84}
H.~Barendregt.
\newblock {\em The Lambda Calculus: its Syntax and Semantics}.
\newblock North-Holland, Amsterdam, revised edition, 1984.

\bibitem{Bellin-Scott'94}
G. ~Bellin and P. J. ~Scott.
\newblock {\em On the pi-Calculus and Linear Logic}.
\newblock {\em TCS}, 135(1), 11--65, 1994. 

\bibitem{Coquand-Huet'88}
T.~Coquand and G.~Huet.
\newblock The {C}alculus of {C}onstructions.
\newblock {\em IAC}, 76(2,3), 1988.

\bibitem{Curien-Herbelin'00}
P.-L. Curien and H.~Herbelin.
\newblock The {D}uality of {C}omputation.
\newblock In {\em ICFP'00}, ACM, 2000.

\bibitem{Curry-Feys'58}
H.B. Curry and R.~Feys.
\newblock {\em Combinatory {L}ogic}, volume~1.
\newblock North-Holland, Amsterdam, 1958.

\bibitem{DeBruijn'78}
N.~G. de~Bruijn.
\newblock A namefree lambda calculus with facilities for internal definition of
 expressions and segments.
\newblock TH-Report 78-WSK-03, University of Eindhoven, 1978.

\bibitem{GentzenG'35}
G.~Gentzen.
\newblock Untersuchungen {\"u}ber das {L}ogische {S}chliessen.
\newblock {\em Math. Zeitschrift}, 39, 1935.

\bibitem{LL}
J.-Y. Girard.
\newblock Linear logic.
\newblock {\em Theoretical Computer Science}, 50:1--102, 1987.

\bibitem{Girard'86}
J.Y. Girard.
\newblock The {S}ystem {F} of {V}ariable {T}ypes, {F}ifteen years later.
\newblock {\em TCS}, 45, 1986.

\bibitem{Girard'91}
J.-Y. Girard.
\newblock A new constrcutive logic: classical logic.
\newblock {\em Mathematical Structures in Computer Science}, 1(3):255--296,
  1991.

\bibitem{Griffin'90}
T.~Griffin.
\newblock A formulae-as-types notion of control.
\newblock In {\em POPL'90}, ACM, 1990.

\bibitem{Herbelin'95}
H.~Herbelin.
\newblock S\'equents qu'on calcule~: de l'interpr\'etation du calcul des
 s\'equents comme calcul de $\lambda$-termes et comme calcul de strat\'egies
 gagnantes.
\newblock Th\`ese d'universit\'e, Paris 7, 1995.

\bibitem{Herbelin'05}
H.~Herbelin.
\newblock C'est maintenant qu'on calcule: au c\oe ur de la dualit\'e.
\newblock M\'emoire de habilitation, Universit\'e Paris 11, D\'ecembre 2005.

\bibitem{Honda-Tokoro'91}
K.~Honda and M.~Tokoro.
\newblock An object calculus for asynchronous communication.
\newblock In {\em ECOOP'91}, LNCS 512, 133--147, 1991.

\bibitem{Honda-Yoshida'95}
K.~Honda and N.~Yoshida.
\newblock On the {R}eduction-based {P}rocess {S}emantics.
\newblock {\em TCS}, 151:437--486, 1995.

\bibitem{Honda-Yoshida-Berger'04}
K.~Honda, N.~Yoshida, and M.~Berger.
\newblock Control in the $\pi$-calculus.
\newblock In {\em CW'04}, 2004.

\bibitem{Klop'92}
J.W. Klop.
\newblock Term {R}ewriting {S}ystems.
\newblock In {\em
 Handbook of Logic in Computer Science}, volume~2, chapter~1, pages 1--116.
 Clarendon Press, 1992.

\bibitem{Milner'92}
R.~Milner.
\newblock Function as processes.
\newblock In {\em MSCS},
 2(2), 1992.

\bibitem{Milner'99}
R.~Milner.
\newblock {\em Communicating and {M}obile {S}ystems: the $\pi$-calculus}.
\newblock Cambridge {U}niversity {P}ress, 1999.

\bibitem{Parigot'92}
M.~Parigot.
\newblock An algorithmic interpretation of classical natural deduction.
\newblock In {\em LPAR'92}, {\em LNCS} 624, 1992.

%\bibitem{Sangiorgi-Walker'01}
%D.~Sangiorgi and D.~Walker.
%\newblock On barbed equivalences in the $\pi$-calculus.
%\newblock In {\em CONCUR'01}, {\em LNCS} 2154, 2001.

\bibitem{Sangiorgi-Walker'03}
D.~Sangiorgi and D.~Walker.
\newblock {\em The {P}i-{C}alculus}.
\newblock Cambridge University Press, 2003.

\bibitem{Summers'08}
A.J. Summers.
\newblock Extending lambda-mu with first class continuations.
\newblock Manuscript, 2007.

\bibitem{Thielecke'97}
H.~Thielecke.
\newblock {\em Categorical {S}tructure of {C}ontinuation {P}assing {S}tyle}.
\newblock PhD thesis, University of Edinburgh, 1997.

\bibitem{Urban'00}
C.~ Urban.
\newblock {\em Classical Logic and Computation}.
\newblock PhD thesis, University of Cambridge, 2000.

\bibitem{Urban'01}
C~Urban.
\newblock Strong {N}ormalisation for a {G}entzen-like {C}ut-{E}limination
 {P}rocedure'.
\newblock In {\em TLCA'01}, {\em LNCS} 2044, 2001.

\bibitem{Urban-Bierman'01}
C.~Urban and G.~M. Bierman.
\newblock Strong normalisation of cut-elimination in classical logic.
\newblock {\em FI}, 45(1,2), 2001.

\bibitem{WadlerDual}
P.~ Wadler.
\newblock {C}all-by-{V}alue is {D}ual to {C}all-by-{N}ame.
\newblock In {\em ICFP'03}, ACM, 2003.

\bibitem{Zunic'07}
D.~\v{Z}uni\'c.
\newblock Computing with {S}equents and {D}iagrams in {C}lassical {L}ogic - {C}alculi $\astX$, $\diagX$, and $\copyX$.
\newblock PhD thesis, ENS Lyon, 2007.

\end{thebibliography}
